\documentclass[twoside,openright,12pt]{CUEDthesisPSnPDF}
\usepackage{graphicx}
\usepackage{amsmath,amssymb}
\usepackage{amscd,amsfonts,amsthm}
\usepackage{mathtools}
\usepackage[pdftex,bookmarks]{hyperref}
\usepackage{float}
\usepackage{times}
\usepackage{comment}
\usepackage{color}
\usepackage{t1enc}
\usepackage{longtable}
\usepackage{xspace}
\usepackage{psfrag}
\usepackage{latexsym,pstricks,rotating,enumerate}
\usepackage{wrapfig}
\usepackage{color}
\usepackage{framed,xcolor}
\usepackage{caption}
\usepackage{tikz}
\usepackage{appendix}
\usepackage{ulem}
\usepackage[numbered]{matlab-prettifier}
\usetikzlibrary{calc,positioning,fit,backgrounds}
\pgfdeclarelayer{background}
\pgfsetlayers{background,main}
\usepackage{watermark}
\newcommand{\ie}{\textit{i.e.~}}
\newcommand{\eg}{\textit{e.g.~}}

\begin{document}
\renewcommand\baselinestretch{1.2}
\baselineskip=18pt plus1pt
\setcounter{secnumdepth}{3}
\setcounter{tocdepth}{3}
\frontmatter

\thispagestyle{empty}
\baselineskip=18pt
\begin{center}
{\Large \bf Generation and Detection of Quantum Correlations and Entanglement on a Spin-Based Quantum Information Processor} \\
\vspace*{1cm}
{\large{\bf Thesis}} \\
\vspace*{0.5cm}
{For the award of the degree of}\\
\vspace{0.5cm}
{\large{\bf DOCTOR OF  PHILOSOPHY}} \\
\vspace{0.25cm}
\end{center}
\vspace*{3cm}
\begin{tabular}{lp{6cm}l}
{{\it Supervised by:}}  &&
{{\it Submitted by:}} \\
\\
{\bf Prof. Kavita Dorai} &&
{\bf Amandeep Singh} \\
{\bf Prof. Arvind} &&\\

\end{tabular}
\begin{center}
\vspace*{1.5cm}
\hspace*{0cm}
\end{center}
\vspace*{-1cm}
\begin{center}
\includegraphics[scale=0.24]{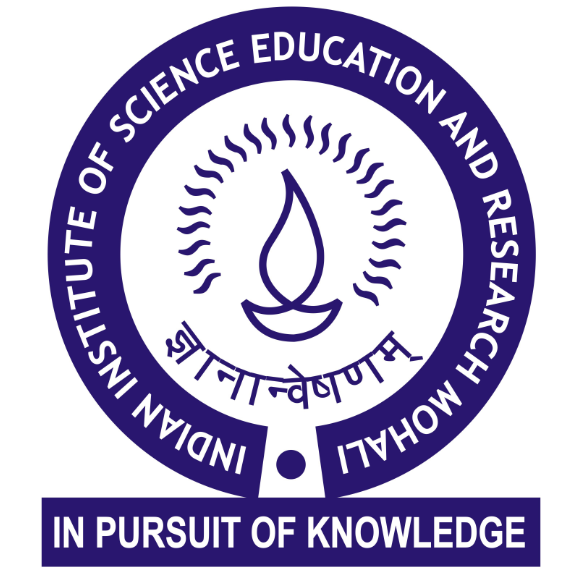}\\
\vspace*{1.5cm}
{\bf Indian Institute of Science Education \& Research Mohali\\
Mohali - 140 306\\
India\\ (February 2019)
 }
\end{center}

\newpage
\thispagestyle{empty}
\begin{center}
\end{center}
\thispagestyle{empty}
\centerline{\Large \bf Declaration}
\vspace{1cm}
\par 
\noindent The work presented in this thesis has been carried out by me under the guidance of
Prof. Kavita Dorai and Prof. Arvind at the Indian Institute of Science Education and Research Mohali.\\

\noindent This work has not been submitted in part or in full for a degree, diploma or a fellowship to any other University or Institute. Whenever contributions of others are involved, every effort has been made to indicate this clearly, with due acknowledgment of collaborative research and discussions. This thesis is a bonafide record of original work done by me and all sources listed within have been detailed in the bibliography.

\vspace*{1in}

\hspace*{-0.25in}
\parbox{8in}{
\noindent {{\bf Amandeep Singh}}
}

\vspace*{0.125in}

\hspace*{-0.25in}
\noindent
\parbox{2.5in}{
\noindent Place~:  \\
\noindent Date~:
}

\vspace*{0.5in}

\noindent
In our capacity as supervisors of the candidate's PhD thesis work, we certify that the above statements by the candidate are true to the best of our knowledge.

\vspace*{2.5cm}

\hspace*{-0.25in}
\parbox{8in}{
\noindent {{\bf Dr. Kavita Dorai} \hspace*{5.5cm}  {\bf Dr. Arvind}} \\
\noindent Professor of Physics \hspace*{5.1cm} Professor of Physics\\
\noindent Department of Physical Sciences \hspace*{3cm} Department of Physical Sciences\\
\noindent IISER Mohali \hspace*{6.2cm}  IISER Mohali
}
\vspace*{0.125in}

\hspace*{-0.25in}
\noindent
\parbox{8in}{
\noindent Place~:  \hspace*{7.4cm} Place~: \\
\noindent Date~:   \hspace*{7.55cm} Date~:
}


\newpage
\thispagestyle{empty}

\begin{center}
\end{center}
\thispagestyle{empty}
\centerline{\Large \bf Acknowledgments}
\vspace{10pt}
First and foremost I would like to express my sincere gratitude to my thesis supervisors Prof. Kavita Dorai and Prof. Arvind for their relentless support, guidance and motivation throughout the course of PhD. I am, and will always be, short of words to express my thanks for their resolute encouragement. More than teaching and guiding, on various projects in this thesis, they inculcated the attitude to understand, formulate and accomplish any given task for which I'll always be thankful to them.\\

I would also like to thank my PhD doctoral committee member Prof. Ramandeep Singh Johal for his continued support and guidance. I am also grateful to all my excellent PhD course instructors Prof. Jasjeet Singh Bagla, Prof. Sudeshna Sinha, Dr. Rajeev Kapri, Dr. Abhishek Chaudhuri and Head of the Physics Department Dr. Sanjeev Kumar for all the support they have given during and after the course work.\\

I am also thankful to all the members of journal club meetings of QCQI and NMR research groups. I have benefited a lot from the discussions of these meeting. I would like to thank my group members Dr. Shruti Dogra, Dr. Harpreet Singh, Dr. Satnam Singh, Dr. Navdeep Gogna, Rakesh Sharma, Jyotsana Ojha, Akshay Gaikwad and Dileep Singh. I am also grateful to Dr. Gopal Verma, Dr. Archana Sangwan, Varinder Singh and Dr. Kavita Mehlawat for their cheerful company. I cherish the memories of discussions during routine tea breaks with Dr. Raju Nanda, Sumit Mishra and Akanksha Gautam.\\

I owe special thank to scientific officer Dr. Paramdeep Singh Chandi for his help and support with software troubleshooting. I am also thankful to Bruker application engineer Dr. Bhawani Shanker Joshi for his help in solving problems and rectifying errors during NMR spectrometer operations. I am thankful to NMR lab scientific staff Mr. Balbir Singh for his generous support.\\

I would like to thankfully acknowledge all the support provided by the NMR research facility, IISER Mohali to carry out my experimental work. I also owe thank to IISER Mohali for the research fellowship as well as financial support to attend an international conference. I am also thankful to Perimeter Institute for Theoretical Physics, Institute for Quantum Computing and the German Physical Society for the travel grants and financial support to attend conferences in Canada and Germany.\\

And finally, last but by no means least, I would like to express my heartfelt gratitude to my family for their continuous encouragement and belief in me!

\baselineskip=15pt

\vspace{1cm}

\rightline{\bf \large{Amandeep Singh}}

\chapter{Abstract}
This thesis focuses on the experimental creation and detection of different types of quantum correlations using nuclear magnetic resonance (NMR) hardware. The idea of encoding computational problems into physical quantum system and then harnessing the quantum evolution to perform information processing is at the core of quantum computing. Quantum entanglement is a striking feature exhibited by composite quantum systems which has no classical analog. It has been shown that quantum entanglement is a key resource to achieve computational speedup in quantum information processing and for quantum communication related tasks.  Creation and detection of such correlations experimentally is a major thrust area in experimental quantum computing. Main goals of the studies undertaken in this thesis were to design experimental strategies to detect the entanglement in a `state-independent' way and with fewer experimental resources. Experimental schemes have been devised which enables the measurement of desired observable with high accuracy and these schemes were utilized in all the investigations. Experimental protocols were successfully implemented to detect the entanglement of random two-qubit states. Further, the schemes for the experimental detection as well as classification of generic and general three-qubit pure states have also been devised and implemented successfully. Detection of quantum correlations possessed by mixed separable states, bound-entanglement for states of $2\otimes4$ systems and non-local nature of quantum systems were also investigated. In all the investigations, results were verified by one or more alternative ways \eg full quantum state tomography, quantum discord, negativity and \textit{n}-tangle.\\
\noindent Content of the thesis has been distributed in
seven chapters and the chapter-wise abstract is as follows.
\subsubsection*{Chapter 1}
This chapter briefly introduces the field of quantum
computation followed by the main features of NMR quantum
processor architecture. Latter part of the chapter describes
the theory of entanglement detection and experimental
realization on various hardware. Chapter concludes with
goals and motivations for the work undertaken in this
thesis. 
\subsubsection*{Chapter 2} This chapter focuses on the
entanglement detection of random two-qubit states. Random
local measurements have recently been proposed to construct
entanglement witnesses and thereby detect the presence of
bipartite entanglement. We experimentally demonstrate the
efficacy of one such scheme on a two-qubit NMR
quantum-information processor. We show that a set of three
random local measurements suffices to detect the
entanglement of a general two-qubit state. We experimentally
generate states with different amounts of entanglement and
show that the scheme is able to clearly witness
entanglement. We perform complete quantum state tomography
for each state and compute state fidelity to validate our
results. Further, we extend previous results and perform a
simulation using random local measurements to optimally
detect bipartite entanglement in a hybrid system of 2$
\otimes $3 dimensionality.
\subsubsection*{Chapter 3} In this chapter the focus is on a
more general kind of quantum correlation possessed by
separable states. A bipartite quantum system in a mixed
state can exhibit non-classical correlations, which can go
beyond quantum entanglement.  While quantum discord is the
standard measure of quantifying such general quantum
correlations, the non-classicality can be determined by
simpler means via the measurement of witness operators.  We
experimentally construct a positive map to witness
non-classicality of two-qubits in an NMR system.  The map
can be decomposed  in terms of measurable spin magnetization
so that a single run of an experiment on an ensemble of
spins suffices to detect the non-classicality in the state,
if present.  We let the state evolve in time and use the map
to detect non-classicality as a function of time.  To
evaluate the efficacy of the witness operator as a means to
detect non-classicality, quantum discord was measured by
performing full quantum state tomography at each time
instant and obtain a fairly good match between the two
methods.
\subsubsection*{Chapter 4} This chapter details the
experimental detection of the entanglement present in
arbitrary three-qubit pure quantum states on an NMR quantum
information processor. Measurements of only four observables
suffice to experimentally differentiate between the six
classes of states which are inequivalent under stochastic
local operation and classical communication (SLOCC).  The
experimental realization is achieved by mapping the desired
observables onto Pauli $z$-operators of a single qubit,
which is directly amenable to measurement. The detection
scheme is applied to known entangled states as well as to
states randomly generated using a generic scheme that can
construct all possible three-qubit states.  The results are
substantiated via direct full quantum state tomography as
well as via negativity calculations and the comparison
suggests that the protocol is indeed successful in detecting
tripartite entanglement without requiring any {\it a priori}
information about the states.
\subsubsection*{Chapter 5} This chapter details the
experimental creation and characterization of a class of
qubit-ququart PPT (positive under partial transpose)
entangled states using three nuclear spins on an NMR quantum
information processor. Entanglement detection and
characterization for systems with a Hilbert space dimension
$ > 2 \otimes 3$ is nontrivial since there are states in
such systems which are both PPT as well as entangled.  The
experimental detection scheme that we employed for the
detection of  this qubit-ququart PPT entanglement was based
on the measurement of  three Pauli operators.  The class of
states considered, in the current study, is an incoherent
mixture of five pure states. Measuring three Pauli
operators, with high precision using our recently devised
method, is crucial to detect entanglement. All the five
states were prepared with high fidelities and the resulting
PPT entangled states were prepared with mean fidelity $ \geq
$ 0.944 using temporal averaging technique.
\subsubsection*{Chapter 6} This chapter presents the
experimental investigations of non-local nature of quantum
correlations possessed by multipartite quantum states. It
has been shown that fewer body correlations can reveal the
non-local nature of the correlations arising from quantum
mechanical description of the nature. Such tests on the
correlations can be transformed to a semi-definite-program
(SDP). This study presents the experimental implementation
of Navascu\'es-Pironio-Ac\'{\i}n (NPA) hierarchy on NMR
hardware utilizing three nuclear spins. The protocol has
been tested on two types of genuine tripartite entangled
states. In both the cases the experimentally measured
correlations were used to formulate the SDP under linear
constraints on the entries of the moment matrix. It has been
observed that in both the cases SDP failed to find a
semi-definite-positive moment matrix consistent with the
experimental data which is indeed the signature that the
observed correlations can not arise from local measurements
on a separable state and hence are non-local in nature. This
also confirms that both the states under test are indeed
entangled. Results were verified by direct full quantum
state tomography in each case.
\subsubsection*{Chapter 7} This chapter summarizes the
results of all the projects constituting this thesis, and
the key findings, with possible future directions of work.
\thispagestyle{empty}
\pagenumbering{roman}
\setcounter{page}{1}
\thispagestyle{empty}

\begin{center}
\end{center}
\thispagestyle{empty}
{\LARGE \bf List of Publications}
\vspace{1cm}
\vspace*{12pt}
\begin{enumerate}
\addtolength{\itemsep}{12pt} 

\item \textbf{Amandeep Singh}, Arvind and Kavita Dorai. \textit{ Entanglement detection on an NMR quantum information processor using random local measurements},   \href{http://journals.aps.org/pra/abstract/10.1103/PhysRevA.94.062309}{\rm Phys. Rev. A \textbf{94}, 062309 (2016)}.
	
	\item \textbf{Amandeep Singh}, Arvind and Kavita Dorai. \textit{Witnessing nonclassical correlations via a single-shot experiment on an ensemble of spins using NMR}, \href{http://journals.aps.org/pra/abstract/10.1103/PhysRevA.95.062318}{\rm Phys. Rev. A \textbf{95}, 062318 (2017)}.
	
	\item Akshay Gaikwad, Diksha Rehal, \textbf{Amandeep Singh}, Arvind and Kavita Dorai. \textit{Experimental demonstration of selective quantum process tomography on an NMR quantum information processor}, \href{https://journals.aps.org/pra/abstract/10.1103/PhysRevA.97.022311}{\rm Phys. Rev. A \textbf{97}, 022311 (2018)}.
	
	\item \textbf{Amandeep Singh}, Harpreet Singh, Kavita Dorai and Arvind. \textit{ Experimental Classification of Entanglement in Arbitrary Three-Qubit Pure States on an NMR Quantum Information Processor}, \href{https://journals.aps.org/pra/abstract/10.1103/PhysRevA.98.032301}{\rm Phys. Rev. A \textbf{98}, 032301 (2018)}.

\item \textbf{Amandeep Singh}, Kavita Dorai and Arvind. \textit{ Experimentally identifying the entanglement class of pure tripartite states},  \href{https://doi.org/10.1007/s11128-018-2105-5}{\rm Quant. Info. Proc. \textbf{17}, 334 (2018)}.

\item \textbf{Amandeep Singh}, Akanksha Gautam, Kavita Dorai and Arvind. \textit{Experimental Detection of Qubit-Ququart Pseudo-Bound Entanglement using Three Nuclear Spins}, \href{https://doi.org/10.1016/j.physleta.2019.02.027}{\rm Phys. Lett. A \textbf{383}(14), 1549-1554 (2019)}.

\item \textbf{Amandeep Singh}, Kavita Dorai and Arvind. \textit{Detection of Non-local Quantum Correlations via Experimental Implementation of NPA Hierarchy on NMR}, (Manuscript in preparation)
\end{enumerate}

\thispagestyle{empty}

\tableofcontents
\listoffigures
\newpage
\thispagestyle{empty}
\pagebreak
\listoftables

\begin{center}
\end{center}
\chapter{Abbreviations used in the Thesis}
\begin{table}[H]
\begin{center}
\begin{tabular}{l l l }
BE    &:& Bound Entangled \\
BS    &:& Bi-Separable \\
CCNR  &:& Computable Cross Norm or Realignment \\
CP    &:& Completely Positive \\
CHSH  &:& Clauser-Horne-Shimony-Holt \\
EM    &:& Entanglement Measure(s) \\
EPR   &:& Einstein-Podolsky-Rosen \\
EW    &:& Entanglement Witness \\
FID   &:& Free Induction Decay \\
FT    &:& Fourier Transform \\
GHZ   &:& Greenberger-Horne-Zeilinger \\
LOCC  &:& Local Operations and Classical Communications \\
NCC   &:& Non Classical Correlations \\
NMR   &:& Nuclear Magnetic Resonance \\
NMRQC &:& NMR Quantum Computing \\
NPT   &:& Negative under Partial Transpose \\
PCC   &:& Properly Classically Correlated \\
PNCP  &:& Positive but Not Completely Positive\\
POVM  &:& Positive Operator Valued Measure\\
PPT   &:& Positive under Partial Transpose \\
PT    &:& Partial Transposition \\
QC    &:& Quantum Computing \\
QD    &:& Quantum Discord \\
QIP   &:& Quantum Information Processing \\
QST   &:& Quantum State Tomography \\
RF    &:& Radio Frequency \\
SDP   &:& Semi Definite Program/Programming \\
SLOCC &:& Stochastic LOCC \\
\end{tabular}
\end{center}
\end{table}
\newpage
\thispagestyle{empty}
\pagebreak
\setcounter{page}{1}
\thispagestyle{empty}
\mainmatter
\chapter{Introduction}
\begin{figure}
\begin{center}
\includegraphics[scale=0.6]{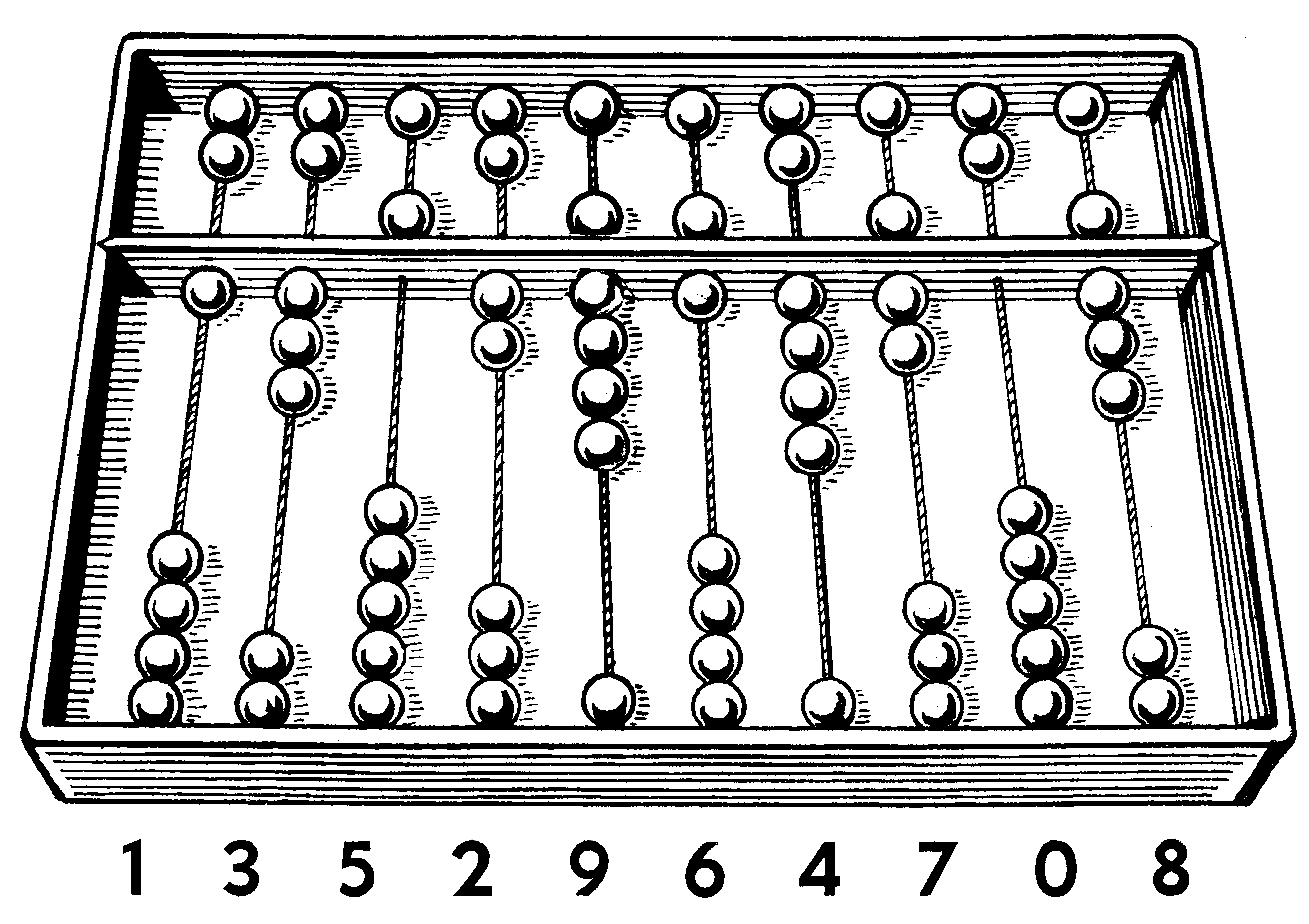}
\caption{Abacus: The First Computer \cite{firkin-online-13}.}
\label{abacus}
\end{center}
\end{figure}

\begin{figure}
\begin{center}
\includegraphics[scale=1.1]{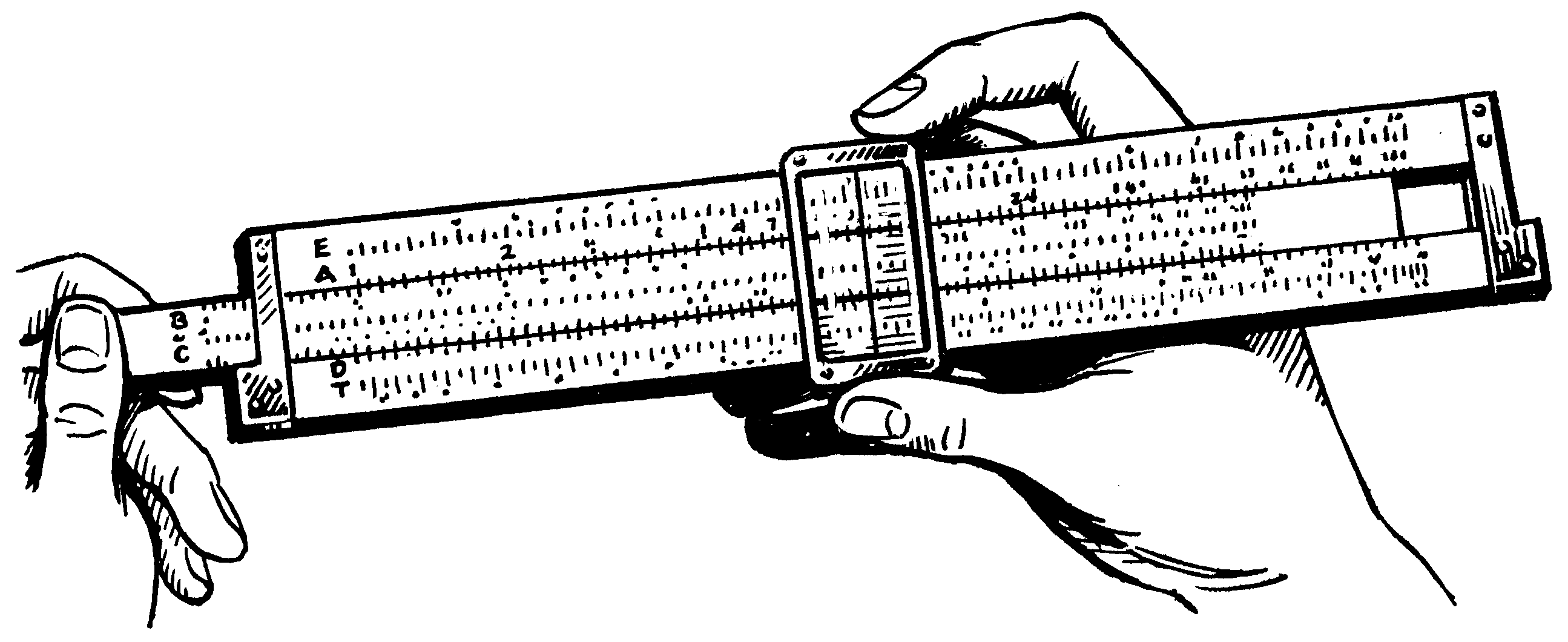}
\caption{The Slide Ruler used in 17$^\mathrm{th}$ century AD for Multiplication \cite{georgebog-online-10}.}
\label{sliderule}
\end{center}
\end{figure}

Being an intelligent species, modern human beings, \ie \textit{Homosapiens}, (meaning a `wise man' in Latin), started using the first computing machine named \textit{the `Abacus'}. Records show that the earliest users of abacus were the Sumerians and the Egyptians back in 2000 BC. The principle is as: a frame holding a series of rods, with ten sliding beads on each. When all the beads had been slid across the first rod, it was time to move one across on the next, showing the number of tens, and thence to the next rod, showing hundreds, and so on (with the ten beads on the initial row returned to the original position), (Fig.(\ref{abacus})). 
That is where technology of computation was stuck for nearly 3600 years until the beginning of the 17$^\mathrm{th}$ century AD, when mechanical calculators started appearing in Europe. Most notably, after John Napier invented logarithms, and Edmund Gunter created the logarithmic scales (lines, or rules) upon which slide rules are based, it was Oughtred who first used two such scales sliding by one another to perform direct multiplication and division; and he is credited as the inventor of the slide rule in 1622, Fig.(\ref{sliderule}). William Oughtred (1574-1660) was an English mathematician born in Eton and he was the person who introduced the symbol `$\times$' for multiplication as well as $sin$ and $cos$ for trigonometric functions \textit{Sine} and \textit{Cosine}, respectively. The slide rule is basically a sliding stick that uses logarithmic scales to allow rapid multiplication and division, (Fig.(\ref{sliderule})). Slide rules evolved to allow advanced trigonometry and logarithms, exponential and square roots.

\begin{figure}
\begin{center}
\includegraphics[scale=0.4]{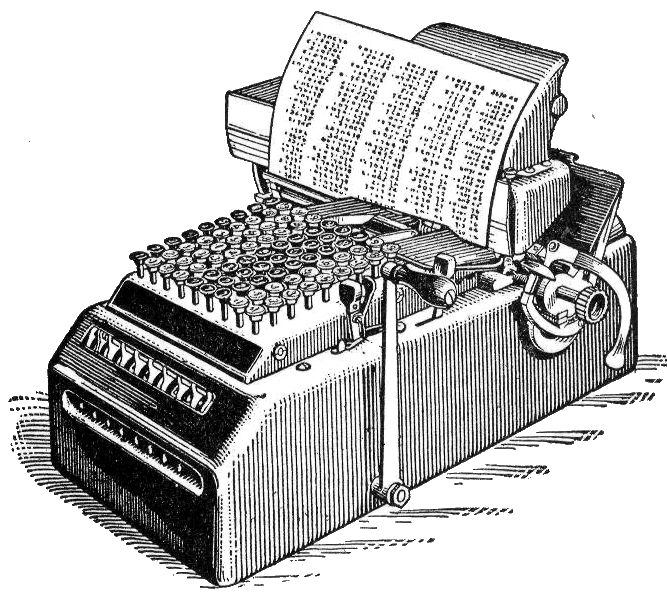}
\caption{Comptometer: The Modern Day Calculator \cite{comptometer-online-06}.}
\label{comptometer}
\end{center}
\end{figure}

A further step forward in computation occurred in 1887 when Dorr. E. Felt's US-patented key driven 'Comptometer' took calculating into the push button age, (Fig(\ref{comptometer})).
Further, the story of electronic calculator began in 1930 when world was preparing for war. Such calculator were much in demand due calculation of trajectories of bombs. During the Second World War, the challenges of code-breaking produced the first all-electronic computer, \textit{Colossus}. But this was a specialized machine that basically performed ``exclusive or'' (XOR) Boolean algorithms. With the advent of semiconductor-devices, in the mid of the 20$^\mathrm{th}$ century, the first generation of modern days computers came into the existence which was a huge leap as compared the huge-inefficient computation machines based on hundred of thermionic valves. Since then the power of computing machines grows exponentially following the then proposed Moore's law. The law was described as early as 1965 by the Intel co-founder Gordon E. Moore after whom it is named \cite{moore-e-65}.\\

There have been a large number of classical algorithms/problems posed which seem unsolvable on these computing machines although the computational power was expanding at an exponential rate. In computer science, the computational complexity of an algorithm is usually ascribed to the time complexity which estimates the time to run the algorithm as a function of order of input strings. Order of the input string is generally denoted by $n$ while the complexity of the computation is represented represented using big $O$. For example,  an algorithm with time complexity $O(n)$ is a linear time algorithm, $O(n^{\alpha})$ with $\alpha>1$ represent a polynomial time algorithm while $O(2^{p(n)})$ represents the time complexity of an exponential time algorithm with $p(n)$ being polynomial of order $n$. Two extreme cases are the constant time (a sub-class of polynomial time) and exponential time, termed as EXPTIME, complexity classes. It is understandable that if an algorithm has its time complexity in EXPTIME class then in order to run such an algorithm the required time scales exponentially with the size of input string $n$. One may not wish to wait too long, \eg few years, to run an algorithm. Some of EXPTIME class of problems are the prime factorization problem, optimization problems with large number of variables, matrix chain multiplication via brute-force search and simulation of quantum systems using classical models. Then in the early eighties Feynman proposed the idea of exploring quantum systems to simulate quantum systems \cite{feynman-ijtp-82}. The idea of encoding computational problems into physical quantum system and then harnessing the quantum evolution to perform information processing is at the core of quantum computing and quantum information (QCQI) processing \cite{nielsen-book-02}. There are various features, \eg  quantum superposition and quantum entanglement, which enable computation utilizing quantum systems to outperform any classical computing machine. These concepts will be briefly introduced in subsequent sections after introducing the basic ideas of QCQI.
\section{Quantum Computing and Quantum Information Processing}
Classical information processing is solely realized by encoding the bit-strings into the classical states of physical systems \cite{cover-book-91}. For example high or low voltages in a digital electronic circuit are used to represent the classical Boolean states `0' or `1', respectively or light or no-light can be used to encode `0' or `1' in an optical computer. Most of the digital information processing hardware available today performs classical information processing by encoding the problems into binary strings and performing logical operations governed by Boolean algebra \cite{boole-book-09}. Classical physical systems can be either in `0' or `1' state at an instant of time, and this limits the information processing achievable on such classical information processors. On the other hand the states of quantum physical systems can exist in superposition of `0' and `1'. allowing new possibilities for computation \cite{nielsen-book-02}. Later subsection of this chapter will address the issue of physical realization of such quantum states.
There a striking similarity between what a classical computer does and how a physical system evolves. A computer performs a computation on some input bit string under certain logical operations to yield the output. Analogously, a physical system evolves from an initial state following the laws of motion to give the final state. The idea of simulating classical as well as quantum systems by encoding the problem as an initial state of the quantum system was put forward by Feynman\cite{feynman-ijtp-82}. Computation can be achieved by the quantum evolution and the results get encoded in the final state of the quantum system which can be read. This was radically a new way of performing the computation.

		\subsection{The Quantum Bit}
\begin{figure}
\begin{center}
\includegraphics[scale=0.6]{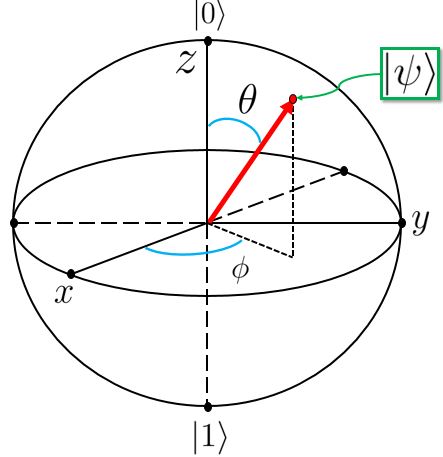}
\caption{The Bloch-sphere representation of a single quantum-bit.}
\label{blochrep}
\end{center}
\end{figure}

		Information can be encoded in the physical state of a quantum system and the minimum dimension of the involved Hilbert space, to represent the states of such system, is two. Such encoding can be achieved by using a two-level quantum system, \eg a spin-half system, generally termed as a quantum-bit or \textit{Qubit}. The two eigenstates of such two-level quantum systems representing the logical states `0' and `1' are $\vert 0 \rangle$ and $\vert 1 \rangle$ which represent the eigenvectors $\begin{bmatrix} 1 \\
		0 
		\end{bmatrix}
		$ and $\begin{bmatrix} 0 \\
		1 
		\end{bmatrix}
		$ respectively. The most general state $\vert \psi \rangle$ of a qubit can be a superposition of basis vectors and can assume the polar form as
	 	\begin{equation}\label{1qubitstate}
		\vert \psi \rangle=cos\left(\frac{\theta}{2}\right) \vert 0 \rangle + e^{i\phi}sin\left(\frac{\theta}{2}\right) \vert 1 \rangle
		\end{equation}
		where the global phase is ignored in writing the above polar form as it does not have any observable effect during quantum evolution or on measurement outcomes. The most contrasting feature of the qubit from a classical bit is that a qubit can simultaneously exist in the basis states $\vert 0 \rangle$ and $\vert 1 \rangle$ and this quantum parallelism of quantum systems gives them tremendous computational power which a classical computer may never match \cite{nielsen-book-02}. Each value of the pair  ($\theta$,\;$\phi$) represents a valid quantum state on the surface of a three-dimensional unit radius sphere, shown in Fig.(\ref{blochrep}), called the Bloch sphere. The radius of the sphere is indeed related to the quantity  $(\vert a_0 \vert ^2 + \vert a_1 \vert ^2)$ which is unit due to normalization of state vector $\vert \psi \rangle$. One can observe that the North ($\theta=0$) and South ($\theta=\pi$) poles of the Bloch sphere represent the basis states $\vert 0 \rangle$ and $\vert 1 \rangle$ respectively.\\
				
		Similarly one may have multiqubit quantum register. For the general case of an $N$-qubit quantum register, the basis vectors of Hilbert space have dimension $2^N$ and can be obtained from the tensor product of individual qubit states as 
		\begin{equation}\label{nqubitsep}
		\vert \Psi \rangle= \vert \psi_1 \rangle \otimes \vert \psi_2 \rangle\otimes... \otimes \vert \psi_N \rangle
		\end{equation}
		It will be seen later that the most general $N$-qubit state can be put in the form
		\begin{equation} \label{nqubitstate}
		\vert \Psi \rangle= \sum_{i=1}^{2^N} \alpha_i \vert \alpha_i \rangle 
		\end{equation}
		here $\vert \alpha_i \rangle$ is a $N$-qubit quantum register of form $\vert b^i_1b^i_2.....b^i_N \rangle$ with $b_j\in [0,1]$ and $\sum_i \vert \alpha_i \vert ^2=1$.  There exist multiqubit states which can be cast in the form of Eq.(\ref{nqubitstate}) and may not assume the form given by Eq.(\ref{nqubitsep}). Such states are called entangled states and they play a major role in QCQI. 
		\subsection{The Density Matrix Formalism}
		As discussed in the previous section, the state of a quantum system has one-to-one correspondence with the vectors in Hilbert space. Consider a quantum system prepared in a state $\vert \psi_1 \rangle$ and we have $n_1$ such systems constituting a pure ensemble. Similarly consider another ensemble composed of $ n_2 $ quantum systems, each of which is in the state $\vert \psi_2 \rangle $. If one mixes these two ensembles then how can one write the quantum state of the resulting ensemble? The total number of quantum systems are $N_T=n_1+n_2$. Another important question is, if we now pick a quantum system from this ensemble and measure it, what is the result? There are two probabilities for the action: (i) The probability with which the chosen quantum system can come from ensemble $\vert \psi_1 \rangle$ or $\vert \psi_2 \rangle$ \ie $p_1=p=\frac{n_1}{N_T}$ and $p_2=1-p=\frac{n_2}{N_T}$ and (ii) the probability with which, the chosen quantum system after measurement collapses to $\vert 0 \rangle$ or $\vert 1 \rangle$. One thing is clear that the state description of form Eq.(\ref{nqubitstate}) is not appropriate for this situation \ie such an ensemble can not be represented by vectors in a Hilbert space.\\
		
		It has been shown that a more suitable state representation is the density operator formalism \cite{reif-book-65,sakurai-book-94,nielsen-book-02,oliveira-book-07}. For a pure state the density operator can be written as
		\begin{equation}\label{pure_ch1}
		\rho=\vert \psi \rangle \langle \psi \vert
		\end{equation}
		In order to write the density matrix one may choose a set of orthogonal basis vectors ${\vert b_i \rangle}$ and the matrix elements can be computed as $\rho_{ij}=\langle b_i \vert \rho \vert b_j \rangle$. It can be shown that
		\begin{equation}
		\sum_i\rho_{ii}=\sum_i\langle b_i \vert \rho \vert b_i \rangle=1
		\end{equation}
		This will lead to the condition Tr$(\rho)=1$ independent of chosen basis. For pure states $\rho^2=\vert \psi \rangle \langle \psi \vert.\vert \psi \rangle \langle \psi \vert=\vert \psi \rangle \langle \psi \vert=\rho$ and hence Tr$(\rho^2)=$Tr$(\rho)=1$. For a mixed ensemble, the density operator of the ensemble can be written as
		\begin{equation}
		\rho_{\mathrm{ensemble}}=p\vert \psi_1 \rangle \langle \psi_1 \vert+(1-p)\vert \psi_2 \rangle \langle \psi_2 \vert
		\end{equation}
		and this correctly incorporates both the probabilities mentioned earlier. The most general density operator for a single-qubit system can be written as
		\begin{equation}\label{mixedrho}
		\rho=\sum_j p_j\vert \psi_j \rangle \langle \psi_j \vert; \;\;\;\;\;\; p_j\geq 0; \;\;\;\;\;\; \sum_j p_j=1
		\end{equation}

		It may be noted that one-qubit mixed states \ie states with Tr$(\rho^2)<1$ can be represented with points inside the Bloch sphere in Fig.(\ref{blochrep}) and the center of the sphere represents the maximally mixed state
		
		\begin{equation}
		\rho^{}_{\mathrm{mixed}_{max}}=\frac{\mathbb{I}}{2^1}=\begin{pmatrix} 1/2 & 0 \\
		0 & 1/2 \end{pmatrix}
		\end{equation}
		For a general $N$-qubit state the condition $\frac{1}{2^N} \leq \mathrm{Tr}(\rho^2) \leq 1$ is valid. The lower and upper bounds on Tr$(\rho^2)$ are achieved by maximally mixed and pure states respectively.

		\subsection{Quantum Evolution}
		In quantum mechanics there are broadly two kinds of time evolution (a) the continuous time evolution of the states of a closed quantum system as governed by Schr{\"o}dinger equation and (b) a discontinuous time evolution during a quantum measurement following Born's rule for the probabilities of the possible outcomes. Following subsections briefly describe both of these time evolutions. A detailed description of quantum evolution is given in Refs. \cite{shankar-book-80,sakurai-book-94}.

		\subsubsection{Continuous Time Evolution: Unitary Evolution}
\textbf{Closed quantum system:}	The continuous time evolution of closed quantum systems is unitary \ie the time evolution operator can be represented by a unitary matrix $U$ obeying $UU^{\dagger}=U^{\dagger}U=\mathbb{I}$. Equivalently one can say that the state of a quantum system at time `$t_o$' transforms to the state at a later time `$t_o+t$' via unitary transformation as
		\begin{equation}
		\vert \psi ' (t_o+t)\rangle =U\vert \psi(t_o) \rangle 
		\end{equation}
		For a closed quantum system, described by the Hamiltonian $\mathit{H}$, the equation for state evolution can be written as 
		\begin{equation}\label{TDSEq}
		i\hbar\frac{\partial\vert \psi \rangle}{\partial t}=\mathit{H}\vert \psi \rangle
		\end{equation}
		With minimal assumptions and considering the case of a time independent Hamiltonian one can solve the above differential equation. The result leads to a time evolution operator of form $U\sim e^{-i\mathit{H} t/\hbar}$. Important point to note here is that, once $\mathit{H}$ is defined and the resulting $U$ is obtained, the state $\vert \psi ' (t+t_o)\rangle$ at all later times evolves continuously in a predictable fashion via $U$. The advantage of the unitary evolution is that such evolutions are always reversible.\\
		
\noindent \textbf{Open quantum system:} There can be situation that a system under consideration interacts with its environment and usually termed as open quantum systems. Such a composite system can be assumed to be in a separable state $\rho_{comp}=\rho_{sys}\otimes\rho_{env}$. There can be three possible energy operators in this scenario (i) system Hamiltonian, $\textit{H}_{sys}$, (ii) environment Hamiltonian, $\textit{H}_{env}$, and (iii) the interaction Hamiltonian, $\textit{H}_{int}$ due to the interaction between system and its environment. Hence the Hamiltonian of the composite system can be written as
\begin{equation}
\textit{H}_{comp}=\textit{H}_{sys}+\textit{H}_{env}+\textit{H}_{int}
\end{equation}
The time evolution of density operator can be obtained from time-dependent Schr{\"o}dinger Eq.(\ref{TDSEq}) as follows:
\begin{equation}\label{TDSE_ch1}
\frac{\partial\vert \psi \rangle}{\partial t}=-\frac{i}{\hbar}\mathit{H}\vert \psi \rangle \;\;\; \Leftrightarrow \;\;\; \frac{\partial \langle \psi \vert}{\partial t}=\frac{i}{\hbar}\langle \psi \vert\mathit{H}
\end{equation}
For a pure state of form Eq.(\ref{pure_ch1})
\begin{eqnarray}
\frac{\partial\rho}{\partial t}&=&\frac{\partial \left[ \vert \psi \rangle\langle \psi \vert \right]}{\partial t}\nonumber\\
&=& \left[\frac{\partial\vert \psi \rangle}{\partial t}\right] \langle \psi \vert + \vert \psi \rangle \frac{\partial\langle \psi \vert}{\partial t} 
\end{eqnarray}
and on using Eq.(\ref{TDSE_ch1})
\begin{eqnarray}
\frac{\partial\rho}{\partial t}&=&-\frac{i}{\hbar}\mathit{H}\vert \psi \rangle\langle \psi \vert + \frac{i}{\hbar}\vert \psi \rangle\langle \psi \vert\mathit{H}\nonumber\\
\frac{\partial\rho}{\partial t}&=& -\frac{i}{\hbar}\left[\mathit{H},\rho\right]
\end{eqnarray}		
Above is the Liouville-Von Neumann equation. Although, the above equation is derived using pure state density operator but it can be shown that it is also valid for mixed states.\\
For the composite system $\rho_{comp}$ the equation of motion can be written as
\begin{equation}
\frac{\partial\rho_{comp}(t)}{\partial t} = -\frac{i}{\hbar}\left[\mathit{H}_{comp},\rho_{comp}(t)\right]
\end{equation}
The solution to the above equation is of form $\rho_{comp}(t)=U(t)\rho_{comp}(0)U^{\dagger}(t)$. It is worth mentioning here that the unitary time evolution operator $U(t)$ acts on the composite system $\rho_{comp}$ and one may be interested in the evolution of the system	state  $\rho_{sys}(t)$ only. The way out is that one may trace out the environment to get $\rho_{sys}(t)=\mathrm{Tr}_{env}(\rho_{comp})$. Evolution of $\rho_{sys}(t)$ can  formally be derived using Lindblad master equation formalism and we will not expand on this here. Another important aspect here is that although the evolution of $\rho_{comp}$ is unitary but the state of the system $\rho_{sys}$ may evolve non-unitarily and irreversibly.
		
		\subsubsection{Discontinuous Time Evolution: Quantum Measurement}
		During a quantum measurement process, the state of a quantum system, \eg Eq.(\ref{nqubitstate}), abruptly collapses to one of the eigenstates, of the observable being measured, in an unpredictable way. Quantum measurement is typically described by a set of measurement operators $\{ M_m \}$. Here the index `$m$' is the label of the measurement outcome after quantum measurement $M$. If the state of the system before measurement is $\vert \psi \rangle$ then the probability of getting the measurement outcome $m$, \ie $p(m)$, is given by
		\begin{equation}
		p(m)=\langle \psi \vert M^{\dagger}_m M_m \vert \psi \rangle
		\end{equation}
		while the renormalized state after obtaining the measurement outcome $m$ can be written as 
		\begin{equation}
		\frac{M_m\vert \psi \rangle}{\sqrt{\langle \psi \vert M^{\dagger}_m M_m \vert \psi \rangle}}
		\end{equation}
		The sum of probabilities \ie $\sum_m p(m)=1$, is equivalently represented by the condition that all the measurement operators sum to identity and usually referred as \textit{The Completeness Condition}:
		\begin{equation}
		\sum_m M^{\dagger}_m M_m =\mathbb{I}
		\end{equation}
		One of the commonly used measurement basis is the computation basis $\{ \vert 0 \rangle,\;\vert 1 \rangle \}$ and in such a scenario $M_0=\vert 0 \rangle \langle 0 \vert$ and $M_1=\vert 1 \rangle \langle 1 \vert$. Hence if one measure the observable \eg $\sigma_z$ on a quantum state given by Eq.(\ref{1qubitstate}), the probability of getting `0' or `1' is given by $p(0)=\langle \psi \vert M^{\dagger}_0 M_0 \vert \psi \rangle=\langle \psi \vert  M_0 \vert \psi \rangle=\vert a_0 \vert ^2$ and $p(1)=\langle \psi \vert M^{\dagger}_1 M_1 \vert \psi \rangle=\langle \psi \vert  M_1 \vert \psi \rangle=\vert a_1 \vert ^2$ respectively. The most unsettling thing here is that there is \textit{no-way} to predict that after a measurement in which eigenstate the quantum system will collapse to! The only thing quantum mechanics predicts is the probability with which a quantum system, after measurement, will collapse to a certain eigenstate of the observable being measured. This is the standard measurement scenario in quantum mechanics and is one of the postulates of the theory. However in functional analysis of quantum measurement theory, the quantum measurements are usually associated with a positive-operator valued measures (POVMs). POVMs are positive operator on Hilbert space and for a given measurement they sum to identity. Projective measurements on a large system \ie, measurements that are performed mathematically by a projection-valued measure (PVM) will act on a subsystem in ways that cannot be described by a PVM on the subsystem alone, the POVM formalism becomes necessary.

\subsection{Expectation Values}
		In quantum formalism, every observable is represented by a Hermitian operator, say $\mathcal{A}$. One may be interested in writing the average of an observable resulted from a large number of measurements of such an observable on a state, say $\vert \psi \rangle$, and can be written as
		\begin{equation}\label{expecvalue}
		\langle \mathcal{A} \rangle=\langle \psi \vert \mathcal{A} \vert \psi \rangle
		\end{equation}		 
One may choose some orthonormal basis $\lbrace \vert \alpha_i \rangle \rbrace $ to expand the state $\vert \psi \rangle$ as $\vert \psi \rangle =\sum_i \alpha_i \vert \alpha_i \rangle $. On using this expansion, Eq.(\ref{expecvalue}) yields
		\begin{eqnarray}\label{expecvalue1}
		\langle \mathcal{A} \rangle=\langle \psi \vert \mathcal{A} \vert \psi \rangle &=& \left( \alpha_1^{\ast} \langle \alpha_1 \vert+\alpha_2^{\ast} \langle \alpha_2 \vert+... \right)\mathcal{A}\left( \alpha_1^{} \vert \alpha_1 \rangle+\alpha_2^{} \vert \alpha_2 \rangle+... \right)\nonumber\\
		&=&\sum_{i,j}\alpha^{\ast}_i\alpha^{}_j\langle \alpha^{}_i \vert \mathcal{A} \vert \alpha^{}_j \rangle\nonumber\\
		&=&\sum_{i,j} \alpha^{\ast}_i\alpha^{}_j \mathcal{A}_{ij}		
		\end{eqnarray}
Here $\mathcal{A}_{ij}$	is the matrix representation of the Hermitian operator $\mathcal{A}$ in the basis $\lbrace \vert \alpha_i \rangle \rbrace $ and the expansion coefficients can be obtained as $\alpha_i=\langle \alpha_i \vert \psi \rangle$ and hence
\begin{equation}
\alpha^{\ast}_i\alpha^{}_j=\langle \psi \vert \alpha_i \rangle \langle \alpha_j \vert \psi \rangle=\langle \alpha_j  \vert \psi \rangle \langle \psi \vert \alpha_i \rangle=\langle \alpha_j \vert  \rho \vert \alpha_i \rangle
\end{equation}
Using above expression, Eq.(\ref{expecvalue1}) further takes the form
		\begin{eqnarray}
		\langle \mathcal{A} \rangle &=&\sum_{i,j} \alpha^{\ast}_i\alpha^{}_j \mathcal{A}_{ij}=\sum_{i,j} \langle \alpha_j  \vert \psi \rangle \langle \psi \vert \alpha_i \rangle \mathcal{A}_{ij}=\sum_{i,j} \langle \alpha_j \vert  \rho \vert \alpha_i \rangle \mathcal{A}_{ij}\nonumber\\
		&=& \sum_{i,j}\langle \alpha_j \vert  \rho \vert \alpha_i \rangle \langle \alpha^{}_i \vert \mathcal{A} \vert \alpha^{}_j \rangle=\sum_{j}\langle \alpha_j \vert  \rho \left\lbrace \sum_{i} \vert \alpha_i \rangle \langle \alpha^{}_i \vert \right\rbrace \mathcal{A} \vert \alpha^{}_j \rangle\nonumber\\
		&=&	\sum_{j}\langle \alpha_j \vert  \rho \mathcal{A} \vert \alpha^{}_j \rangle=\mathrm{Tr}(\rho\mathcal{A})
		\end{eqnarray}
while the fact that an orthonormal basis follows the completeness property, \ie
\linebreak
$\sum_{i} \vert \alpha_i \rangle \langle \alpha^{}_i \vert=\mathbb{I}$, is used in writing the last line of the above equation. Similarly, for a mixed state of form Eq.(\ref{mixedrho}), it can also be shown that $\langle \mathcal{A} \rangle=\mathrm{Tr}(\rho\mathcal{A})$.
		\subsection{Quantum Gates}
		In classical computation, there are logical gates to realize the logical Boolean operations \eg OR, AND and NOT gate. There are other gates, composed of the three basic logic gates, \eg Exclusive-OR (XOR), NOR, NAND, bubbled-AND ($\sim$ OR) and bubbled-OR ($\sim$ AND) gates. \textit{Universal logic gates} are those gates from which any arbitrary logic gate can be realized. There are many universal gates available to achieve arbitrary Boolean logic operations, with NAND and NOR being common examples. Most of the multi-bit gates are irreversible in nature \ie given the output of the logic gate one may not always predict the input with certainty. Analogously, similar logic gates can be constructed using quantum systems, via unitary evolution. Due to the unitary nature, in principle, one can always retrieve the input state of the quantum system if the output state after gate implementation is known. Similar to universal gates in classical computation there are universal quantum gates. It was shown that a set of gates that consists of all one-bit quantum gates $[U(2)]$ and the two-bit exclusive-OR gate which maps Boolean values $(x,y)$ to $(x,x\oplus y)$, is universal in the sense that all unitary operations on arbitrarily many bits $n$ $[U(2^n)]$ can be expressed as compositions of one-bit and two-bit XOR quantum unitary gates \cite{barenco-pra-95}. 
		
		\subsubsection*{The Pauli $X$-Gate}
		The Pauli $X$-gate or NOT gate is the quantum analog of the classical NOT gate. This single-qubit gate inverts the state of the logical qubit. The matrix representation of NOT gate is
		\begin{equation}
		X=\begin{bmatrix} 0 & 1\\
		1 & 0 \end{bmatrix}
		\end{equation}
		One may observe the action of the NOT gate on the single qubit state of Eq.(\ref{1qubitstate}) as
		\begin{equation}
		X\vert \psi \rangle=\begin{bmatrix} 0 & 1\\
		1 & 0 \end{bmatrix}
		\begin{bmatrix} a_0\\
		a_1 \end{bmatrix} = \begin{bmatrix} a_1\\
		a_0 \end{bmatrix} 
		\end{equation}
		Hence the resulting state is $(a_0\vert 1 \rangle+a_1\vert 0 \rangle)$ which on comparison with Eq.(\ref{1qubitstate}) clearly reflects the inverting effect of the NOT gate.

		\subsubsection*{The Pauli $Y$-Gate and $Z$-Gate}
		Pauli $Y$ and $Z$ gates have the following matrix representation.
		\begin{equation}
		Y=\begin{bmatrix} 0 & -1\\
		1 & 0 \end{bmatrix} \;\;\;\;\;\;\; Z=\begin{bmatrix} 1 & 0\\
		0 & -1 \end{bmatrix}
		\end{equation}
		The Pauli $Z$-gate acting on $\vert \psi \rangle$ results in $(a_0\vert 0 \rangle-a_1\vert 1 \rangle)$ while the action of $Y$-gate yields $(-a_0\vert 1 \rangle+a_1\vert 0 \rangle)$. Hence the $Z$-gate introduces a relative phase between the basis states while the $Y$-gate introduces a relative phase as well as inverts the basis states. 
		\subsubsection*{The Hadamard Gate} Another very important single-qubit quantum gate is the Hadamard gate or $ H $-gate. The matrix representation of $H$-gate is 
		\begin{equation}
		H=\frac{1}{\sqrt{2}}\begin{bmatrix} 1 & 1\\
		1 & -1 \end{bmatrix}
		\end{equation}
		The action of $H$-gate on the computational basis states produces an equal superposition of all the basis states \eg
		\begin{equation}
		H\vert 0 \rangle = \frac{\vert 0 \rangle + \vert 1 \rangle}{\sqrt{2}}\;\;\;\;\; \mathrm{and} \;\;\;\;\; H\vert 1 \rangle = \frac{\vert 0 \rangle - \vert 1 \rangle}{\sqrt{2}}
		\end{equation}
		\subsubsection*{The Most General Single Qubit Gate}
		The action of any arbitrary single qubit can be simulated using only three unitary operations parametrized by four real numbers $\alpha,\;\beta,\;\gamma$ and $\delta$ as following \cite{nielsen-book-02}
		\begin{equation}
		U(2)=e^{i\alpha}\begin{bmatrix} e^{-i\beta/2} & 0\\
		0 & e^{i\beta/2} \end{bmatrix}\begin{bmatrix} cos(\frac{\gamma}{2}) & -sin(\frac{\gamma}{2})\\
		sin(\frac{\gamma}{2}) & cos(\frac{\gamma}{2}) \end{bmatrix}\begin{bmatrix} e^{-i\delta/2} & 0\\
		0 & e^{i\delta/2} \end{bmatrix}
		\end{equation}
		Here the parameter $\alpha$ introduces a global phase and has no observable effect on the state of the quantum system.

		\subsubsection*{The Controlled Not Gate}
		The controlled-$\mathrm{NOT}$ or $\mathrm{CNOT}$ gate is a two-qubit quantum gate. Action of this gate is to perform a $\mathrm{NOT}$ operation on the target qubit depending upon the logical state of the control qubit. The matrix representation of the $\mathrm{CNOT}$ gate is:
		\begin{equation}
		\mathrm{CNOT}=\begin{bmatrix} 1 & 0 & 0 & 0 \\
	0 & 1 & 0 & 0 \\
		0 & 0 & 0 & 1 \\
		0 & 0 & 1 & 0 \end{bmatrix}
		\end{equation}
		One can verify that the action of $\mathrm{CNOT}$ gate on two-qubit basis states, $\vert 0 \rangle \otimes \vert 0 \rangle \sim \vert 00 \rangle$ and $\vert 01 \rangle$, is to leave them unaltered while interconvert the basis states $\vert 10 \rangle$ and $\vert 11 \rangle$. Another example of a two-qubit gate is the SWAP gate which maps the state $\vert 01 \rangle \Leftrightarrow \vert 10 \rangle $ onto each other while leaving the states $\vert 00 \rangle$ and $\vert 11 \rangle$ unaltered. It can be shown that the SWAP gate can be achieved using three $\mathrm{CNOT}$ gates as $\mathrm{SWAP\equiv CNOT_{12}.}$ $\mathrm{CNOT_{21}.CNOT_{12}}$, where $\mathrm{CNOT}_{ij}$ represents a $\mathrm{CNOT}$ gate with $i$ being the control qubit and $j$ being the target.

\subsection{Quantum Computing}
Quantum computing is a radical way of computation that utilizes the quantum mechanical phenomena, such as entanglement and superposition, for computation. Classical computation, in principle, can be thought of in terms of circuits built from universal logic gates and one can analogously construct quantum circuit using various quantum gates to process quantum information and hence quantum computation. Information can be encoded in the physical states of the quantum systems. The requirement to perform quantum computation is to implement the quantum gates efficiently. For the physical realization of quantum computation there are certain benchmarks laid down \cite{DiVincenzo-fdp-00} and are briefly discussed as follows.
	
\subsubsection{The DiVincenzo Criterion} \label{DiVincenzo_criterion}
 The DiVincenzo criterion were devised as a set of requirements for any physical realization of a quantum information processor \cite{DiVincenzo-fdp-00}. 
\begin{itemize}
\item[(i)] A scalable physical system with well characterized qubits.
\item[(ii)] The ability to initialize the state of the qubits to a fiducial state, such as $\vert 000... \rangle$.
\item[(iii)] Long relevant decoherence times, much longer than the gate operation time.
\item[(iv)] A ``universal'' set of quantum gates.
\item[(v)] A qubit-specific measurement capability.
\end{itemize}
There are two additional requirements specifically on the physical realization of quantum information processor to be utilized for quantum communications and quantum cryptography.
\begin{itemize}
\item[(vi)] The ability to interconvert stationary and flying qubits and
\item[(vii)] The ability to faithfully transmit flying qubits between specified locations.
\end{itemize}

\subsubsection{Physical Realization}
For the physical implementation of a quantum information processor, various hardware have been tried. None of them satisfies DiVincenzo criterion	completely. Following is the list of a few physically realized quantum processor.
\begin{itemize}
\item[$\bullet$] Superconducting Josephson junctions as qubits \cite{clarke-nature-08}.

\item[$\bullet$] Quantum dot computer, spin-based: qubit given by the spin states of trapped electrons \cite{imamoglu-prl-99}.

\item[$\bullet$] Quantum dot computer, spatial-based: qubit given by electron position in double quantum dot \cite{fedichkin-qcc-00}.

\item[$\bullet$] Optical lattices: qubit implemented by internal states of neutral atoms trapped in an optical lattice.

\item[$\bullet$] Trapped ion quantum computer: qubit implemented by the internal state of trapped ions.

\item[$\bullet$] Coupled Quantum Wire: qubit implemented by a pair of Quantum Wires coupled by a quantum point contact \cite{bertoni-prl-00}.

\item[$\bullet$] Nuclear magnetic resonance quantum computer: (NMRQC) implemented with the nuclear magnetic resonance of molecules in solution, where qubits are provided by nuclear spins \cite{cory-pnas-97,chuang-sc-97,vandersypen-rmp-05}.

\item[$\bullet$] Solid-state NMR quantum computers: qubit realized by the nuclear spin state of phosphorus donors in silicon.

\item[$\bullet$] Diamond-based quantum computer: qubit realized by the electronic or nuclear spin of nitrogen-vacancy centers in diamond \cite{neumann-sc-08}.

\item[$\bullet$] Fullerene-based ESR quantum computer: qubit based on the electronic spin of atoms or molecules encased in fullerenes.

\item[$\bullet$] Cavity quantum electrodynamics: qubit provided by the state of trapped atoms coupled to high-finesse cavities.

\item[$\bullet$] Molecular magnet: qubit given by spin states \cite{leuenberger-nature-01}.

\item[$\bullet$] Electrons-on-helium quantum computers: qubit is the electron spin.

\item[$\bullet$] Linear optical quantum computer: qubits realized by processing states of different modes of light through linear elements \eg mirrors, beam splitters and phase shifters \cite{knill-nature-01}.

\item[$\bullet$] Bose-Einstein condensate-based quantum computer.

\end{itemize}

In this thesis the physical realization of the quantum information processor is achieved utilizing nuclear spins of a molecule on an NMR hardware.

\section{Basics of NMR Spectroscopy}\label{nmrbasics}
\begin{figure}
\begin{flushleft}
\includegraphics[scale=0.5]{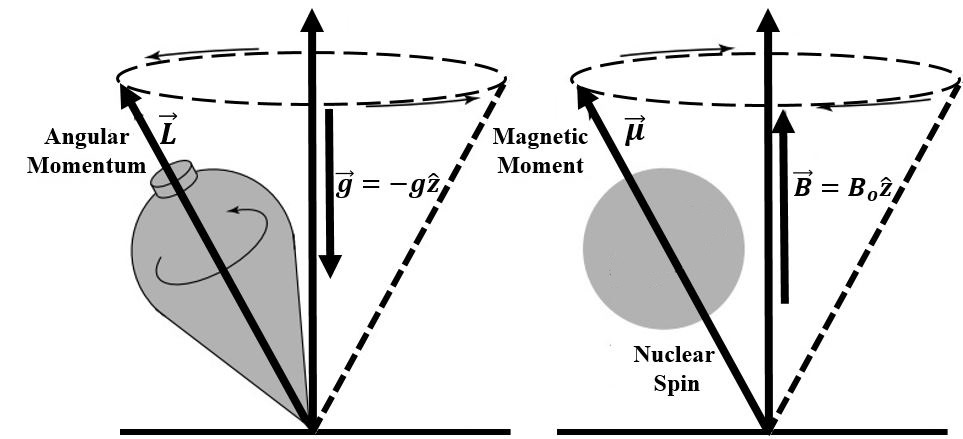}
\caption{The precession of a top spinning in the gravitational field analogous to the nuclear spin precession in a magnetic field.}
\label{lar}
\end{flushleft}
\end{figure}

\begin{figure}[t]
\begin{center}
\includegraphics[scale=0.4]{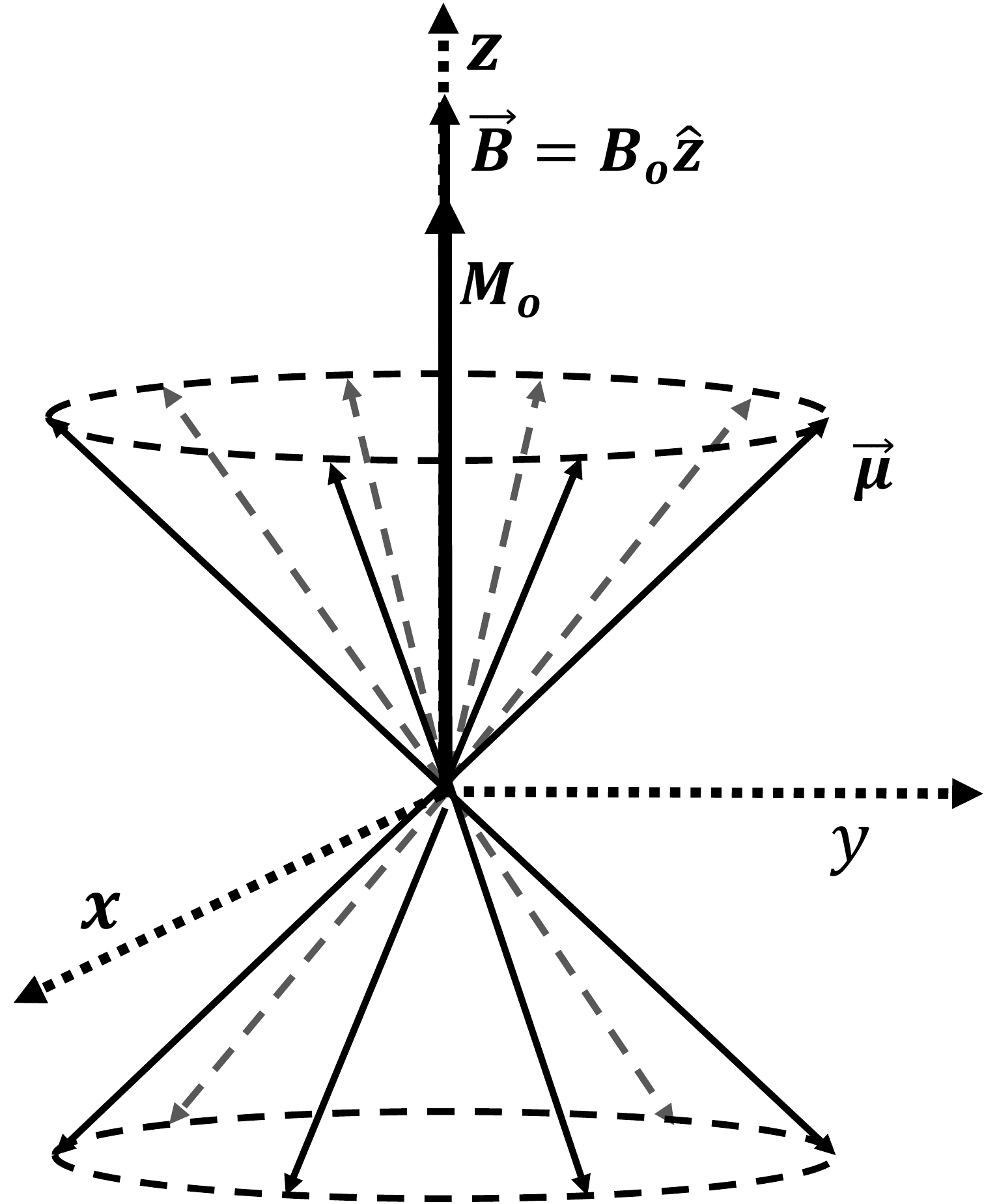}
\caption{In an external magnetic field, more spins will be precessing around the direction parallel to the field than against it. This imbalance creates a macroscopic magnetization which points in the direction of the field.}
\label{z_mag}
\end{center}
\end{figure}

The magnetic properties of an atomic nucleus forms the basis of NMR spectroscopy. A nuclear spin as well as magnetic moment are quantized \ie they exhibit discrete values when measured. The total spin angular momentum of the nucleus is the vector sum of all the spin and orbital angular momentum, of the constituting nucleons, and in general complex to compute \cite{krane-book-88}. Similar to the electronic configuration used to distribute electrons in atomic shells one can utilize the nuclear shell model to arrange the nucleons (\ie protons and neutrons) in nuclear energy levels \cite{gapon-dn-32}. From the knowledge of the nucleon arrangement in the nucleus one can obtain the nuclear spin angular momentum $ \hbar \mathbf{I} $. In addition to nuclear spin, the nuclei also possess magnetic moment $\mathbf{\mu}$ which is related to the spin angular momentum as
\begin{equation}
\mathbf{\mu} = \gamma \hbar \mathbf{I}
\end{equation} 
where $ \gamma $ is the \textit{nuclear gyromagnetic ratio} and is a characteristic feature of the nucleus. For an intuitive picture, such nuclei can be considered as tiny magnets and interact with external magnetic fields in quite a similar way. The Hamiltonian for a magnetic moment $ \mathbf{\mu} $ placed in a magnetic field applied in $z$-direction, \ie $ \mathbf{B}=B_o\hat{z} $ can be written as 
\begin{equation}\label{hamil_Ch1}
\mathit{H}=-\mathbf{\mu}.\mathbf{B}=-\gamma \hbar \mathbf{I}.\mathbf{B}=-\gamma \hbar B_o I_z= -\hbar \omega_L I_z
\end{equation}
$I_z$ being the $z$-component of the spin angular momentum. The nuclear spin experiences a torque due to the interactions of magnetic moment with the $ B_o $ and  dictates the precession of the spin angular momentum, about $\hat{z}$, at a characteristic \textit{Larmor frequency} \ie $ \omega_L = \gamma B_o $. z-direction is defined by $ \mathbf{B} $ and all the operators acts on the vector space spanned by $ \vert m \rangle $ where $m=-I,\;-I+1,...,0,...,\;I-1,\;I$ is the magnetic spin quantum number. The Cartesian components of the spin angular momentum in the transverse direction \ie $\langle I_x \rangle$ and $\langle I_y \rangle$ exhibit oscillatory motion with frequency $\omega_L$ while longitudinal component, \ie $\langle I_z \rangle$, stays stationary \cite{oliveira-book-07}. In this sense the nuclear magnetic moment is analogous to the classical magnetic dipole. See Fig.(\ref{lar}) for the analogy \cite{oliveira-book-07}.
The energy eigenvalues of the Hamiltonian, Eq.(\ref{hamil_Ch1}), acting on the state space $\vert m \rangle$ can be computed as
\begin{equation}\label{zeeman}
\mathit{H}\vert m \rangle = E_m \vert m \rangle=-m\hbar \omega \vert m \rangle
\end{equation}
One may observe from the quantum formalism of angular momentum and above energy eigen equation that for a nucleus, with $I \neq 0$, the nuclear energy spectrum is composed of $(2I+1)$ equally spaced energy levels and the energy gap between two consecutive levels is $\hbar\omega$. The lowest energy level is given by $m=I$ while the highest is given by $m=-I$. The population distribution for an ensemble of identical nuclear spin at high temperature is governed by the Boltzmann distribution \cite{reif-book-65}. For example, for $I=\frac{1}{2}$ there are two energy levels correspond to $m=\pm\frac{1}{2}$ and the population of the energy levels characterized by $m=\frac{1}{2}$ and $m=-\frac{1}{2}$, denoted by $n_+$ and $n_-$ respectively, is governed by Boltzmann factor as
\begin{equation}
\frac{n_-}{n_+}=e^{-(E_{-I}-E_{+I})/k_BT}=e^{-\hbar\omega_L/k_BT}
\end{equation}
$k_B$ is the Boltzmann constant and $T$ is the absolute temperature of the spin ensemble. For $^1H$ ensemble placed inside a magnetic field of 14.1 Tesla, the Boltzmann factor is $\approx 10^{-5}$ which implies only one in $10^5$ spins is aligned in the external field direction which makes the ensemble weakly paramagnetic in nature, (Fig.(\ref{z_mag})). Nevertheless, this slight difference between the populations of the energy levels gives rise to a total magnetization in $z$-direction given by
\begin{equation}
M_z=\frac{\mu_o\gamma^2\hbar^2B_o}{4k_BT}
\end{equation}
\noindent where $\mu_o$ is the magnetic susceptibility and not to be confused with magnetic moment $\mu$. 

		\subsection{Interaction of the Nuclear Spin with Radio Frequency: The Nuclear Magnetic Resonance Phenomenon }\label{rfint}
The undisturbed spin ensemble in the presence of an external static magnetic field will stay in thermal equilibrium with population of various energy levels following Boltzmann distribution. However, transition between the energy eigenstates of the Hamiltonian defined by Eq.(\ref{hamil_Ch1}) can be induced using an oscillating magnetic field of appropriate Larmor frequency. For nuclear spins in a static magnetic field of few Tesla, the Larmor frequency is of the order of MHz, and hence to induce the transition between various energy levels a radio frequency (RF) field is required. In order to excite the population one can consider the transverse RF magnetic field $\mathbf{B}_1(t)$ perpendicular to the static magnetic field $\mathbf{B}$ as 
\begin{equation}
\mathbf{B}_1(t)=2B_1cos(\Omega t+\phi)\hat{i}
\end{equation}
where $\Omega$ and $\phi$ are the frequency and phase of the RF field and $\hat{i}$ is the unit vector in $x$-direction. The interaction Hamiltonian between nuclear spin and the RF field can be written as
\begin{equation}\label{rf_hamil}
\mathit{H}_{RF}=-\mathbf{\mu}.\mathbf{B}_1(t)=-\gamma\hbar I_x [2B_1cos(\Omega t+\phi)]
\end{equation}
 The Hamiltonian due to RF field can be considered as a perturbation to the Zeeman Hamiltonian, Eq.(\ref{hamil_Ch1}), as the magnitude of $\mathbf{B}_1(t)$ field is a few Gauss as compared to the $\mathbf{B}$ field magnitude. Under this consideration the effect of $\mathit{H}_{RF}$ can be investigated using time-dependent perturbation theory \cite{sakurai-book-94}. To understand the key features of the results of time-dependent perturbation theory one can assume that the linearly oscillating magnetic field $\mathbf{B}_1(t)$ is composed of two circularly polarized fields, with same amplitude and phase as that of $\mathbf{B}_1(t)$, precessing about $z$-axis in opposite direction \ie
\begin{eqnarray}
\mathbf{B}_1(t) &=& \mathbf{B}^+_1(t)+\mathbf{B}^-_1(t) \nonumber\\
\mathbf{B}^+_1(t) &=& B_1[cos(\Omega t+\phi)\hat{i}+sin(\Omega t+\phi)\hat{j}] \nonumber\\
\mathbf{B}^-_1(t) &=& B_1[cos(\Omega t+\phi)\hat{i}-sin(\Omega t+\phi)\hat{j}] 
\end{eqnarray}
For $\Omega=\omega$, \ie on resonance, the $\mathbf{B}^-_1(t)$ component rotates around $z$-axis in sync with the nuclear spin. In a coordinate system rotating with angular velocity $\mathbf{\Omega}=-\Omega \hat{k}$, \ie \textit{rotating frame} the component $\mathbf{B}^-_1(t)$ will appear stationary to the nuclear spins and spins experience a torque. By controlling the RF exposure time to the spins they can be excited from low energy eigenstate to higher energy eigenstates and this forms the basis of the NMR signal \cite{ernst-book-90}.

\section{NMR Quantum Information Processing}\label{NMRQIP}
At the turn of twentieth century, NMR was proposed as a potential platform for the physical realization of quantum information processor \cite{cory-pnas-97,chuang-sc-97,vandersypen-rmp-05}. The NMR quantum information processor utilizes the spin ensemble to encode and process quantum information and the results of the computation are obtained via expectation values of the observables. Since then NMR has been a useful testbed for the experimental demonstrations of  quantum algorithms as well as quantum information processing. NMR has been utilized for the experimental demonstration of Grover's search algorithm \cite{jones-nature-98}, Shor's algorithm \cite{vandersypen-nature-01}, Deutsch-Jozsa algorithm utilizing non-commuting selective pulses \cite{dorai-pra-00}, Order-Finding algorithm \cite{vandersypen-prl-00}, adiabatic quantum-optimization algorithm \cite{steffen-prl-03} and many more \cite{lu-bookchapter-16}.\\

The following subsections reviews the capabilities of NMR as demanded by the DiVincenzo criterion \cite{DiVincenzo-fdp-00} discussed in Sec-(\ref{DiVincenzo_criterion}).

		\subsection{Nuclear Spins as Qubits}
		As discussed earlier, the atomic nuclei with non vanishing nuclear spin placed in a static magnetic field exhibit nuclear Zeeman effect, see Eq.(\ref{zeeman}) \ie degeneracy in various energy eigenstates of the spin Hamiltonian is lifted in the presence of static magnetic field. This generates an energy spectrum of $(2I+1)$ levels. Minimum possible nuclear spin angular momentum \ie $I$ is one-half. Examples of spin-1/2 nuclei are $^{1}H$, $^{13}C$, $^{15}N$, $^{19}F$ and $^{31}P$. All such spin-1/2 nuclei are two-level quantum systems and can encode the quantum information as a qubit. Although nuclear spins with spin $>$ 1/2 were utilized for NMR quantum information processing but in general controlling higher-dimensional quantum systems in liquid state NMR is much more complicated due to their very low coherence times. Nevertheless, NMR qubits have been utilized extensively in the physical realization of quantum information \cite{cory-pnas-97,chuang-sc-97,jones-nature-98,vandersypen-rmp-05,dorai-pra-00,vandersypen-prl-00,vandersypen-nature-01,steffen-prl-03,lu-bookchapter-16}. \\
		\begin{figure}
\begin{center}
\includegraphics[scale=0.5]{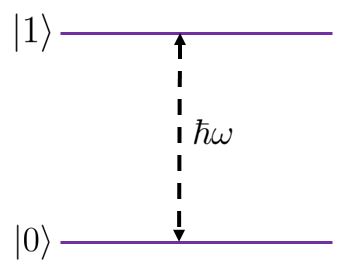}
\caption{Energy level diagram of a single spin-1/2 nucleus as a two-level quantum system.}
\label{NMR_qubit}
\end{center}
\end{figure}

\begin{figure}
\begin{center}
\includegraphics[scale=0.35]{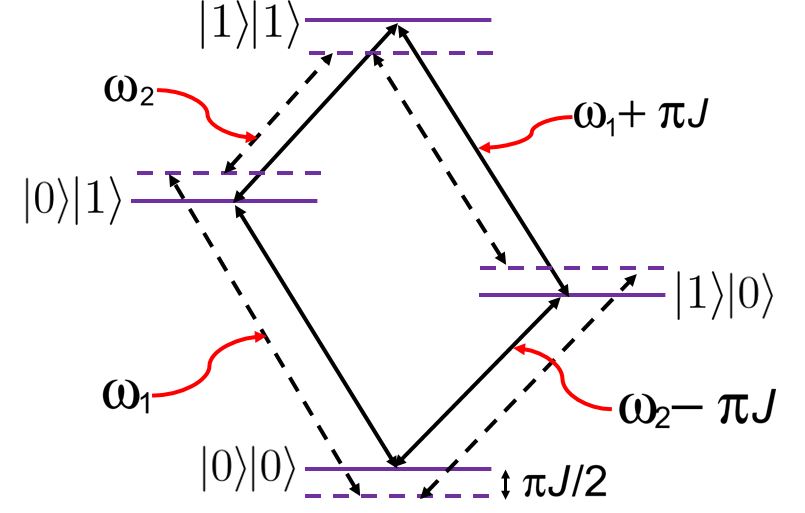}
\caption{Energy level diagram for two $J$-coupled spins. Dashed lines are the energy levels in the absence of $J$-coupling while solid lines are the energy levels modified by $J$-coupling.}
\label{two_coupled_spins_ELD}
\end{center}
\end{figure}

The energy eigenvalues of the spin Hamiltonian, given by Eq.(\ref{hamil_Ch1}), are $-\hbar\omega/2$ and $\hbar\omega/2$  and corresponding eigenvectors are $\vert I_z : (m=+\frac{1}{2}) \rangle$ and $\vert I_z : (m=-\frac{1}{2}) \rangle$ respectively. The eigenvectors, $\vert I_z : (m=+\frac{1}{2}) \rangle$ and $\vert I_z : (m=-\frac{1}{2}) \rangle$, of operator $\sigma_z=2I_z$ serve as computational basis and usually denoted by $\vert 0 \rangle$ and $\vert 1 \rangle$ respectively. See Fig.(\ref{NMR_qubit}) for a schematic of NMR qubit. Further, there can be more than one spin-1/2 nuclei in a molecule. Such spins can interact via direct \textit{magnetic dipole-dipole} interaction or indirectly via covalent bonds termed as \textit{scalar-coupling} or \textit{J-coupling} interactions. Dipole-dipole interactions are direct interactions of the nuclear magnetism and need no medium while J-coupling is through the interaction of nucleus with the electronic environment of the bonded electron cloud to the other nuclei \cite{ernst-book-90}. The Hamiltonian for $n$ such weakly interacting spin-1/2 systems is given by
		\begin{equation}\label{NMR_hamil_Ch1}
		\mathit{H}=-\sum_{i=1}^{n}\hbar \omega^i I^i_{z}+\sum_{\underset{i<j}{i,j=1}}^{n}2\pi\hbar J_{ij}I^i_{z}I^j_{z}
		\end{equation}
where $J_{ij}$ is the scalar coupling constant between $i^{\rm th}$ and $j^{\rm th}$ spins. Usually the NMR Hamiltonian, Eq.(\ref{NMR_hamil_Ch1}), is written in frequency units by letting Plank constant $h=1$. Intuitively one can interpret the second term of the Hamiltonian, Eq.(\ref{NMR_hamil_Ch1}), as additional magnetic field created by surrounding spins which further shifts the energy levels of $i^{\rm th}$ spin by $-\frac{J_{ij}}{2}$ if $j^{\rm th}$ spin is in $\vert 0 \rangle$ state or by $\frac{J_{ij}}{2}$ if $j^{\rm th}$ spin is in $\vert 1 \rangle$ state. Fig.(\ref{two_coupled_spins_ELD}) depicts the modification of energy levels in the presence of scalar J-coupling between two spins.
In this scenario, each spin transition splits up into two transitions at frequencies $\omega^i \pm \frac{J}{2}$ and results in doublet in NMR frequency spectrum. For a given system the exact energy level diagram can be obtained by diagonalizing the Hamiltonian given in Eq.(\ref{NMR_hamil_Ch1}).

		\subsection{Ensemble State Initialization}\label{NMR_ensemble_state}
		The next requirement of a quantum information processor is to initialize the quantum register in a fiducial state \ie a pure state. NMR deals with a large ensemble and inherently the ensemble is in a mixed state. Although mixed states are inadequate for QIP, elegant procedures were devised independently by Cory \textit{et al.} \cite{cory-pnas-97} and Chuang \textit{et al.} \cite{chuang-sc-97} whereby they have demonstrated that the spin magnetization can be manipulated to prepare spin ensemble in an effective pure state termed as \textit{pseudo pure state} (PPS). The motivation behind such a construct is that in NMR, one can interact with the deviation part of the ensemble density operator by means of RF fields. So once such deviation density part of the ensemble is initialized similar to the deviation part of pure state $\vert\psi\rangle\langle\psi\vert$ then such an PPS ensemble can mimic the pure state evolution under unitary transformations achieved via RF fields.\\
		
		As discussed earlier in Sec-(\ref{nmrbasics}), the NMR spin ensemble follows Boltzmann distribution law and thermal equilibrium state of the spin ensemble at temperature $T$ in the presence of magnetic field $B$ can be written as follows
		\begin{equation}
		\rho_{th}=\frac{e^{-\textit{H}/k_BT}}{\sum_m e^{-E_m/k_BT}}
		\end{equation}
		The term in the denominator, \ie $Z=\sum_m e^{-E_m/k_BT}$ is the spin ensemble partition function. For the Zeeman Hamiltonian Eq.(\ref{zeeman}) and basis formed by eigenstates of $I_z$ the diagonal entries, which are proportional to the energy level population, of the thermal state can be simplified as
		\begin{equation}
		[ \rho_{th} ]_{mm}=\frac{e^{m\hbar\omega/k_BT}}{\sum_{s=-I}^{I} e^{s\hbar\omega/k_BT}}
		\end{equation}
		Under high temperature limit \ie $\hbar\omega << k_BT$ following approximation can be used to simplify the expression of $\rho_{th}$ 
		\begin{eqnarray} 
		e^{m\Delta} &\approx& 1+m\Delta \nonumber\\
		\mathrm{and}\;\;\; \sum_{s=-I}^{I} e^{s\Delta} &\approx& 2I+1
		\end{eqnarray}
		where $\Delta=\frac{\hbar\omega}{k_BT}<<1$ is a measure of thermal magnetization of spin ensemble at temperature $T$ in the presence of magnetic field $B$. So in high temperature limit the thermal equilibrium state can be recast as
		\begin{equation}\label{nmrthrm}
		\rho_{th}=\frac{1}{2I+1}\mathbb{I}+\frac{\Delta}{2I+1}I_z
		\end{equation}
		One can observe that first the term on the right hand side of Eq.(\ref{nmrthrm}) is the uniform background represented by identity operator $\mathbb{I}$ and only a tiny part ($\Delta\approx 10^{-5}$) is in the state  having deviation part $I_z$. Similarly the NMR ensemble can be initialized in PPS of form Eq.(\ref{nmrthrm}) and the state can be put in the form
		\begin{equation}
		\rho_{\mathrm{PPS}} = \frac{(1-\Delta)}{2^n}+\frac{\Delta}{2^n} \vert\psi\rangle\langle\psi\vert
		\end{equation}
		There are a number of techniques to prepare PPS in NMR \eg temporal averaging \cite{knill-pra-98}, spatial averaging \cite{cory-physD-98,oliveira-book-07}, logical labeling \cite{chuang-sc-97},  state initialization utilizing long-lived singlet states \cite{roy-pra-10} and NMR line-selective pulses\cite{peng-cpl-01}. Generally such methods of PPS preparation suffer magnetization loss due to non-unitary evolution achieved by gradient pulses and remedies have been proposed to circumvent such difficulties \cite{kawamura-ijqc-04}. Nevertheless, it is well established that the NMR ensemble can be initialized in the PPS which mimics the pure state behavior and can be used for QCQI \cite{cory-pnas-97,chuang-sc-97,vandersypen-rmp-05}. The next subsection details the type of evolution feasible with NMR and methods to implement unitary operation utilizing RF fields.

		\subsection{NMR Unitary Gate Implementation}\label{nmrgates}
		In the computational basis the three spin angular momentum operators for spin 1/2 , in $\hbar$ units, can be written as 
		\begin{equation}
		I_x=\frac{\sigma_x}{2}=\frac{1}{2}
		\begin{bmatrix}
		0 & 1 \\
		1 & 0
		\end{bmatrix}
		\;\;\;\;
		I_y=\frac{\sigma_y}{2}=\frac{1}{2}
		\begin{bmatrix}
		0 & -i \\
		i & 0
		\end{bmatrix}
		\;\;\;\;
		I_z=\frac{\sigma_z}{2}=\frac{1}{2}
		\begin{bmatrix}
		1 & 0 \\
		0 & -1
		\end{bmatrix}
		\end{equation}
		where $\sigma$'z are the Pauli spin operators. From Eq.(\ref{nmrthrm}), one can write the deviation density matrix as
		\begin{equation}\label{deviation_rho}
		\Delta \rho_{th}=\frac{\hbar \omega}{4k_BT}\begin{bmatrix}
		1 & 0 \\
		0 & -1
		\end{bmatrix}
		\end{equation}
		As discussed in Sec.(\ref{rfint}), the interaction of RF fields can be understood utilizing rotating frame considerations. The lab frame density operator, $\rho^{\mathrm{lab}}$ can be transformed to rotating frame density operator using rotation operator $e^{-i\Omega t I_z}$ and can be written as 
		\begin{equation}\label{rottherm}
		\rho^{\mathrm{rot}}= e^{-i\Omega t I_z}.\rho^{\mathrm{lab}}.e^{i\Omega t I_z}
		\end{equation}
		The deviation density operator $\Delta \rho_{th}$ is invariant under the above transformation and the superscript ``rot'' can be dropped for convenience. The total Hamiltonian for spin-1/2 ensemble in the presence of magnetic field $\mathbf{B}$ being acted upon RF field $\mathbf{B_1}$ is the sum of Zeeman Hamiltonian in Eq.(\ref{hamil_Ch1}) and RF Hamiltonian in Eq.(\ref{rf_hamil}) which can be further transformed to rotating frame using the rotation operator. The resulting effective Hamiltonian in rotating frame can be computed as
		\begin{equation}\label{H_eff}
		\mathit{H}_{\mathrm{eff}}=-\hbar(\omega-\Omega)I_z-\hbar\omega_1I_x
		\end{equation}
		The striking feature of the above rotating frame Hamiltonian is its time independence. For the case when RF frequency, (also termed as nutation frequency), $\omega_1>>(\omega-\Omega)$, rotating frame Hamiltonian can be approximated as $H_{\mathrm{eff}}\approx -\hbar\omega_1I_x$. This approximation becomes true for the on resonance excitation \ie for $\omega=\Omega$. This approximation becomes particularly suitable for small resonance offsets, \ie $\vert\omega-\Omega\vert\approx 0$, and strong RF pulse with irradiation time say $t_p$. One can write explicitly the evolution operator for the RF pulse as
		\begin{equation}
		U_p=e^{-i\mathit{H}_{\mathrm{eff}}t_p/\hbar}=e^{i\omega_1 t_p I_x}=R_x(-\theta_p)
		\end{equation}
		Here $R_x(-\theta_p)$ is the rotation operator, in rotating frame, about $x$-axis through an angle $-\theta_p$ dictated by RF irradiation time $t_p$. One can write explicit forms of rotation operators achievable in NMR as
		
		\begin{equation}
		R_x(\theta_p)=
		\begin{bmatrix}
		cos\left( \frac{\theta_p}{2}\right) & -i\;sin\left( \frac{\theta_p}{2}\right) \\
		-i\;sin\left( \frac{\theta_p}{2}\right) & cos\left( \frac{\theta_p}{2}\right)
		\end{bmatrix}
		\end{equation}
		\begin{equation}
		R_y(\theta_p)=
		\begin{bmatrix}
		cos\left( \frac{\theta_p}{2}\right) & sin\left(\frac{\theta_p}{2} \right) \\
		-sin\left( \frac{\theta_p}{2}\right) & cos\left( \frac{\theta_p}{2}\right)
		\end{bmatrix}
		\end{equation}
		\begin{equation}
		R_{\phi_p}(\theta_p)=
		\begin{bmatrix}
		cos\left( \frac{\theta_p}{2}\right) & -i\;sin\left( \frac{\theta_p}{2}\right)e^{-i\phi_p} \\
		-i\;sin\left( \frac{\theta_p}{2}\right)e^{i\phi_p} & cos\left( \frac{\theta_p}{2}\right)
		\end{bmatrix}
		\end{equation}
		The last rotation operator is for RF pulse having a phase $\phi_p$ with $x$-axis in the rotating frame. Using above formulated rotation operators one can compute the resulting state of the ensemble after the action of $\mathit{H}_{\mathrm{eff}}$, Eq.(\ref{H_eff}) on deviation density operator, Eq(\ref{deviation_rho}). Considering $\omega_1 t_p=\theta_p=\frac{\pi}{2}$ it can be shown that
		\begin{equation}
		\Delta\rho(t_p)=R_x\left(-\frac{\pi}{2} \right).\Delta\rho_{th}.R_x\left(\frac{\pi}{2} \right)=\frac{\hbar\omega}{2k_BT}I_y
		\end{equation}
		A semi-classical description gives an intuitive picture that the initial deviation density operator proportional to $I_z$ evolve to a density operator proportional to $I_y$ under a rotation by an angle $-\pi/2$ about $x$-axis in rotating frame which strikingly appears to be a classical behavior!\\
		
		Its interesting to note that a $R_y(\pi/2)$ achieves the effect of Hadamard(H)-gate while the NOT(X)-gate can be achieved by $R_y(\pi)$ rotation. Similarly a $\mathrm{CNOT}$ gate can be achieved by exploiting the scalar $J$-couplings of the spins and a typical NMR pulse sequence of $\mathrm{CNOT}$ gate, for two coupled spins, is as follows
		\begin{equation}
		U_{\mathrm{CNOT}}=R^1_z\left( -\frac{\pi}{2} \right)R^2_x\left(\frac{\pi}{2} \right)R^2_y\left(- \frac{\pi}{2} \right)\frac{1}{2J}R^2_y\left( \frac{\pi}{2} \right)
		\end{equation}
		Here the time period $1/2J$ is the free evolution for which spin system evolve under NMR weak field Hamiltonian, Eq.(\ref{NMR_hamil_Ch1}), to effectively achieve the non-local operation of $\mathrm{CNOT}$ gate. Superscript on various rotations denote spin label. Also the $z$-rotation can be achieved by cascading three $x$, $y$ rotations. In nutshell the NMR technique is equipped to achieve, in principle, any arbitrary unitary operator.
		
		\subsection{Measurements in NMR and State Tompgraphy}
As mentioned earlier, NMR generally deals with the ensemble of spin-1/2 nuclei and a typical NMR liquid state sample, having volume $\sim$ 500-600 $\mu l$, contains $\sim 10^{18}$ spins. As governed by the Boltzmann distribution, these spins generate a bulk magnetization in the presence of an external magnetic field, Fig-(\ref{z_mag}). External RF field can be used to manipulate this bulk magnetization. The net magnetization in the thermal equilibrium state, \ie $\rho_{th}$, is along $z$-direction and  can be brought in $xy$-plane by applying RF field of appropriate duration, using a rotation operator $R_x (-\pi/2)$.
\begin{figure}
\begin{center}
\includegraphics[scale=0.5]{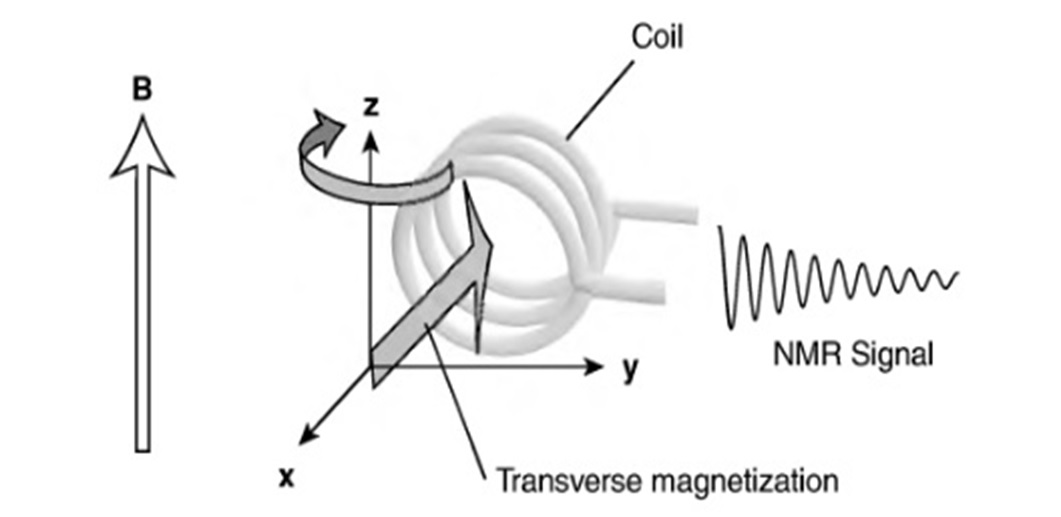}
\caption{Precession of bulk magnetization in the presence of an external static magnetic field induces current in the pick-up coils which further amplified and stored as a time domain signal termed as free induction decay (FID) \cite{fid-online-12}.}
\label{nmrsignal}
\end{center}
\end{figure}
In the transverse plane, the net magnetization undergoes Larmor precession about the $-z$-direction and can be detected by induction coils. Changing magnetic flux induces an electromotive force in the pick-up coils which in turn produces a detectable current. This induced current is then digitized and stored as time-domain NMR signal and termed as free induction decay (FID). FID typically has an oscillatory decaying nature due to various NMR relaxation processes and one can obtain the frequency-domain NMR signal by performing discrete Fourier transform (FT) on the digitized time-domain signal. Such processing in NMR results in \textit{Lorentzian} peaks correspond to transitions, (Fig-(\ref{two_coupled_spins_ELD})), between various energy eigenstate of NMR Hamiltonian, (Eq.(\ref{hamil_Ch1})). Schematic of NMR signal acquisition and processing is depicted in the Fig.(\ref{nmrsignal}).\\

The normalized intensities/amplitudes of NMR peaks are proportional to the respective spin ensemble magnetization which in turn is proportional to the expectation value of operator $I_z$ in the state of spin ensemble \cite{ernst-book-90} \ie $\langle I_z \rangle _{\rho}$. Its worth noting here that NMR enables the measurement of ensemble average in a single experiment as the net effect of the NMR measurement is equivalent of measuring, \eg $I_z$, on individual spins one-by-one and take the average.\\

Any single-qubit density matrix, in the computaional basis, can be brought into the form \cite{sakurai-book-94}
\begin{equation}\label{1Qdm}
\rho=\frac{\mathbb{I}}{2}+a\sigma_x+b\sigma_y+c\sigma_z
\end{equation}
which can further be written as
\begin{equation}
\rho=
\begin{bmatrix}
\frac{1}{2}-c & a-i\;b \\
a+i\;b & \frac{1}{2}+c
\end{bmatrix}
\end{equation}
One may observe that $\langle \sigma_z \rangle _{\rho}=\mathrm{Tr}(\rho\sigma_z)=c$. Further, with an appropriate choice of rotation operators one can measure the unknown parameters $a$ and $b$ as well and hence can reconstruct the density operator $\rho$ in Eq.(\ref{1Qdm}). The process of reconstructing the density operator from several experimental settings is known as quantum state tomography (QST) and there have been numerous studies on developing schemes for QST \cite{long-job-01,leskowitz-pra-04}. In this thesis, methods have been developed for accurately measuring the expectation values of two- and three-qubit Pauli operators in a given ensemble state which can be generalized to higher-dimensional Hilbert spaces \cite{singh-pra-16,singh-pra-18,singh-qip-18}.\\

Typical hallmark of QCQI, the projective measurements, generally is not possible in NMR, although there have been few experimental studies reporting projective measurements in NMR \cite{lee-apl-06,khitrin-qip-11} utilizing non-unitary evolutions by means of gradient pulses.

\section{Quantum Entanglement}\label{entanglementtheory}

Quantum entanglement first described by Erwin Schr{\"o}dinger \cite{schrodinger-Naturwissenschaften-35,schrodinger-mpc-35} in 1935. Quantum entanglement \cite{schrodinger-mpc-35} is a counter-intuitive feature exhibited by quantum particles which has no analog in classical mechanics. Quantum entanglement arrises when for a composite quantum system we are not able to describe the state of the quantum system in terms of quantum states of the parts. It has been shown \cite{aditi-cs-17} that quantum entanglement is a key resource to achieve computational speedup in quantum information processing (QIP) \cite{horodecki-rmp-09} and for quantum communication related tasks \cite{sibasish-njp-02,aditi-pra-09, pankaj-epd-15}. Being a fragile resource, prone to decoherence \cite{gedik-pla-06}, there have been proposals and demonstrations to store and protect entanglement \cite{mahesh-pra-11,aditi-pra-15} as well as environment-assisted enhancement \cite{archan-pra-06} of the entanglement.\\

Entanglement detection and characterization is of utmost importance for the physical realization of quantum information processors \cite{guhne-pr-09}. There have been a large number of measures proposed for the detection of quantum correlations \cite{aditi-pra-07} and in particular quantum entanglement \cite{guhne-pr-09}. In recent years, enormous experimental efforts have gone in the creation of entanglement. Typically in such an experimental scenario one would always be interested in queries like does entanglement actually get created in the experiment and can one detect and quantify the entanglement. In general, these questions are difficult to answer. There are many proposals to address such queries such as positivity under partial transposition (PPT)\cite{horodecki-pla-96} criterion, permutation based measures of quantum correlations \cite{arul-pra-16}, correlations in successive spin measurements \cite{sibasish-ijqi-07}, entanglement measures based on no-local-cloning and deleting\cite{aditi-pra-04} and isotropic spin lattice entanglement characterization \cite{aditi-prl-13}. Entanglement detection utilizing entanglement witnesses \cite{archan-qip-13,archan-pra-14a} is also a well developed field where such detections are explored from teleportation capabilities \cite{archan-prl-11}. Tripartite quantum states were characterized \cite{aditi-pra-12} using monogamy scores as well as mutual information in permutation symmetric states \cite{arul-pre-18}. Universal bipartite entanglement detection using two copies \cite{sibasish-arxiv-18} as well as extent of entanglement by sequential observers \cite{aditi-pra-18a} have recently been explored in a measurement-device-independent way \cite{sibasish-pra-17}.\\

Further, some of the experiments create entanglement between more than two subsystems and there are different classes \cite{dur-pra-00, verstraete-pra-02} of entanglement that exist in such cases. So entanglement characterization should be capable of distinguishing different classes. It should be noted that entanglement characterization is much more challenging rather than mere detection \cite{guhne-pr-09}. The situation is even more complex \cite{pankaj-epd-18a} in case of mixed states \cite{aditi-jmo-03} where geometric measures \cite{aditi-pra-16} have been resorted to for quantification of quantum correlations in multipartite and mutidimensional \cite{pankaj-pra-17} cases.\\

Entanglement witnesses \cite{lewenstein-pra-00,bourennane-prl-04} and approximation of positive maps \cite{rahimi-jpamg-06, rahimi-pra-07} proved to be experiment friendly but lack generalization, since most of the experiments focus on creation of a specific entangled state and witness-based detection protocols usually require the state information beforehand. In this thesis, the goals are to explore the entanglement detection as well as characterization protocols and experimentally implement them in a \textit{state-independent} manner using nuclear magnetic resonance (NMR). NMR has been proposed as a promising candidate for realizing quantum processors \cite{jones-jcp-98,laflamme-qic-02}. NMR has been the testbed for the demonstration of the Deutsch$-$Jozsa algorithm \cite{anil-pra-01, mahesh-jmr-01}, quantum No-hiding theorem \cite{anil-prl-11} and parallel search algorithm \cite{rangeet+anil-pra-05} as well as of foundational aspects such as delayed choice experiments \cite{mahesh-pra-12} and querying Franck-Condon factor \cite{mahesh-pra-14}. Control of 5 to 8 qubits for quantum information processing was achieved \cite{rangeet-jmr-04,rangeet-aipcp-06} and bench-marking of quantum controls on a 12 qubit quantum processor was demonstrated \cite{mahesh-prl-06} using NMR systems. Highly accurate control via radio frequency pulses made initialization of NMR system \cite{cory-physD-98,anil-pra-00} and read out using quantum state tomography \cite{leskowitz-pra-04,mahesh-pra-13a} accessible, in contrast to other hardware.\\

As discussed above, quantum entanglement is a striking feature of quantum mechanical description of nature and was quickly followed by the demand of a physical and more intuitive interpretation by Einstein-Podolsky-Rosen \cite{EPR-pr-35}. Quantum entanglement proved to be a physical resource\cite{nielsen-book-02} which can be utilized to accomplish quantum computational tasks \cite{horodecki-rmp-09}, which are impossible to perform using classical resources.

\subsection{Bipartite Entanglement}
Consider a quantum system consisting of two subsystems $A$ and $B$. Quantum states of $A$ and $B$ can be defined in respective Hilbert spaces $\mathcal{H}_A$ and $\mathcal{H}_B$ having dimension $d_A$ and $d_B$ respectively. The states of the composite system are defined by vectors in tensor-product of the Hilbert spaces $\mathcal{H}_A \otimes \mathcal{H}_B$ having dimension $d_Ad_B$. Any vector in the joint Hilbert space $\mathcal{H}_A \otimes \mathcal{H}_B$ can be expressed as
\begin{equation}
\vert \psi \rangle = \sum_{i,j=1}^{d_A,d_B}{c_{ij} \vert a_i \rangle \otimes \vert b_i \rangle } \;\; \in \mathcal{H}_A \otimes \mathcal{H}_B
\end{equation}
where $\lbrace \vert a_i \rangle \rbrace $ is basis in $\mathcal{H}_A$, $\lbrace \vert b_i \rangle \rbrace $ is basis in $\mathcal{H}_B$ while $ c_{ij} \in \mathbb{C} $ and normalization of $ \vert \psi \rangle $ requires $ \sum_{i,j=1}^{d_A,d_B}{\vert c_{ij} \vert ^2=1} $. Now if $ \vert \psi \rangle \in \mathcal{H} $ represents any general pure state of the composite system $AB$ and one can express it as
\begin{equation}
\vert \psi \rangle = \vert \phi^A \rangle \otimes \vert \phi^B \rangle
\end{equation}
where states $\vert \phi^A \rangle \in \mathcal{H}_A$ and $\vert \phi^B \rangle \in \mathcal{H}_B$, then the state $\vert \psi \rangle $ is separable else it is entangled. Physically, the separable states are uncorrelated from statistics of measurement outcomes perspective. In a more general case when the system can be in any one of the states $ \vert \phi_i \rangle \in \mathcal{H}$ with probability $p_i$ then mixed state of the system can be expressed as
\begin{equation}
\rho=\sum_i{p_i \vert \phi_i \rangle \langle \phi_i \vert}
\end{equation}
with $ \sum_i{p_i}=1$ and $p_i\geq 0$. If the state of a composite system can expressed as a convex mixture, of the product states $\rho^A \otimes \rho^B$, as
\begin{equation}\label{mixedsep_ch1}
\rho=\sum_i{w_i\rho_i^A \otimes \rho_i^B}
\end{equation}
then $\rho$ is separable otherwise it is entangled. These definitions of bipartite entanglement can be generalized to multipartite cases as well.

		\subsection{Entanglement Detection and Characterization}
		We review here the commonly used entanglement detection criteria for bipartite systems.
			\subsubsection{The Positive Under Partial Transposition (PPT) Criterion}
			The density operator of a composite bipartite system can be expanded in a chosen product basis as
			\begin{equation}
			\rho=\sum_{i,j=1}^{N}\sum_{k,l=1}^{M} \rho_{ij,kl}\vert i \rangle  \langle j \vert \otimes \vert k \rangle  \langle l \vert
			\end{equation}
			here $N$ and $M$ are the dimensions of local Hilbert spaces of the bipartite state. Having defined the above decomposition, one can write the partial transposition of the density operator with respect to subsystem A as
			\begin{equation}
			\rho^{T_A}=\sum_{i,j=1}^{N}\sum_{k,l=1}^{M} \rho_{ji,kl}\vert i \rangle  \langle j \vert \otimes \vert k \rangle  \langle l \vert
			\end{equation}
			Partial transposition with respect to subsystem B \ie $\rho^{T_B}$ can be obtained by interchanging the indices $k$ and $l$ instead $i$ and $j$. One may also use the fact that the usual transposition $\rho^{T}=(\rho^{T_A})^{T_B}=(\rho^{T_B})^{T_A}$ to obtain $\rho^{T_B}=(\rho^{T_A})^T$. Partial transposition depends upon the basis in which it is performed but the spectrum is independent of the basis which is also true for matrix transposition. The density matrix in a basis has positive partial transposition \ie $\rho$ is PPT, if its partial transposed density matrix does not have negative eigenvalues and hence is positive semi-definite. If a density operator is not PPT then it is NPT. Based on PPT there are two strong conditions satisfied by separable states described as follows:\\
			
			\noindent\textbf{PPT Criterion:} If $\rho$ is a bipartite separable state then $\rho$ is PPT.\\
			
\noindent PPT criterion has an intuitive description. As mentioned earlier, the information regarding the state of an entangled composite quantum system is stored in the joint state of the system rather than its parts. State of a bipartite separable composite system can be cast in the form of Eq.(\ref{mixedsep_ch1}). Hence and partial transposition operation will independently transpose the state of either subsystem $A$ or $B$ and always result in PPT $\rho$. Here the fact, that the transpose of a positive semi-definite density operator will also be a positive semi-definite, is used. Hence any state which is NPT is always entangled \cite{horodecki-pla-96} but a PPT state may be separable or entangled.\\
			
			\noindent\textbf{Horodecki Theorem:} If $\rho$ is the density operator acting on a 2$\otimes$2 or 2$\otimes$3 Hilbert space, then $\rho^{T_A}\geq 0$ implies $\rho$ is separable.
			\begin{equation}
			\rho=\sum_{i,j=1}^{N}\sum_{k,l=1}^{M} \rho_{ij,kl}\vert i \rangle  \langle j \vert \otimes \vert k \rangle  \langle l \vert
			\end{equation}
For higher dimensional Hilbert spaces, \ie $>$ 2$\otimes$3, this may not be the case \cite{horodecki-pla-96}. There are entangled states which do not violate Horodecki's theorem \ie PPT entangled states. Such PPT entangled states makes an important class of states falls in the class of bound entanglement, Sec-\ref{boundEnt}. 		
			
		\subsubsection{The Computable Cross Norm or Realignment (CCNR) criterion}
		The PPT criterion in not a necessary condition for the density matrices acting on Hilbert spaces having dimension greater than six. Although many stronger criteria were proposed but its worth mentioning the CCNR criterion \cite{rudolph-qip-05}. To describe the CCNR criterion the concept of Schmidt decomposition is utilized. For a density matrix $\rho$, the Schmidt decomposition can be written as
			\begin{equation}\label{sch-decom}
			\rho=\sum_{k} \lambda_k G^A_k\otimes G^B_k
			\end{equation}
		where $\lambda_k\geq0$ and $G^A_k$ and $G^B_k$ are orthonormal bases of the observable spaces $ H_A $ and $ H_B $. Such a basis consists of $d^2$ Hermitian observables satisfying Tr$(G^A_iG^A_j)$=Tr$(G^B_iG^B_j)$ =$\delta_{ij}$.\\
			
			\noindent\textbf{CCNR Criterion:} If the state $\rho$ is separable, then the sum of all $\lambda_k$ in Eq.(\ref{sch-decom}) is smaller than 1 \ie
			\begin{equation}\label{ccnr}
			\sum_k \lambda_k \leq 1
			\end{equation}
			Hence if $ \sum_k \lambda_k > 1 $ then $ \rho $ is entangled.

		\subsubsection{The Positive Map Method}	
			PPT criterion is an example of positive but not completely positive (PNCP) maps. One may define PNCP as follows: Let the $\mathcal{B}(\mathcal{H}_i)$ be linear operators acting on the Hilbert spaces $\mathcal{H}_B$ and $\mathcal{H}_C$. A positive linear map, \ie $\Lambda:\mathcal{B}(\mathcal{H}_B)\rightarrow \mathcal{B}(\mathcal{H}_C) $, will map the Hermitian operators onto the Hermitian operators and satisfies $\Lambda (X^{\dagger})=\Lambda (X)^{\dagger}$ and $ \Lambda(X) \geq 0 $ for $X\geq 0$. A map $\Lambda$ is completely positive (CP) if for an arbitrary Hilbert space $\mathcal{H}_A$, the map $\mathbb{I}_A\otimes \Lambda$ is also positive otherwise $\Lambda$ is PNCP. For example, transposition is PNCP map: as transpose of a positive operator is positive but partial transposition (equivalent to transposition after including extended operator space) may result in negative operator. Having defined PNCP maps, the separability criterion can be described as follows: For any separable state $\rho$ and any positive map $\Lambda$ following is always satisfied \cite{horodecki-pla-96}
			\begin{equation}
			\mathbb{I}_A\otimes \Lambda \geq 0
			\end{equation}
			Above condition is a sufficient criterion \ie a state violating above criterion is entangled but a state which doesn't show violation may also be entangled.
		\subsubsection{The Majorization Criterion}
			For a general bipartite state $\rho$, one can obtain the reduced state of subsystem B by tracing out the state of subsystem A \ie $\rho_B=\mathrm{Tr}_B(\rho)$. Let $\mathcal{P}=(p_1,\;p_2,...)$ and $\mathcal{Q}=(q_1,\;q_2,...)$ denote the sets of decreasingly ordered eigenvalues of $\rho$ and $\rho_B$ respectively. The majorization criterion states that if $\rho$ is separable then 
			\begin{equation}
			\sum_{i=1}^k p_i \leq \sum_{i=1}^k q_i
			\end{equation}
			holds for all $k$ \cite{nielsen-prl-01}. For separable $\rho$ above condition also holds for the reduced state of subsystem A \ie for $\rho_A=\mathrm{Tr}_A(\rho)$.
		\subsection{Bound Entanglement}\label{boundEnt}
		Generally maximally entangled two qubit states, \ie singlet states, are needed to accomplish many tasks in quantum information theory \eg teleportation, superdense coding and cryptography. But generally in an experiment, due to inadvertent noise, one ends up with mixed states. It is a practical question that how one can create singlet state from some given mixed states. This process of creating singlet or maximally entangled state from given mixed states is called \textit{entanglement distillation}. Entanglement distillation can be described as follows: Consider two parties Alice and Bob, sharing an arbitrary, but finite, number of copies of the entangled state $\rho$. Entanglement distillation is the process of transforming available quantum resources/states, by performing local operations and classical communications (LOCC), to a singlet state \ie
			\begin{equation}
			\underbrace{\rho\otimes \rho\otimes ... \otimes \rho}_{k-\rm copies}\stackrel{\rm LOCC}{\longrightarrow} \frac{1}{\sqrt{2}}(\vert 00 \rangle - \vert 11 \rangle)
			\end{equation}
		If Alice and Bob can achieve the above task with $k$-copies of $\rho$ then $\rho$ is distillable  else $\rho$ is \textit{bound entangled}. Although there is no protocol which ensures entanglement distillability but the sufficient conditions for undistillability \cite{horodecki-pra-99, tohya-prl-03} as well as distillability \cite{deutsch-prl-96, horodecki-prl-98} have already been proposed. A special kind of undistillable entanglement is PPT entanglement. It has been shown that \\
		
		\textit{ If a bipartite state is PPT, then the state is undistillable. If a state violates the reduction criterion (\eg, due to a violation of the majorization criterion) then the state is distillable.}\\
		
		One of the first PPT entangled class of states was proposed in Ref. \cite{horodecki-pla-97} and later more classes were discovered \cite{bennet-prl-99, dagmar-pra-00, piani-pra-06, piani-pra-07}. Entangled states that are undistillable are called bound entangled states and PPT entangled states is the most important class of such states. Characterization of bound entanglement is an interesting as well as challenging task in entanglement theory.

		\subsection{Entanglement Witnesses} 
		All of the entanglement detection criteria discussed above require knowledge of the density operator. However there is a sufficient entanglement criterion in terms of a measurable observable termed as \textit{Entanglement Witness} (EW) \cite{horodecki-pla-96,tehral-pla-00,lewenstein-pra-00,dagmar-jmo-02}.\\
		An observable $\mathcal{W}$ is an \textit{Entanglement Witness} iff:
		\begin{itemize}
		\item $\mathrm{Tr}(\rho_s\mathcal{W})\geq 0$ for all separable states $\rho_s$ and
		\item $\mathrm{Tr}(\rho_e\mathcal{W})< 0$ for at least one entangled state $\rho_e$
		\end{itemize}
		holds. Thus entanglement of $\rho_e$ is witnessed by measuring $\mathcal{W}$ and establishing $\mathrm{Tr}(\rho_e\mathcal{W})$
\linebreak
$<0$. It is worth mentioning here that constructing an EW is, in general, a difficult task. There may be the cases when a given EW unable to witness the entanglement. $\mathrm{Tr}(\rho\mathcal{W})< 0$ confirms the presence of entanglement but for the case when $\mathrm{Tr}(\rho\mathcal{W})\geq 0$, $\rho$ may be separable or entangled. EW is one of the most utilized concept for the entanglement detection in experiments. Following has been proved \cite{horodecki-pla-96} as a strong criterion for entanglement detection in experiments.\\
		
		\textit{Completeness of Witnesses: For each entangled state $\rho_e$ there exists an entanglement witness detecting it}.\\
		
				A few methods to construct an EW are:\\
		\begin{itemize}
				\item Consider an entangled NPT state $\rho_e$  whose partial transpose, \ie $\rho_e^{T_A}$, has at least one negative eigenvalue $\lambda_{-}$ and let $\vert \eta \rangle $ be the corresponding eigenvector. It can be shown that
				\begin{equation}
				\mathcal{W}=\vert \eta \rangle \langle \eta \vert ^{T_A}
				\end{equation}
				can act as an EW for the detection of entanglement of $\rho_e$. It can be proved as Tr$(\rho_e\mathcal{W})$=Tr$(\rho_e \vert \eta \rangle \langle \eta \vert ^{T_A})$=Tr$(\rho_e^{T_A}\vert \eta \rangle \langle \eta \vert)$=$\lambda_{-}<0$ hence $\rho_e$ is entangled.
				\item Consider a state $\rho_e$ violating CCNR criterion. Then by definition, there exists a Schmidt decomposition given by Eq.(\ref{sch-decom}) with $\lambda_k\geq 0$. In such cases, the EW can be formulated \cite{yu-prl-05,guhne-pra-06} as
				\begin{equation}
				\mathcal{W}=\mathbb{I}-\sum_{k} \lambda_k G^A_k\otimes G^B_k
				\end{equation}
				with $G^{A/B}_k$ are the local observables in Schmidt decomposition, (Eq. (\ref{sch-decom})). One can see that Tr$(\rho_r\mathcal{W})$=1-$\sum_{k} \lambda_k<0$ and hence detect the entanglement in $\rho_e$.
				\item To construct the entanglement witnesses one can consider the states close to an entangled state, which must also be entangled depending upon their overlap with the original entangled state. For a pure entangled state $\vert \psi \rangle $ the projector based EW can be written as
				\begin{equation}
				\mathcal{W}=\alpha\mathbb{I}-\vert \psi \rangle \langle \psi \vert
				\end{equation}
				Motivation for the above construct is that the quantity $\mathrm{Tr}(\rho \vert \psi \rangle \langle \psi \vert) $ = $ \langle \psi \vert \rho \vert \psi \rangle $ is the fidelity of the state $ \vert \psi \rangle $ in the mixed state $ \rho $ and if this fidelity exceeds the threshold value $ \alpha $ then above EW $ \mathcal{W} $ detects the entanglement in $\rho$. $\alpha$ can be computed \cite{bourennane-prl-04} such that expectation value of $\mathcal{W}$ is non-negative for all the separable states and given as follows: 
				\begin{equation}
				\alpha=\underset{\rho\;\rm{is\;separable}}{\mathrm{max}}   \mathrm{Tr}(\rho \vert \psi \rangle \langle \psi \vert)	= \underset{\vert \phi \rangle = \vert a \rangle \otimes \vert b \rangle}{\mathrm{max}} \vert \langle \psi \vert \phi \rangle \vert^2
				\end{equation}
				The fact that a linear function takes its maximum on a convex set in one of the extremal points has been used, and for the convex set of the separable states these extremal points are just the pure product states. It has been shown \cite{bourennane-prl-04} that the above maximum can be directly computed and is given by the square of the maximal Schmidt coefficient of $\vert \psi \rangle$. 
		\end{itemize}
		\subsection{Entanglement Measures}
		Above discussed methods enables the detection of entanglement in a given state however one may be interested in quantifying the entanglement in the state. In order to do so there exist a number of entanglement measures (entanglement monotone) \cite{horodecki-rmp-09}. Its worth mentioning the requirements of an entanglement measure (EM). First and basic requirement for EM is that it should quantify the entanglement present in a given state \cite{vedral-prl-97}:
		\begin{itemize}
		\item[(i)] An entanglement measure $E(\rho)$ should vanish for all separable states.
		
		\item [(ii)] An entanglement measure should be invariant under local change of basis \ie it should be invariant under local unitary transformation of form:
		
		\begin{equation}
		E(\rho)=E(U_A\otimes U_B \rho U_A^{\dagger}\otimes U_B^{\dagger}).
		\end{equation}
		
		\item[(iii)] Entanglement cannot be created or increased under LOCC so $E(\rho)$ should not increase under LOCC. If $\Lambda^{\rm LOCC}$ is positive map that can be implemented using only LOCC then 
		
		\begin{equation}\label{EM_3}
		E(\Lambda^{\rm LOCC}(\rho)) \leq E(\rho)
		\end{equation}
		
		A stronger version of the above requirement is that $E(\rho)$ should not increase on an average under LOCC \ie if LOCC operations maps $\rho$ to $\rho_k$ with probabilities $p_k$ then
		\begin{equation}
		\sum_kp_k E(\rho_k)\leq E(\rho)
		\end{equation}
		The monotonicity under LOCC in Eq. (\ref{EM_3}), implies invariance under local unitary transformations.
		
		\item[(iv)] Entanglement decreases on mixing two or more states \ie
		\begin{equation}
		E\left(\sum_k p_k \rho_k\right) \leq \sum_kp_k E(\rho_k)
		\end{equation}
		This condition requires that if one starts with an ensemble of states $\rho_k$, and loses the information about the single instance of $\rho_k$, then entanglement should decrease.

		\item[(v)] For the case when one have access to two or more copies of the states then additivity of EM should obey
		\begin{equation}
		E(\rho^{\otimes n})=n E(\rho)
		\end{equation}
		Here Alice and Bob shares $n$-copies of the same state $\rho$. In case Alice and Bob share different states, say $\rho_A$ and $\rho_B$ then even a stronger requirement of additivity requirement can be written as
		\begin{equation}
		E(\rho_A\otimes\rho_B)=E(\rho_A)+E(\rho_B)
		\end{equation}
		Above additivity requirement is in general difficult to prove and satisfied by few EMs \cite{plenio-qic-07}.\\
	\end{itemize}	
		Various EMs have been proposed which satisfy partially the above listed requirements. Commonly used EMs are entanglement cost \cite{bennett-pra-96a}, entanglement of formation \cite{wootters-prl-98,chen-prl-05}, concurrence \cite{wootters-prl-98}, Negativity \cite{zyczkowski-pra-98,vidal-pra-02}, relative entropy of entanglement \cite{vedral-pra-98} and $n$-tangle \cite{coffman-pra-00}. Many of these measures are used in this thesis and their details are given in the subsequent chapters where they have been introduced first.\\
		
		Quantum discord (QD) captures the fact that even separable states can possess quantum correlations \cite{zurek-adp-00,ollivier-prl-02}. There have been intense theoretical and experimental advancements \cite{aditi-rpp-18, aditi-pre-16} utilizing QD to capture nonclassicality \cite{rahimi-pra-10} and quantum-to-classical transition \cite{sibasish-pra-10}. Dynamics of QD was studied \cite{mahesh-pra-12} and used as a quantifier of nonclassicality \cite{singh-pra-17}. Quantum correlation dynamics in a hybrid qubit-qutrirt system was explored \cite{gedik-pla-11} utilizing QD. The interplay between entanglement and nonclassicality was explored \cite{ivan-pra-11} in multimode radiation states. Invariant QD in hybrid qubit-qutrit quantum system and tradeoff with entanglement display interesting features \cite{gedik-ps-13}.  This thesis also explores the detection of nonclassical correlations possessed by separable mixed states using positive maps and QD.  Certain quantum states possess non-local nature of quantum correlations and violation \cite{aditi-pra-01} of a Bell type inequality \cite{ bell-ppf-64} may reveal such non-localities. It has been shown that W class of states possess stronger non-locality than GHZ class of states\cite{aditi-pra-03}. Such non-local correlations need to be investigated from \textit{ease of experimental implementation} as well as \textit{state independence} perspective.\\

		\subsection{Entanglement in NMR}
As discussed in Sec-{\ref{NMR_ensemble_state}}, typically the state of NMR ensemble at room temperature remains in the vicinity of maximally mixed state and hence it is not possible to create a genuine entangled state of the nuclear spins in small thermally polarized molecules in liquid state \cite{braunstein-prl-99}. This  has initiated a debate on the quantumness of the states in a typical NMRQC experiments which argued that all the states produced by NMR are classical. On the contrary any simulation of the dynamics of coupled nuclear spins using any classical model has been proved unsuccessful and it is conjectured that although the states in NMR may be in the vicinity of maximally mixed state but the dynamics is truly quantum mechanical \cite{schack-pra-99}. This can be well observed from the discussion in Sec-(\ref{nmrbasics}) that dynamics of NMR ensemble follows the laws of quantum evolution and as conjectured that the PPS perfectly mimics the behaviors of pure state and indeed generates the observable NMR signal. The the state of the sub-ensemble, \eg in NMR, truly possesses all the quantum features like superposition and entanglement. So with this understanding ensemble can be prepared in any desired state and is generally termed as \textit{pseudo} to make a distinction from \textit{pure} state.

\section{Motivations and Organization of the Thesis}
This thesis focuses on the experimental creation and detection of different types of quantum correlations using nuclear spins and NMR hardware. Quantum entanglement, being the most important and counter-intuitive, is one of the main types of correlation considered in this thesis. One of the main goals of the studies undertaken in this thesis was to design experimental strategies to detect the entanglement in a \textit{state-independent} way that are low on experimental resources. Core of all the detection protocols is a novel method which enables the measurement of any observable with high accuracy. Although these methods have been implemented on NMR hardware but they were developed in a hardware-independent manner and hence can be utilized on other QCQI hardware. Experimental protocols have been successfully implemented to detect the entanglement of random two-qubit states utilizing semi definite programming to construct an entanglement witness and thereby detect the entanglement. This random measurement based scheme to detect entanglement is also extended to a bipartite hybrid qubit-qutrit quantum system. It is shown via simulations that a two parameter class of qubit-qutrit entangled states get detected using only four local observables. Further, schemes for the experimental detection as well as classification of generic and general three-qubit pure states have also been devised and implemented successfully. Protocol to detect and classify three-qubit generic entangled states utilized only four observables and scheme require no prior state information. Three-qubit entanglement detection scheme is further extended to the general case of three-qubit pure states and the effect of mixedness present in the state is also investigated. This thesis also explores the quantum correlations possessed by mixed and/or separable states \eg non-classical correlations. A positive map-based witness is used to detect the non-classicality in the experimentally created non-classically correlated states. Results of non-classicality detection are compared with the quantum correlation measure QD while the state is evolving under free NMR Hamiltonian. The salient feature of the developed experimental scheme is that a single-shot experiment is able to detect the non-classicality present in the state under investigation. Schemes to detect pseudo-bound entanglement in a qubit-ququart system are also explored. Only three observables suffice to detect the entanglement in such PPT entangled states. In all the investigations, results were verified by one or more alternative ways \eg full quantum state tomography, QD, negativity and \textit{n}-tangle.\\

Thesis is organized as follows: Chapter \ref{chap2} describes the creation and detection of entanglement in arbitrary two-qubit states. Experimental creation and detection of non-classical correlations possessed by mixed separable states is discussed in Chapter \ref{chapter_ncc}. Chapter \ref{chapter_3QEntDet} details the experimental entanglement detection as well as characterization of three-qubit random states. In Chapter \ref{BoundEnt} a more subtle type of entanglement possessed by mixed states undetectable by PPT criterion is discussed. The experimental explorations of quantum non-local nature of the quantum correlations are reported in Chapter \ref{NPA}. Thesis concludes with Chapter \ref{Summary} briefing the main results and future directions of work.

\chapter{Bipartite Entanglement Detection on an NMR Quantum Processor}\label{chap2} 
\section{Introduction}
Quantum entanglement \cite{schrodinger-mpc-35} is a striking feature exhibited by quantum systems and have no analog in classical mechanics. It has been shown \cite{aditi-cs-17} that quantum entanglement is a key resource to achieve computational speedup in quantum information processing (QIP) \cite{horodecki-rmp-09} and for other quantum communication related tasks \cite{sibasish-njp-02,aditi-pra-09,pankaj-epd-15}.  In general, the detection of entanglement is a hard problem in quantum mechanics \citep{horodecki-rmp-09}. It has been demonstrated that the entanglement classification and
detection are daunting tasks \citep{guhne-pr-09}. There have
been attempts to do so utilizing methods based on Bell-type
inequalities
\citep{marcus-pra-11,bastian-prl-11,wallman-pra-12}, quantum
state tomography \citep{white-prl-99,thew-pra-02}, dynamic
learning tools and numerical schemes \citep{behrman-qic-13},
entanglement witnesses
\citep{lewenstein-pra-00,guhne-jmo-03,arrazola-pra-12},
positive-partial-transpose mixtures
\citep{novo-pra-13,bartkiewicz-pra-15}, and expectation
values of Pauli operators
\citep{zhao-pra-13,miranowicz-pra-14}. The negativity under
partial transpose (NPT) of the density operator is a
necessary and sufficient condition for the existence of
entanglement in 2$ \otimes $2 and 2$ \otimes $3
dimensional quantum systems
\citep{horodecki-pla-96,peres-prl-96}. For quantum states in
higher dimensional Hilbert spaces there are sufficient
conditions available but complete entanglement
characterization is still an open problem
\citep{guhne-pr-09}.

The creation of entanglement in an experiment and then
detecting the same is an important theme in experimental quantum computing. Experimental
detection and characterization of entanglement was
demonstrated using optical hardware
\citep{qin-lsa-15,wang-sb-16, gou-s-china-15}. Entanglement
was explored in an NMR scenario using an entanglement
witness \citep{filgueiras-qip-12}, by measuring quantum
correlations of an unknown quantum state,
\citep{silva-prl-13} as well as by a multiple-quantum
coherence based entanglement witness
\citep{feldman-jetpl-08}. It was shown that tomography is
necessary for universal entanglement detection using
single-copy observables in a system of two qubits
\citep{lu-prl-16}. Single-copy here implies that not more
than one copy of the state was used in a single run of the
experiment designed for entanglement detection. Three NMR
qubits were used to simulate the entanglement dynamics of
two interacting fermions \citep{lu-sb-15}. Three-qubit
entanglement was characterized on an NMR quantum-information
processor \citep{dogra-pra-15,das-pra015}, and the evolution
of multiqubit entangled states was studied with a view to
control their decoherence
\citep{kawamura-ijqc-06,singh-pra-14}.

To study entanglement, at least two quantum systems are
required which can be entangled, and to begin with one can
choose two two-level quantum systems. The resultant Hilbert space, of this 2$ \otimes $2 system, is 4 dimensional. Although the NPT criterion is
necessary and sufficient to detect entanglement in this case, the pre-requisite is that the full density operator
is known \citep{horodecki-pla-96,peres-prl-96}.

A promising direction of research in the detection of
quantum entanglement is the use of local observables to find
an optimal decomposition of entanglement witnesses
\citep{guhne-pra-02}. Although the method assumes some prior
knowledge of the density matrix, it can detect entanglement
by performing only a few local measurements
\citep{guhne-ijtp-03,toth-prl-05,guhne-njp-10}. Entanglement
detection schemes were designed for pure states with totally
uncorrelated measurement settings that use only two copies
of the state \citep{tran-pra-15}. These entanglement
detection schemes were recently extended
\citep{szangolies-njp-15} to the case of completely unknown
states with no prior information. This scheme uses a set of
random local measurements and optimizes over the space of
possible entanglement witnesses that can be constructed
thereof \citep{szangolies-njp-15}.

In the current chapter, focus is on experimental use of a
set of random local measurements to detect bipartite
entanglement of unknown pure entangled states. Particularly
interest from an experimental point of view is that can one
be able to detect entanglement using a minimum number of
experimental settings. Current experiments demonstrate the
optimality of using random local measurements to detect
entanglement in a system of two qubits on an NMR quantum
information processor. The expectation values of a set of
local measurement operators are obtained and used in
semi-definite programming to thereby construct the witness
operator to detect the presence of entanglement. It is shown
that a set of three local measurements is sufficient to
unequivocally detect entanglement of most entangled states
of two qubits. States with different amounts of entanglement
are generated experimentally and  their entangled (or
separable) nature is evaluated by performing this optimal
set of local measurements. Further, results are validated by
constructing experimental tomographs of each state and
negativity is computed as a measure of entanglement from
them. With a view to generalize these methods to larger
Hilbert spaces, simulations were performed to detect
bipartite entanglement of unknown pure entangled states in a
2$ \otimes $3 dimensional system, using a set of random
measurements acting locally on the qubit and the qutrit. It
is observed that by performing a few measurements, the
entanglement of most states gets characterized.

\section{Entanglement Detection in a $ 2\otimes2 $ Dimensional Quantum System by Sub-System Measurements}\label{det_protocol2Q}

Several protocols~\citep{guhne-pr-09, dagmar-jmp-02} have been put forward to detect the entanglement  and most of them are based on full knowledge of the quantum state. Most of the proposed protocols are not readily implementable \ie they can not be measured directly in an experiment. Entanglement witnesses are the observables which can give a `yes/no' answer on the entanglement present in a state when measured in an experiment. For witness-based experimental entanglement detection, knowledge about the state is required beforehand. One may argue that if the state is already known or has been tomographed, then one can calculate its entanglement properties by using the witness. In the present study a different approach is followed, where \textit{a priori} state information is not required but instead local measurements were strategically chosen. Semidefinite programming (SDP) is used to obtain the relative weights of the expectation values of these local measurements which are then used to build the entanglement witness for the unknown state. Procedure outlined by Szangolies \textit{et al} \citep{szangolies-njp-15} is followed to construct a class of decomposable entanglement witness operators for an unknown state using random local measurements. In this protocol, once the set of measurements got fixed, the witness is optimized to increase the possibility of detecting entanglement.

Consider a composite system in a Hilbert space $
\mathcal{H}_{AB}= \mathcal{H}_A \otimes \mathcal{H}_B $. A
witness operator is a Hermitian operator $ W $ acting on the
composite Hilbert space, such that Tr$ (W\rho) > $0 for all
separable $ \rho $ and Tr$ (W\rho) \leq $0 for at least one
entangled $ \rho $. A witness is called decomposable if it can be written as linear a combination of two positive operators $P$  and $Q$ such that 

\begin{eqnarray}
W = P + Q^{T_A}
\end{eqnarray}
where the operation $ T_A $ represents the partial transposition with
respect to subsystem $A$. Further, since one would like to
build the witness operator out of local measurements,
consider local Hermitian operators $ A_i $  and $ B_j $
acting on $ \mathcal{H}_A $ and $ \mathcal{H}_B $
respectively. Indices $ i,\; j $ are the measurement labels
that one wish to carry out for each local system. Range of
the indices $ i $ and $ j $ depends upon the number of
orthogonal operators spanning an arbitrary operator acting
locally on the respective Hilbert spaces, (see
Sec-\ref{Exp_Measure_NMR}). One would therefore like the
witness operator to be given as

\begin{equation}\label{Wdecomp}
W=\sum_{i,j}c_{ij}A_i\otimes
B_j 
\end{equation}
with $ \rm c_{ij}\in \mathbb{R} $. It
should be noted here that if one were to allow $\rm A_i $ and $ \rm
B_j $ to run over a complete set of bases in the local
operator spaces, then by Bloch decomposition, every
Hermitian operator can be written in the form given in Eq.~(\ref{Wdecomp}) \citep{szangolies-njp-15}. However, in
present case first the measurements will be chosen which one
wants to experimentally perform and then witness operator
optimized in such a way that the chances of entanglement
detection get maximized.  \subsection{Semi Definite Program
(SDP) for Entanglement Detection}

Finding the expectation value of the entanglement witness
operator W (given the set of local observables  $ A_i $ and
$ B_j $) is equivalent to finding the coefficients $ c_{ij}
$ subject to the trace constraints on the witness operator.
Let us define a column vector $ \boldsymbol{c} $ where one
can take the columns of $ c_{ij} $ and stack them one below
the other and similarly define a vector $ \boldsymbol{m} $
in which experimentally measured expectation values $
\langle A_i\otimes B_j \rangle $ are stacked to form  a long
column vector such that \begin{eqnarray} \rm Tr(W\rho) =
\sum_{i,j} c_{ij}\langle A_i\otimes B_j\rangle =
\boldsymbol{\rm c^{T}.m} \end{eqnarray} The SDP looks for
the class of entanglement witness operators with unit trace
that are decomposable as $ P + Q^{T_A} $. This is the most
general witness capable of detecting states with
non-positive partial transpose as Tr$(\rho W) < 0$  implies
$ \rho^{T_A}\ngeqq 0$, since Tr$(\rho
W)=$Tr$(P\rho)+$Tr$(Q\rho^{T_A}) $, which can only be
smaller than zero if $ \rho^{T_A} $ is not positive. Hence
this decomposition ensures the detection of bipartite NPT
states. The corresponding SDP can be constructed as
\citep{szangolies-njp-15} \begin{eqnarray}\label{SDP_Def}
\rm Minimize &:& \boldsymbol{c^{T}.m} \nonumber\\ \rm s.t.
\nonumber\\ W &=& \rm P+Q^{T_A} \nonumber\\ P &\geq & 0
\nonumber\\ Q &\geq & 0 \nonumber\\ Tr(W) &=& 1
\end{eqnarray} SDP is implemented using MATLAB
\citep{MATLAB} subroutines that employed SEDUMI
\citep{strum-oms-99} and YALMIP \citep{Lofberg2004} as SDP
solvers. MATLAB script and detailed code explanation, using
data in Sec-\ref{SDP_example} as an example, is given in
Appendix-\ref{Append-A}. 

\subsection{Measuring Expectation Values via NMR Experiments}\label{Exp_Measure_NMR}

This section describe the procedure followed to
experimentally measuring the expectation values of various observables using NMR for a system of two weakly interacting spin-1/2 particles. The density operator for this system can be decomposed as a linear combination of products of Cartesian spin angular momentum operators $ I_{ni} $, with $n=1,2$ labeling the spin and $i = x,y$ or $z$ \citep{ernst-book-90}. A total of 16 product
operators completely span the space of all 4$ \times $4
Hermitian matrices. The four maximally entangled Bell states
for two qubits and their corresponding entanglement witness
operators can always be written as a linear combination of
the three product operators $ 2I_{1x}I_{2x} $, $
2I_{1y}I_{2y} $, $ 2I_{1z}I_{2z} $ and the identity
operator. The symbols $ O_i $ (1 $ \leq  i \leq $ 15) is
used to represent product operators, with the first three
symbols $ O_1 $, $ O_2 $ and $ O_3 $ representing the
operators $ 2I_{1x}I_{2x} $, $ 2I_{1y}I_{2y} $, and $
2I_{1z}I_{2z} $ respectively. Need is to experimentally
determine the expectation values of these operators $ O_i $
in state $ \rho $ whose entanglement is to be characterized.
The expectation values of these operators are mapped to the
local $z$ magnetization of either of the two qubits by
specially crafted unitary operator implemented  and are
summarized in Table~\ref{mapping2Q}. $CNOT$ is the
controlled-NOT gate with first qubit as the control qubit
and second qubit as the target qubit. $X$, $\overline{X}$,
$Y$ and $\overline{Y}$ represent local $\frac{\pi}{2}$
unitary rotations with phase $x$, $-x$, $y$ and $-y$
respectively. Subscript on $ \frac{\pi}{2} $ local unitary
rotations denotes the qubit number. The expectation values
are obtained by measuring the $z$ magnetization of the
corresponding qubit. The unitary operations given in
Table~\ref{mapping2Q}, implemented via NMR, transform the
state via a single measurement, which is completely
equivalent to the originally intended measurement of local
operators, and considerably simplifies the experimental
protocol.

\begin{table}[t]
\caption{\label{mapping2Q} All 15
observables for two qubits, mapped to the local z
magnetization of one of the qubits. This mapping allows a
simpler method to measure the expectation values of the
operators $ O_i $ and is completely equivalent to the
measurement of the original local operators.} \centering
\begin{tabular}{l l} \hline Observable & Initial state
mapped to\\
\hline
\hline
$ \langle O_{1} \rangle$ = Tr[$
\rho_{1}.I_{2z}$] & $
\rho_1=CNOT.Y_2.Y_1.\rho_0.Y_1^{\dagger}.Y_2^{\dagger}.CNOT^{\dagger}$
\\ 

$ \langle O_{2} \rangle$ = Tr[$ \rho_{2}.I_{2z}$] & $
\rho_2=CNOT.\overline{X}_2.\overline{X}_1.\rho_0.\overline{X}_1^{\dagger}.\overline{X}_2^{\dagger}.CNOT^{\dagger}$
\\ 

$ \langle O_{3} \rangle$ = Tr[$ \rho_{3}.I_{2z}$] & $\rm
\rho_{3}=CNOT.\rho_0.CNOT^{\dagger}$ \\ 

$ \langle O_{4} \rangle$ = Tr[$ \rho_{4}.I_{2z}$] & $\rm
\rho_4=CNOT.\overline{X}_2.Y_1.\rho_0.Y_1^{\dagger}.\overline{X}_2^{\dagger}.CNOT^{\dagger}$
\\ 

$ \langle O_{5} \rangle$ = Tr[$ \rho_{5}.I_{2z}$] & $\rm
\rho_5=CNOT.Y_1.\rho_0.Y_1^{\dagger}.CNOT^{\dagger}$ \\ 

$ \langle O_{6} \rangle$ = Tr[$ \rho_{6}.I_{2z}$] & $\rm
\rho_6=CNOT.\overline{Y}_2.X_1.\rho_0.X_1^{\dagger}.\overline{Y}_2^{\dagger}.CNOT^{\dagger}$
\\ 

$ \langle O_{7} \rangle$ = Tr[$ \rho_{7}.I_{2z}$] & $\rm
\rho_{7}=CNOT.X_1.\rho_0.X_1^{\dagger}.CNOT^{\dagger}$  \\ 

$ \langle O_{8} \rangle$ = Tr[$ \rho_{8}.I_{2z}$] & $\rm
\rho_{8}=CNOT.\overline{Y}_2.\rho_0.\overline{Y}_2^{\dagger}.CNOT^{\dagger}$
\\ 

$ \langle O_{9} \rangle$ = Tr[$ \rho_{9}.I_{2z}$] & $\rm
\rho_{9}=CNOT.X_2.\rho_0.X_2^{\dagger}.CNOT^{\dagger}$ \\ 

$ \langle O_{10} \rangle$ = Tr[$ \rho_{10}.I_{1z}$] & $\rm
\rho_{10}=\overline{Y}_1.\rho_0.\overline{Y}_1^{\dagger}$ \\ 

$ \langle O_{11} \rangle$ = Tr[$ \rho_{11}.I_{1z}$] & $\rm
\rho_{11}=X_1.\rho_0.X_1^{\dagger}$ \\ 

$ \langle O_{12} \rangle$ = Tr[$ \rho_0.I_{1z}$] & $\rm
\rho_0$ is initial state \\ 

$ \langle O_{13} \rangle$ = Tr[$ \rho_{13}.I_{2z}$] & $\rm
\rho_{13}=\overline{Y}_2.\rho_0.\overline{Y}_2^{\dagger}$
\\ 

$ \langle O_{14} \rangle$ = Tr[$ \rho_{14}.I_{2z}$] & $\rm
\rho_{14}=X_2.\rho_0.X_2^{\dagger}$ \\ 

$ \langle O_{15} \rangle$ = Tr[$ \rho_{0}.I_{2z}$] & $\rm
\rho_0$ is initial state  \\ \hline \hline \end{tabular}
\end{table}

\begin{figure}[t] \begin{center}
\includegraphics[angle=0,scale=1.3]{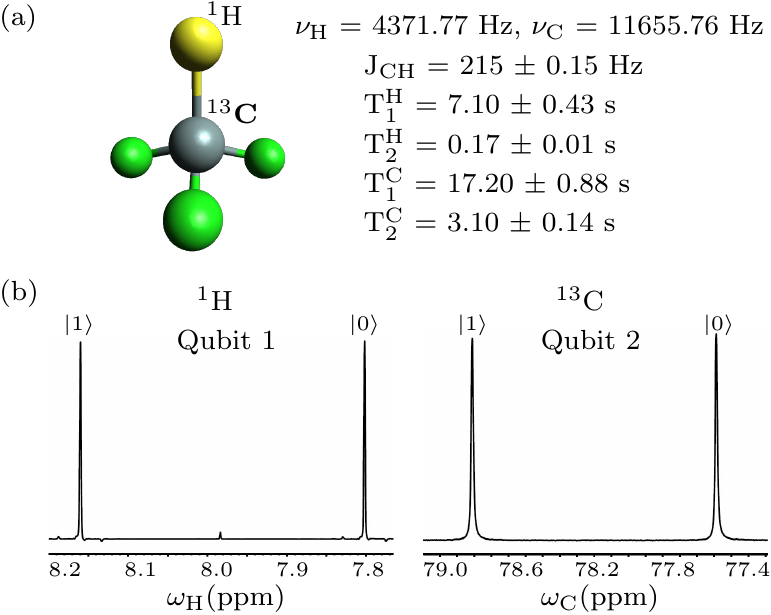} \caption{(a)
Structure of the $ ^{13} $C enriched chloroform molecule
with the two qubits labeled as $ ^1 $H and $ ^{13} $C.
Tabulated experimental NMR parameters with chemical shifts $
\nu_i $ and spin-spin coupling $J_{ij} $ in Hertz and
relaxation times T$ _1 $ and T$ _2 $ in seconds and (b) $ ^1
$H and $ ^{13} $C NMR spectra obtained at thermal
equilibrium after a $  \frac{\pi}{2} $ readout pulse. The
spectral resonances of each qubit are labeled by the logical
state  $ \lbrace \vert 0 \rangle $, $ \vert 1 \rangle
\rbrace $ of the passive qubit.} \label{mol2Q} \end{center}
\end{figure}

Description of the quantum system used for experimental
demonstration of the entanglement detection protocol is as
follows.  The two NMR qubits were encoded in a molecule of $
^{13} $C enriched chloroform, with the $ ^1 $H and $ ^{13}
$C nuclei and encodes the first and second qubits,
respectively. The molecular structure, experimental
parameters and NMR spectrum of the thermal initial state are
shown in Fig~\ref{mol2Q}.

Experiments were performed at room temperature (293K) on a
Bruker Avance III 600 MHz NMR spectrometer equipped with a
quadruple resonance inverse (QXI) probe. The Hamiltonian of
this weakly interacting two-qubit system in the rotating
frame \citep{ernst-book-90} is \begin{equation} H = \nu_H
I_z^H + \nu_C I_z^C + J_{CH}I_z^HI_z^C \end{equation} where
$  \nu_H $, $ \nu_C $ are the Larmor resonance frequencies;
$ I_z^H $, $ I_z^C $ are the $z$ components of the spin
angular momentum operators for the proton and carbon nuclei,
respectively; and $ J_{CH} $ is the spin-spin coupling
constant.

The two-qubit system was initially prepared in the
pseudo-pure state $ \vert 00 \rangle $ using the spatial
averaging technique \citep{cory-physD-98}, with the density
operator given by
\begin{equation}\label{pps2Q_eq}
\rho_{00}=\frac{1}{4}(1-\epsilon)\mathbb{I}+\epsilon\vert
00\rangle\langle00\vert
\end{equation}
where $ \epsilon
\approx 10^{-5} $ is an estimate of the thermal
polarization. One may note here that NMR is an ensemble
technique that can experimentally observe only deviation
density matrices (with zero trace). The state fidelity was
calculated from the Uhlmann-Jozsa relation
\citep{uhlmann-rpmp-76, jozsa-jmo-94}
\begin{equation}\label{fidelity_eq} F=\left[Tr\left(
\sqrt{\sqrt{\rho_{{\rm th}}}\rho_{{\rm ex}} \sqrt{\rho_{{\rm
th}}}}\right)\right]^2 \end{equation} where $ \rho_{th} $
and $ \rho_{ex} $ represent the theoretical and
experimentally measured density operators, respectively. The
experimentally prepared pseudo-pure state was tomographed
using full quantum state tomography
\citep{leskowitz-pra-04}, and the state fidelity was
computed to be 0.98 $ \pm $ 0.01.

\begin{figure}[t]
\begin{center}
\includegraphics[angle=0,scale=1.3]{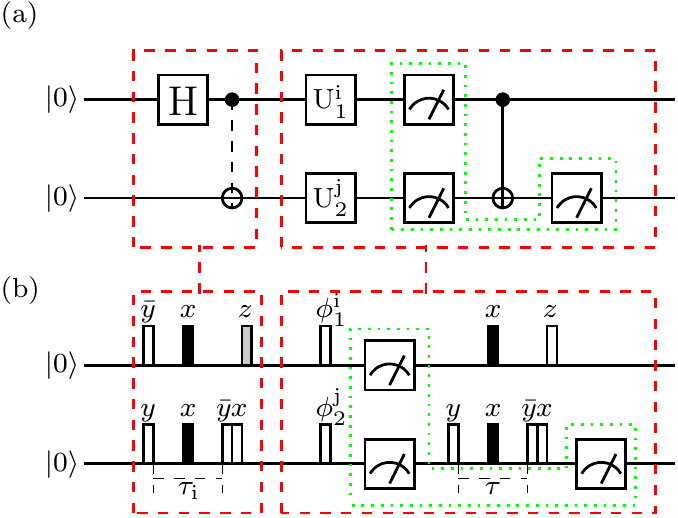} \caption{(a) Quantum circuit to implement the entanglement detection
protocol. The first red box creates states with different
amounts of entanglement. The second red box maps the
observables $ O_i $ to the $z$ magnetization of either
qubit. Only one $z$ magnetization is finally measured in an
experiment (inner green box). (b) NMR pulse sequence for the
quantum circuit. Unfilled rectangles represent $
\frac{\pi}{2} $ pulses, while solid rectangles represent $
\pi $ pulses. Tuning of the interaction between qubits is
controlled by varying the $ \tau_i $ time period and the
z-pulse rotation angle (gray rectangle). Pulse phases are
written above each pulse, with a bar indicating negative
phase. The $ \tau $ evolution period was fixed at $
\frac{1}{2J_{CH}} $, where $ J_{CH} $ is the strength of the
scalar coupling.}
\label{ckt2Q}
\end{center}
\end{figure}

The quantum circuit to implement the two-qubit entanglement
detection protocol is shown in Fig. \ref{ckt2Q}(a). The
first block in the circuit (enclosed in a dashed red box)
transforms the $ \vert 00 \rangle $ pseudo-pure state to an
entangled state with a desired amount of entanglement.
Control of the entanglement present in the state was
achieved by controlling the time evolution under the
nonlocal interaction Hamiltonian. A controlled-NOT (CNOT)
gate that achieves this control is represented by a dashed
line. The next block of the circuit (enclosed in a dashed
red box) maps any one of the observables $ O_i $ (1 $ \leq i
\leq $ 15) to the local $z$ magnetization of one of the
qubits, with $ U_1^i $ and $ U_2^j $ representing local
unitaries (as represented in Table~\ref{mapping2Q}). The
dashed green box represents the measurement. Only one
measurement is performed in a single experiment. The NMR
pulse sequence to implement the quantum circuit for
entanglement detection using random local measurements,
starting from the pseudo-pure state $ \vert 00 \rangle $, is
shown in Fig.~\ref{ckt2Q}(b). Unfilled rectangles represent
$ \frac{\pi}{2} $ pulses, while solid rectangles denote $
\pi $ pulses. Refocusing pulses were used in the middle of
all J-evolution periods to compensate for undesired chemical
shift evolution. Composite pulses are represented by z in
the pulse sequence, where each composite z rotation is a
sandwich of three pulses: $ xy\overline{x} $. The CNOT gate
represented by the dashed line in Fig.~\ref{ckt2Q}(a) was
achieved experimentally by controlling the evolution time $
\tau_i $ and the angle of $z$ rotation (the gray shaded
rectangle); $ \phi_1^i $ and $ \phi_2^j $ are local
rotations and depend upon which $ \langle O_i \rangle $
value is being measured, and the $ \tau $ time interval was
set to $ \tau=\frac{1}{2J_{CH}} $. 

\subsection{An Example To Demonstrate Entanglement Detection via SDP}\label{SDP_example}

Following is an explicit example to demonstrate how the SDP
can be used to construct an entanglement witness. Consider
the Bell state $ \vert
\phi^-\rangle=\frac{1}{\sqrt{2}}(\vert 00 \rangle - \vert 11
\rangle) $. The corresponding density matrix can be written
as a linear superposition of two spin product

\begin{figure}[H]
\begin{center}
\includegraphics[angle=0,scale=1.3]{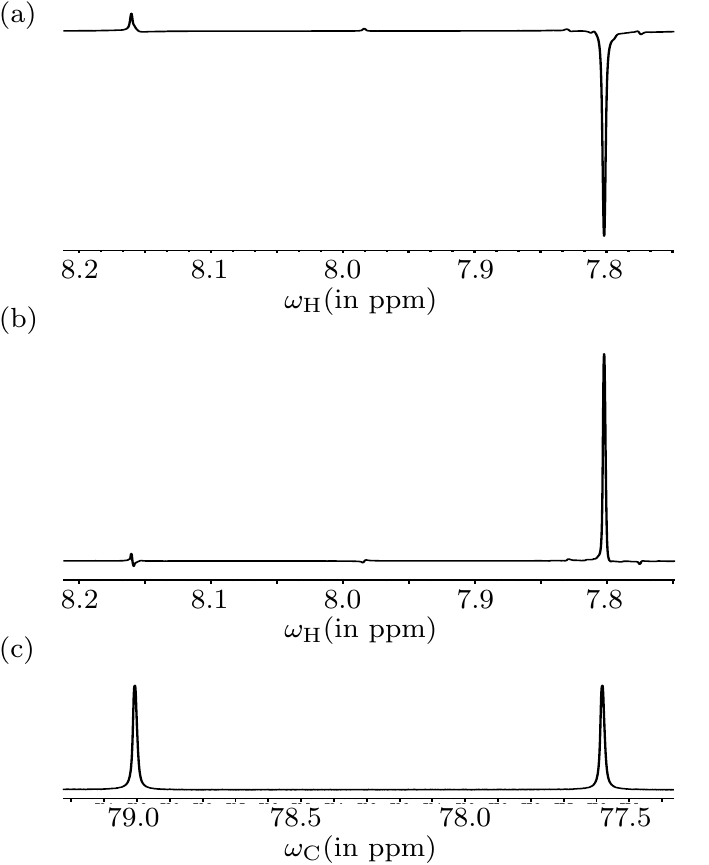}
\caption{NMR spectra of $ ^1 $H and $ ^{13} $C nuclei, showing the experimentally measured expectation values of (a) $ O_1 $, (b) $ O_2 $, and (c) $ O_3 $ in the Bell state $ \vert \phi^-\rangle=\frac{1}{\sqrt{2}}(\vert 00 \rangle - \vert 11 \rangle) $. The expectation values have been measured by the NMR pulse sequences given in Fig.~\ref{ckt2Q} corresponding to respective unitary mapping operator in Table~\ref{negTab}.}
\label{oiz}
\end{center}
\end{figure}
\noindent operators \citep{ernst-book-90} as
\begin{equation} \rho =
\frac{\mathbb{I}}{4}+a2I_{1x}I_{2x}+b2I_{1y}I_{2y}+c2I_{1z}I_{2z}
\end{equation} where $b\; =\; c\; =\; -a\; =\; \frac{1}{1}$.
Since it is known that the given state is entangled, the
corresponding entanglement witness can be constructed as
\citep{guhne-pr-09} \begin{eqnarray} W_{\phi^-} &=&
c_{opt}\mathbb{I}-\rho \nonumber\\ &=&
\frac{\mathbb{I}}{4}-a2I_{1x}I_{2x}-b2I_{1y}I_{2y}-c2I_{1z}I_{2z}
\end{eqnarray} where $ c_{opt} $ is the smallest possible
value such that the witness is positive on all separable
states; for Bell states $ c_{opt}=\frac{1}{2} $. Noting that
Tr($ \rho W_{\phi^-} ) = -\frac{1}{2} < 0$ , by definition
W$ \phi^- $ detects the presence of entanglement in $ \rho
$. However, the detection protocol discussed in
Sec-\ref{det_protocol2Q} has to deal with the situation
where the state is unknown. The question now arises whether
the SDP method is able to find the minimum value of $
\boldsymbol{c^T.m} $ such that the correct $W _{\phi^-} $ is
constructed.

For the Bell state $ \vert \phi^-\rangle $, the expectation
values $ \langle O_1 \rangle $, $  \langle O_2 \rangle $ and
$ \langle O_3 \rangle $ yield $-$ 1/2, 1/2, and 1/2 ,
respectively. The experimental NMR spectra obtained after
measuring $  \langle O_1 \rangle $, $ \langle O_2 \rangle $
and $ \langle O_3 \rangle $ in state $ \vert \phi^-\rangle $
are shown in Fig.~\ref{oiz}, with measured expectation
values of $-$0.490 $ \pm $ 0.021, 0.487 $ \pm $ 0.030, and
0.479 $ \pm $ 0.015, respectively (these values correspond
to the area under the absorptive peaks normalized with
respect to the pseudo-pure state). These experimental
expectation values are used to construct the vector
\textbf{m}. The SDP protocol performs minimization under the
given constraints and, for this Bell state, is indeed able
to construct W$ _{\phi^-} $ as well as the exact values of
$a$, $b$, and $c$ which make up the vector \textbf{c}. Since
the minimum value of $ \boldsymbol{c^T.m} <$ 0 is achieved,
it confirms the presence of entanglement in the state. See
Appendix-\ref{Append-A} for MATLAB code used for
entanglement detection and the SDP result.

\subsection{Entanglement Detection in Unknown 2$\otimes$2 States} This section details the detection of entanglement
in states with varying amounts of entanglement. The
entanglement detection protocol is implemented
experimentally on several different states: four maximally
entangled states (labeled as B1, B2, etc.), two separable
states (labeled as S1 and S2), and 14 non-maximally
entangled states (labeled as E1, E2, E3, . . . ).

To prepare the 14 entangled states E1 to E14 (having
different amounts of entanglement), the control on the
amount of entanglement in the state was achieved by varying
the time interval $ \tau_i $ and the angle $ \theta $ of the
$z$ rotation [Fig. \ref{ckt2Q}(b)]. $ \theta
=n\frac{\pi}{30} $ and $ \tau_i = n\frac{30}{J_{CH}} $ was
used, with 1 $ \leq n  \leq $ 14. These choices for $ \theta
$ and $ \tau_i $ represent a variation of the rotation angle
in a two-qubit controlled-rotation NMR gate and led to a
wide range of entanglement in the generated states (as
tabulated in Table~\ref{negTab}). 

To characterize the amount of entanglement, the entanglement
measure negativity $ \mathcal{N} $ \citep{horodecki-pla-96}
is used and is given as:

\begin{equation} \mathcal{N}=\Vert \rho^{PT} -1 \Vert
\end{equation}
where $ \rho^{PT} $ denotes partial
transposition with respect to one of the qubits and $ \Vert
\cdot \Vert $ represents the trace norm. A nonzero
negativity confirms the presence of entanglement in $ \rho $
and can be used as a quantitative measure of entanglement.
The states prepared ranged from nearly separable (E1, E2
with a low value of negativity) to nearly maximally
entangled (E13, E14 with high negativity values). The
experimental results of the entanglement detection protocol
for two qubits are tabulated in Table~\ref{negTab}.

\begin{figure}[H]
 \begin{center}
\includegraphics[angle=0,scale=1.3]{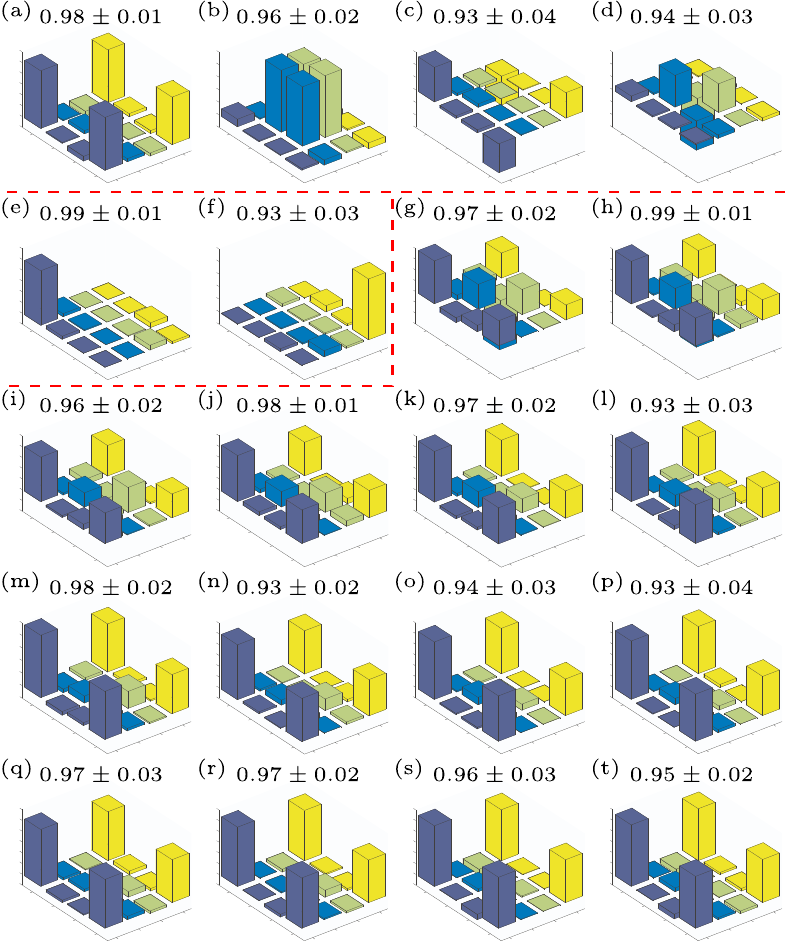} \caption{Real part of the tomographed density matrix for the states
described in Table~\ref{negTab}. (a) to (d) Maximally
entangled Bell states and (e) and (f) separable states. (g)
to (t) The tomographs represent states of different
quotients of entanglement. The state fidelity is written
above  each tomograph. The arrangement of the rows and
columns of the bar graphs is per the computational basis of
the two-qubit system $\lbrace \vert 00 \rangle $, $ \vert 01
\rangle $, $ \vert 10 \rangle $, $ \vert 11 \rangle
\rbrace$.}
\label{tomo}
\end{center}
\end{figure}
For some of the non-maximally entangled states, more than
three local measurements had to be used to detect
entanglement. For instance, SDP required six local
measurement to build the vector $ \boldsymbol{m} $ for the
E$ _8 $ state in Table~\ref{negTab} and to establish that
min($ \boldsymbol{c^T.m}) <$  0.

As is evident from Table~\ref{negTab}, this method of making
random local measurements on an unknown state followed by
SDP to construct an entanglement witness is able to
successfully detect the presence of quantum entanglement in
almost all the experimentally created states. 

\begin{table}[h]
\caption{\label{negTab} Results of entanglement detection via local measurements followed by SDP. States are labeled as B, S, and E, indicating maximally entangled, separable, and non-maximally entangled, respectively. The second and third columns contain the theoretically expected and experimentally obtained values of the entanglement parameter negativity $ \mathcal{N} $. The $ \surd $ in the last column indicates the success of the experimental protocol in detecting entanglement.}
\centering
\begin{tabular}{c | c c c}
\hline \scriptsize & \multicolumn{2}{c}{$ \mathcal{N} $} &  \\
& \multicolumn{2}{c}{$ \mathclap{\rule{3.2cm}{0.4pt}} $}&
Entanglement \\ State & Theo. & Expt. &  Detected \\
\hline
\hline
B$_1$ & 0.500 & 0.486 $ \pm $ 0.011 & $\surd $ \\
B$_2$ & 0.500 & 0.480 $ \pm $ 0.013 & $\surd $ \\
B$_3$ & 0.500 & 0.471 $ \pm $ 0.021 & $\surd $ \\
B$_4$ & 0.500 & 0.466 $ \pm $ 0.025 & $\surd $ \\
S$_1$ & 0.000 & 0.000 $ \pm $ 0.000 & $\surd $ \\
S$_2$ & 0.000 & 0.000 $ \pm $ 0.000 & $\surd $ \\
E$_1$ & 0.052 & 0.081 $ \pm $ 0.005 & $\times$ \\
E$_2$ & 0.104 & 0.088 $ \pm $ 0.024 & $\times$ \\
E$_3$ & 0.155 & 0.177 $ \pm $ 0.015 & $\surd $ \\
E$_4$ & 0.203 & 0.182 $ \pm $ 0.031 & $\surd $ \\
E$_5$ & 0.250 & 0.212 $ \pm $ 0.029 & $\surd $ \\
E$_6$ & 0.294 & 0.255 $ \pm $ 0.033 & $\surd $ \\
E$_7$ & 0.335 & 0.297 $ \pm $ 0.045 & $\surd $ \\
E$_8$ & 0.372 & 0.351 $ \pm $ 0.039 & $\surd $ \\
E$_9$ & 0.405 & 0.400 $ \pm $ 0.033 & $\surd $ \\
E$_{10}$ & 0.433 & 0.410 $ \pm $ 0.040 & $\surd $ \\
E$_{11}$ & 0.457 & 0.430 $ \pm $ 0.037 & $\surd $ \\
E$_{12}$ & 0.476 & 0.444 $ \pm $ 0.029 & $\surd $ \\
E$_{13}$ & 0.489 & 0.462 $ \pm $ 0.022 & $\surd $ \\
E$_{14}$ & 0.497 & 0.473 $ \pm $ 0.025 & $\surd $ \\
\hline
\hline
\end{tabular}
\end{table}

The protocol failed to detect entanglement in states E$ _1 $
and E$ _2 $, a possible reason for this being that these
states have a very low negativity value (very little
entanglement), which is of the order of the experimental
error. In order to validate the experimental results full
quantum state tomography of all experimentally prepared
states is also performed. The resulting tomographs and
respective fidelities are shown in Fig.~\ref{tomo}, and the
negativity parameter obtained from the experimental
tomographs in each case is tabulated in Table~\ref{negTab}.
Figures ~\ref{tomo}(a) to ~\ref{tomo}(d) correspond to
the maximally entangled Bell states B$ _1 $ to B$ _4 $,
respectively, while Figs.~\ref{tomo}(e) and
~\ref{tomo}(f) are tomographs for the separable states S$
_1 $ and S$ _2 $, respectively, and Figs. ~\ref{tomo}(g)
to ~\ref{tomo}(t) correspond to states E$ _1 $ to E$ _{14}
$, respectively. The fidelity of each experimentally
prepared state is given above its tomograph in the figure.
Only the real parts of the experimental tomographs are
shown, as the imaginary parts of the experimental tomographs
turned out to be negligible.

\section{Entanglement Detection in a 2$ \otimes $3 Dimensional Quantum System}

The orthonormal basis states for a 2$ \otimes $3 dimensional
qubit-qutrit system $  \lbrace \vert ij \rangle: \; i\; =\;
0,\; 1 ,\; j\; =\; 0,\; 1,\; 2 \rbrace $ can be written in
the computational basis for the qubit $\lbrace \vert 0
\rangle $, $ \vert 1 \rangle \rbrace$ and the qutrit
$\lbrace \vert 0 \rangle $, $ \vert 1 \rangle $ and $ \vert
2 \rangle \rbrace$, respectively. It has been previously
shown that any arbitrary pure state of a hybrid qubit-qutrit
2$ \otimes $3 system can be transformed to one of the states
of a two-parameter class (with two real parameters) via
local operations and classical communication (LOCC) and that
states in this class are invariant under unitary operations
of the form U$ \otimes $U on the 2$ \otimes $3 system
\citep{chi-jpamg-03}. The state for such a bipartite 2$
\otimes $3 dimensional system can be written as
\citep{chi-jpamg-03} \begin{equation}\label{qubitqutritEqn}
\rho=\alpha \left[  \vert 02 \rangle\langle 02 \vert + \vert
12 \rangle\langle 12 \vert \right] + \beta\left[\vert \phi^+
\rangle\langle \phi^+ \vert + \vert \phi^- \rangle\langle
\phi^- \vert +\vert \psi^+ \rangle\langle \psi^+ \vert
\right] + \gamma \vert \psi^- \rangle\langle \psi^- \vert
\end{equation} where $ \vert \phi^{\pm}
\rangle=\frac{1}{\sqrt{2}}(\vert 00 \rangle \pm \vert 11
\rangle ) $ and $ \vert \psi^{\pm}
\rangle=\frac{1}{\sqrt{2}}(\vert 01 \rangle \pm \vert 10
\rangle ) $ are the maximally entangled Bell states. The
requirement of unit trace places a constraint on the real
parameters $ \alpha $, $ \beta $ and $ \gamma $ as
\begin{equation}
2\alpha + 3\beta + \gamma = 1
\end{equation}
This constraint implies that one can
eliminate one of the three parameters, and one can rewrite $
\beta $ in terms of $ \alpha $ and $ \gamma $ ; however, the
entire analysis is valid for the other choices as well. The
domains for $ \alpha $ and $ \gamma $ can be calculated from
the unit trace condition and turn out to be 0 $ \leq $ $
\alpha $ $ \leq $ 1/2 and 0 $ \leq $ $ \gamma $ $ \leq $ 1.
The Peres-Horodecki positive-partial-transposition (PPT)
criterion is a necessary and sufficient condition for 2$
\otimes $3 dimensional systems and can hence be used to
characterize the entanglement of $ \rho $ via the
entanglement measure negativity $ \mathcal{N} $ . The
partial transpose with respect to the qubit for the
two-parameter class of states defined in
Eq.~(\ref{qubitqutritEqn}) can be written as \begin{eqnarray}
\rho^{PT} = \alpha \left[  \vert 02 \rangle\langle 02 \vert
+ \vert 12 \rangle\langle 12 \vert \right] &+& \frac{(\beta
+ \gamma)}{2}\left[ \vert \phi^- \rangle\langle \phi^- \vert
+\vert 10 \rangle\langle 10 \vert +\vert 01 \rangle\langle
01 \vert \right] \nonumber\\ &+& \frac{(3\beta - \gamma)}{2}
\vert \phi^+ \rangle\langle \phi^+ \vert \end{eqnarray} The
negativity $ \mathcal{N}(\rho) $ for the two-parameter class
of states can be calculated from its partial transpose and
is given by \citep{chi-jpamg-03}
\begin{equation}
\rm \mathcal{N}(\rho)=max\lbrace (2\alpha+2\gamma-1),0\rbrace
\end{equation}
Clearly, states with 0.5 $ < \alpha+\gamma
\leq $ 1 have nonzero negativity (\ie, are NPT) and are
hence entangled. To extend the Bloch representation for
qubits to a qubit-qutrit system described by a 2$ \otimes $3
dimensional hybrid linear vector space an operator O
operating on this joint Hilbert space can be written as
\citep{jami-ijp-05}

\begin{eqnarray}\label{qubitqutritDecmpEqn}
O=\frac{1}{6}\left[
\mathbb{I}_2\otimes\mathbb{I}_3+\sigma^A.\overrightarrow{u}\otimes\mathbb{I}_3+\sqrt{3}\mathbb{I}_2\otimes\lambda^B.\overrightarrow{v}+\sum_{i=1}^3\sum_{j=1}^8\beta_{ij}(\sigma_i^A\otimes\lambda_j^B)\right]
\end{eqnarray} where $ \overrightarrow{u} $ and $
\overrightarrow{v} $ are vectors belonging to linear vector
spaces of dimension 3 and 8 respectively, $ \mathbb{I}_2 $
and $ \mathbb{I}_2 $ are identity matrices of dimensions 2
and 3 respectively and $  \sigma_i $ are the Pauli spin matrices used to span operators acting on the Hilbert space of the qubit. $ \rm \lambda_j $ are the Gell-Mann matrices \citep{gelmann-osti-61}, used to span operators acting on the Hilbert space of the qutrit; other isomorphic choices are equally valid. A Hermitian witness operator can be constructed for every entangled quantum state, and the
expectation value of the witness operator can be locally
measured by decomposing the operator as a weighted sum of
projectors onto product-state vectors
\citep{horodecki-pla-96,bourennane-prl-04,brando-pra-05}.
The $ \rho $ for the 2$ \otimes $3 system given in
Eq.~(\ref{qubitqutritEqn}) is NPT for 0.5 $ < (\alphaα +
\gamma ) \leq $ 1. The eigenvalues for $ \rho^{PT} $ (where
PT represents partial transposition with respect to the
qubit) are $ \alpha $, $ \frac{1}{2}(1  - 2 \alpha  -  2
\gamma )  $, and $\frac{1}{6}(1 - 2 \alpha + 2 \gamma )$.
The eigenvalue $ \frac{1}{2}(1  - 2 \alpha  -  2 \gamma ) $
remains negative for NPT states, and the corresponding
eigenvector is denoted by $ \vert \eta \rangle $. The
corresponding entanglement witness operator can be written
as $ \rm W = (\vert \eta \rangle \langle \eta \vert )^{PT} $
with it's matrix representation

\begin{gather} W=\frac{1}{2} \begin{bmatrix} 1 & 0 & 0 & 0 &
0 & 0 \\ 0 & 0 & 0 & 1 & 0 & 0 \\ 0 & 0 & 0 & 0 & 0 & 0 \\ 0
& 1 & 0 & 0 & 0 & 0 \\ 0 & 0 & 0 & 0 & 0 & 0 \\ 0 & 0 & 0 &
0 & 0 & 1 \end{bmatrix}
\end{gather}
The entanglement
witness $W$ is capable of detecting entanglement of the 2$ \otimes $3 dimensional $ \rho $ given in
Eq.~(\ref{qubitqutritEqn}). Once can explore the decomposition of the entanglement witness $W$ in terms of local observables, so that it can used to detect entanglement of the two-parameter class of states of the 2$ \otimes $3 dimensional $ \rho $. The explicit decomposition of $W$ as
per Eq.~(\ref{qubitqutritDecmpEqn}) results in the following:
\begin{gather}
\label{uvBetaEqn}
\overrightarrow{u}=
\begin{bmatrix}
0 \\ 0 \\ 0
\end{bmatrix} ,
\;\;\;
\overrightarrow{v}=
\begin{bmatrix}
0 \\ 0 \\ 0 \\ 0 \\ 0 \\
0 \\ 0 \\ 1
\end{bmatrix}
\;\;\;
 \rm and,\;\;\;
\beta=\frac{1}{2}
\begin{bmatrix}
1 & 0 & 0 & 0 & 0 & 0 & 0 & 0 \\
0 & 1 & 0 & 0 & 0 & 0 & 0 & 0 \\
0 & 0 & 1 & 0 & 0 & 0 & 0 & 0
\end{bmatrix}
\end{gather}
\noindent The components of $ \overrightarrow{u} $ and $
\overrightarrow{v} $, \ie, $ u_i(i = 1,2,3) $ and $  v_j(j
=1,2,3, . . . ,8) $ can be obtained from $  u_i =Tr[
W(\sigma_i \otimes \mathbb{I}_3)] $ and $  v_j = Tr [
W(\mathbb{I}_2\otimes \lambda_j)] $. Similarly, the elements
of the matrix $ \beta $ can be obtained from $ \beta_{ij} =
Tr[ W(\sigma_i \otimes \lambda_j)] $. There are 35 real
coefficients in the expansion in
Eq.~(\ref{qubitqutritDecmpEqn}), of which 3 coefficients
constitute $ \overrightarrow{u} $, 8 coefficients constitute
$ \overrightarrow{v} $, and the remaining 24 are contained
in the $ \beta $ matrix. Each nonzero entry in $
\overrightarrow{u} $, $  \overrightarrow{v} $, or $ \beta $
matrix is the contribution of the corresponding qubit-qutrit
product operator \citep{ernst-book-90} used in the
construction of operator $W$. Hence one can infer by
inspection of the nonzero matrix entries in
Eq.~(\ref{uvBetaEqn}) that one requires the expectation values
of at least four operators in a given state in order to
experimentally construct the witness operator W. While the
maximum number of expectation values required to be measured
is four, the question remains if this is an optimal set or
if one can find a smaller set which will still be able to
detect entanglement.

\begin{figure}[H] \begin{center}
\includegraphics[angle=0,scale=1.6]{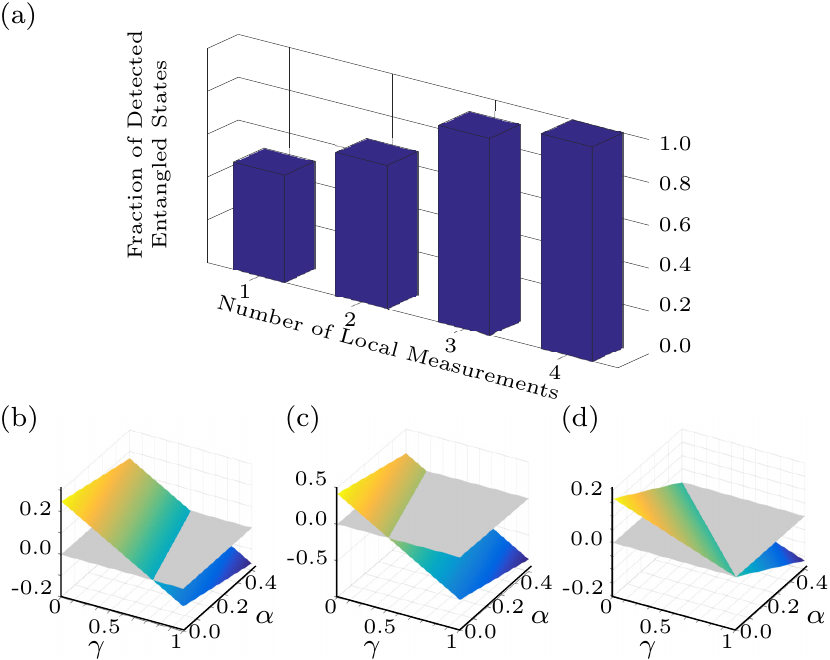} \caption{(a) Bar
graph showing the fraction of the detected entangled states
plotted as a function of the number of local measurements
from the simulation on the qubit-qutrit system. Plots of (a)
$ Tr(W_{I}\rho_{AB}) $, (b) $ Tr(W_{II}\rho_{AB}) $, and (c)
$ Tr(W_{III}\rho_{AB}) $ for 0 $ \leq \alpha \leq $ 0.5 and
0 $ \leq \gamma \leq $ 1. The entanglement witness operators
$ W_{I} $, $ W_{II} $, $  W_{III} $ are constructed by
choosing sets of one, two, and three random local
measurements at a time, respectively. A reference plane
(gray shaded) at vanishing trace is also plotted to better
differentiate positive and negative values.}
\label{qubitQutritResFig} \end{center} \end{figure}

Fraction of entanglement was computationally detected by gradually increasing the number of local observations, and the results
of the simulation are depicted in Fig. 5(a) as a bar chart.
One may note here that even if only one observable [one element of the $ \beta $ matrix in Eq.~(\ref{qubitqutritDecmpEqn})] is measured, half of the
randomly generated entangled states are detected. As the
number of measured observables is increased, the fraction of
detected entangled states improves, as shown in
Fig.~\ref{qubitQutritResFig}(a). To generate the bar plots
in Fig.~\ref{qubitQutritResFig}(a), only one random local
measurement is selected out of the maximum 35 possible
measurements. Only those choices which will establish a
decomposable entanglement witness of unit trace are valid.
For one such choice (denoted by $ W_I $), $ Tr(W_I\rho) $ is
plotted in Fig.~\ref{qubitQutritResFig}(b) in the range 0 $
\leq \alpha \leq $ 0.5 and 0$ \leq \gamma \leq  $1. As is
evident, this $ \rm W_I $ (based on only one random local
measurement) does not detect all the entangled states which
were detected by W. The fraction of entangled states
detected by $ W_I $ can be computed from geometry, \ie\;,
how much area that is spanned by the parameters $ \alpha $
and $ \gamma $ represents entangled states and how much of
that area is detected by the corresponding entanglement
witness operator.

If one consider two random local measurements at a time to
construct a valid entanglement witness [$  W_{II} $ in
Fig.~\ref{qubitQutritResFig}(c)], the detected fraction of
entangled states improves from 0.50 to 0.67 [the second bar
in Fig.~\ref{qubitQutritResFig}(a)]. One can observe from
the geometry that $ \rm W_{II} $ detects more entangled
states than $ \rm W_I $, but this fraction is still smaller
than those detected by W. The result of choosing three
random local measurements (denoted by witness operator $ \rm
W_{III} $) is plotted in Fig.~\ref{qubitQutritResFig}(d),
and it detects 83.3 \% of the total entangled states [the
third bar in Fig.~\ref{qubitQutritResFig}(a)]. Increasing
the set of random local measurements hence increases the
probability of detecting entanglement. The worst-case
detection fraction is shown in
Fig.~\ref{qubitQutritResFig}(a) when choosing random local
measurements. A fraction of 1 in
Fig.~\ref{qubitQutritResFig}(a) implies that the
corresponding set of four random local measurements will
always be able to detect entanglement in the state if it
exists.

\section{Conclusions}
This chapter was aimed at detecting the entanglement of a two-qubit state without any prior state information and with minimum experimental efforts. It has been successfully demonstrated in an actual experiment that the scheme based on local random measurements is able to detect the presence of entanglement, on a two-qubit NMR quantum-information processor. An optimal set of random local measurements was arrived at via semi-definite programming to construct entanglement witnesses that detect bipartite entanglement. The local measurements on each qubit were converted into a single measurement on one of the qubits by transforming the state. This was done to simplify the experimental scheme and is completely equivalent to the originally intended local measurements. Scheme based on random local measurements have been extended to hybrid systems, where qudits of different dimensionality are involved. For the particular case of a qubit-qutrit system, a simulation is performed to demonstrate the optimality of the detection scheme. Characterization of entangled states of qudits is a daunting task, and this work holds promise for further research in this direction. Results of this chapter are published in \href{http://journals.aps.org/pra/abstract/10.1103/PhysRevA.94.062309}{\rm Phys. Rev. A \textbf{94}, 062309 (2016)}.
\chapter{Non-Classical Correlations and Detection in a Single-Shot Experiment}\label{chapter_ncc}

\section{Introduction}

In this chapter, the focus is on more generalized quantum correlations possessed
even by separable states. Such quantum correlations are best captured by a nonclassicality quantifier ``quantum discord''(QD)~\cite{ollivier-prl-02}.
Quantum correlations are those correlations which are not present in classical
systems, and in bipartite quantum systems are associated with the presence of
QD~\cite{nielsen-book-02,ollivier-prl-02,modi-rmp-12}. In quantum information
theory, QD is a measure of nonclassical correlations between the subsystems of a
quantum system. It includes correlations that are due to quantum physical
effects and may not involve quantum entanglement. QD utilizes the concept of
mutual information. In a bipartite system the mutual information is a measure of
knowledge gained, by measuring one variable, about the other and involves entropy in the observed statistics. In classical information, theory there are two inequivalent expressions for the mutual information  which are equivalent for all classical probability distributions or statistics observed by measuring
systems possessing only classical correlations. In case, the system involved possesses quantum correlations then one gets different mutual information, from two expressions defined in the classical information theory, by virtue of the way they are defined. This difference is indeed the QD as it arises due to the quantum nature of the involved system to generate the statistics which is further used to compute QD. Formal definition is given in Sec-\ref{qd-dyn}.
Quantum correlations captured by QD in a bipartite mixed state can go beyond quantum entanglement and therefore can be present even  in separable states~\cite{ferraro-pra-10}.  The threshold between classical and quantum
correlations was investigated in linear-optical systems by observing the emergence of QD~\cite{karo-pra-13}.  QD was experimentally measured in systems such as NMR, that are described by a deviation density
matrix~\cite{pinto-pra-10,maziero-bjp-13,passante-pra-11}. Further,
environment-induced sudden transitions in QD dynamics and their preservation
were investigated using NMR~\cite{auccaise-prl-11-2, harpreet-discord}.

It has already been demonstrated that even with very low (or no) entanglement,
quantum information processing  can still be performed using nonclassical
correlations~\cite{datta-pra-05,fahmy-pra-08}, which are typically characterized
by the presence of QD.  However, computing and measuring QD typically involves
complicated numerical optimization and  furthermore it has been shown that
computing QD is  NP-hard~\cite{bryan-arx-10,huang-njp-14,cable-njp-15}. It is
hence of prime interest to find other means such as witnesses to detect the
presence of quantum correlations captured by QD~\cite{saitoh-qip-11}. While there have been
several experimental implementations of entanglement
witnesses~\cite{rahimi-jpamg-06,rahimi-pra-07,filgueiras-qip-11}, there have
been fewer proposals to witness nonclassicality. A nonlinear classicality
witness was constructed for a class of two-qubit systems~\cite{maziero-ijqi-12}
and experimentally implemented using NMR~\cite{auccaise-prl-11,pinto-ptrsa-12}
and was estimated in a linear optics system via statistics from a single
measurement~\cite{aguilar-prl-12}.  It is to be noted that as the state space
for classical correlated systems is not convex, a witness for nonclassicality is
more complicated to construct than a witness for entanglement and is necessarily
nonlinear~\cite{saitoh-pra-08}.

In the this chapter two qubits were used to demonstrate the experimental
detection of nonclassicality through a recently proposed positive map
method~\cite{rahimi-pra-10}. Two NMR qubits have been recently used to
demonstrate very interesting QIP phenomena such as the quantum simulation of the
ground state of a molecular Hamiltonian~\cite{du-prl-10}, the quantum simulation
of the Avian compass~\cite{pearson-sc-2016}, observing time-invariant coherence
at room temperature~\cite{silva-prl-16} and preserving QD~\cite{silva-prl-13}.
The map is able to witness nonclassical correlations going beyond entanglement,
in a mixed state of a bipartite quantum system. The method requires much less
experimental resources as compared to measurement of QD using full state
tomography and therefore is an attractive alternative to demonstrating the
nonclassicality of a separable state.  The map implementation involves two-qubit
gates and single-qubit magnetization measurements and can be achieved in a
single experimental run using NMR.  Our implementation of the nonclassicality
witness involves the sequential measurement of different free induction decays
(FIDs, corresponding to the NMR signal) in a single run of the same experiment.
This is possible since NMR measurements are nondestructive, thus allowing
sequential measurements on the same ensemble. This feature was exploited to
implement the single-shot measurement of the map value. The NMR pulse sequence
used on Bruker Avance-III spectrometer is given Appendix-\ref{SeqFID}.
Experiments were performed on a two-qubit separable state (non-entangled) which
contains nonclassical correlations.  Further, the state was allowed to freely
evolve in time under natural NMR decohering channels, and the amount of
nonclassicality present was evaluated at each time instant by calculating the
map value. Results were compare using the positive map witness with those
obtained by computing the QD via full state tomography, and a good match was
obtained.

Further, it was observed that beyond a certain time, the map was not able to
detect nonclassicality, although the QD measure indicated that nonclassicality
was present in the state. This implies that while the positive map
nonclassicality witness is easy to implement experimentally in a single
experiment and is a good indicator of nonclassicality in a separable state, it
is not able to characterize nonclassicality completely.  In our case this is
typified by the presence of a small amount of QD when the state has almost
decohered or when the amount of nonclassicality present is small. This of course
leaves open the possibility of constructing a more optimal witness.

\section{Experimental Detection of Non-Classical Correlation (NCC)} \label{fullexptl}
\subsection{Nonclassicality Witness Map Construction} \label{mapvalue} For pure quantum states of a bipartite quantum system which are
represented by one-dimensional projectors $\vert \psi\rangle \langle \psi\vert$
in a tensor product Hilbert space $ \rm {\cal H}_A \otimes {\cal H}_B $, the
only type of quantum correlation is
entanglement~\cite{guhne-pr-09,oppenheim-prl-02}. However, for mixed states the
situation is more complex and quantum correlations can be present even if the
state is separable \ie it is a classical mixture of separable pure states and
can be written as
\begin{equation}
\rm \rho_{sep}=\sum_i{w_i\rho_i^A\otimes\rho_i^B}
\label{sep}
\end{equation}
where $ \rm w_i $ are positive weights and $\rm \rho_i^A,\rho_i^B $ are pure states in
Hilbert spaces ${\cal H}_A$ and  ${\cal H}_B$ respectively~\cite{peres-prl-96}.
A separable state  is called a properly classically correlated state(PCC) if it
can be written in the form~\cite{horodecki-rmp-09}
\begin{equation}
\rm \rho_{PCC}=\sum_{i,j}{p_{ij}\vert e_i\rangle^{A}\langle e_i\vert\otimes\vert e_j\rangle^{B}\langle e_j\vert}
\label{pccform}
\end{equation}
where $ \rm p_{ij}$ is a joint probability distribution and $\rm \vert e_i\rangle^{A}$ and $
\rm \vert e_j\rangle^{B}$ are local orthogonal eigenbasis in local spaces ${\cal
H}_{\rm A}$ and  ${\cal H}_{\rm B}$ respectively.  A state that cannot be
written in the form given by Eq.~(\ref{pccform}) is called a nonclassically
correlated (NCC) state. An NCC state can be entangled or separable. The correlations in NCC states can go beyond those present in PCC states and are due to the fact that the eigenbasis for the respective subsystems may not be orthogonal~\cite{vedral-found-10}. A typical example of a bipartite two-qubit NCC state has been discussed in reference~\cite{streltsov-prl-11} and is given by: \begin{equation}\label{sigma} 
\sigma=\frac{1}{2}\left[\vert00\rangle\langle
00\vert+\vert1+\rangle\langle1+\vert\right] 
\end{equation} with $\vert + \rangle = \frac{1}{\sqrt{2}} \left(\vert 0 \rangle+\vert 1
\rangle\right)$.  In this case the state has no product eigenbasis as the eigenbasis for subsystem B, since $\vert 0 \rangle$ and $\vert + \rangle$ are
not orthogonal to each other. The state is separable (not entangled) as it can
be written in the form given by Eq.~(\ref{sep}); however since it is an NCC
state, it has non-trivial quantum correlations and has non-zero  QD. How to pin
down the nonclassical nature of such a state with minimal experimental effort
and without actually computing QD is something that is desirable. It has been
shown that such  nonclassicality witnesses can be constructed using a positive
map~\cite{rahimi-pra-10}.

The map $\mathcal{W}$ over the state space ${\cal H}={\cal H}_{\rm A} \otimes
\cal{H}_{\rm B}$ takes a state to a real number $\mathbb{R}$ \begin{equation}
\mathcal{W}:{\cal H}\longrightarrow \mathbb{R} \label{map} \end{equation} This
map is a nonclassicality witness map \ie it is capable of detecting NCC states
in ${\cal H}$ state space if and only if~\cite{rahimi-pra-10}: \begin{itemize}
\item[(a)] For every bipartite state $\rho_{{\rm PCC}}$ having a product
eigenbasis, $\mathcal{W}(\rho_{{\rm PCC}}) \geq0$.  \item[(b)] There exists at
least one bipartite state $\rho_{{\rm NCC}}$ (having no product eigenbasis) such
that  $\mathcal{W}(\rho_{{\rm NCC}})<0$.  \end{itemize} A specific non-linear
nonclassicality witness map proposed by~\cite{rahimi-pra-10} is defined in terms
of expectation values of positive Hermitian operators  $ \rm \hat{A}_1$, $ \rm
\hat{A}_2\ldots\hat{A}_m $: \begin{equation} \mathcal{W}(\rho)= \rm
c-\left(Tr(\rho\hat{A}_1)\right)\left(Tr(\rho\hat{A}_2)\right)\ldots\ldots\left(Tr(\rho\hat{A}_m)\right)
\end{equation} where $ \rm c \geq 0 $ is a real number. For the case of
two-qubit systems using the operators $ \rm A_1=\vert 00 \rangle\langle 00 \vert
$ and $ \rm A_2 = \vert 1+\rangle\langle 1+\vert $ a nonclassicality witness map
can be obtained for state in Eq.~(\ref{sigma}) as: \begin{equation}
\mathcal{W}_{\sigma}(\rho) = \rm c-\left(Tr(\rho\vert00\rangle\langle
00\vert)\right)\left(Tr(\rho\vert1+\rangle\langle 1+\vert)\right)
\label{sigmamap} \end{equation} The value of the constant $ \rm c$ in the above
witness map has to be optimized  such that for any PCC state $\rho$ having a
product eigenbasis, the condition $\mathcal{W}_{\sigma}(\rho) \ge 0 $ holds
\cite{rahimi-pra-10}. In order to optimize value of $ c $ in Eq.~\ref{sigmamap}
let us set $ \tau= \rm \vert s \rangle \langle s \vert \otimes \rho^B $ with $
\rm \vert s \rangle = \frac{1}{\sqrt{2}}( \vert 0 \rangle + e^{i \theta} \vert 1
\rangle)$; $ \theta $ be any angle and $ \rm \rho^B $ is a single qubit state.
Then, $ \rm f(\rho_{PCC}) = \left(Tr(\rho \vert 00 \rangle \langle 00 \vert)
\right)\left(Tr(\rho\vert 1+\rangle\langle 1+\vert )\right)$ is maximized for a
state written in the form of $\tau $ \ie \begin{center} $ \rm
c_{opt}=\displaystyle\ \max_{\substack{\tau}} f(\tau)= \displaystyle\
\max_{\substack{\rho^B}}
\frac{1}{4}\langle0|\rho^B|0\rangle\langle+|\rho^B|+\rangle $ \end{center}
Further using \[ \rm \rho^B= \left[ \begin{array}{cc} \rm a & b  \\ \rm b^* &
1-a  \end{array} \right]\] with $ \rm 0\leq a\leq 1$ and $\rm b$ being a complex
number. $ \rm \rho^B$ is positive requiring $ \rm |b|\leq \sqrt{a(1-a)}$. So\\
$$
{\rm c}_{\rm opt}=\displaystyle\ \max_{\substack{a,b}}
\frac{a[2Re(b)+1]}{8}=\displaystyle\ \max_{\substack{a}}
\frac{a[1+2\sqrt{a(1-a)}]}{8}
$$ 
After this step it is only a
maximization process and the above expression takes its maximum value when $ \rm
a$ is $ \rm \hat{a}=\frac{2+\sqrt{2}}{4}$, which results in 
$$
{\rm
c}_{\rm opt}=\frac{\hat{a}[1+2\sqrt{\hat{a}(1-\hat{a})}]}{8}=0.182138...
$$
The map given by Eq.~(\ref{sigmamap}) does indeed witness the
nonclassical nature of the state $\sigma$ as $ \rm
\left(Tr(\rho\vert00\rangle\langle 00\vert)\right)\left(Tr(\rho\vert
1+\rangle\langle 1+\vert)\right)$ for $\rho\equiv\sigma$ has the value 0.25,
which suggests that the state $\sigma$ is an NCC state~\cite{rahimi-pra-10}. The
value of a nonclassicality map, which when negative implicates the nonclassical
nature of the state, is  defined as its map value (MV). 
\subsection{NMR Experimental Setup For NCC Detection} \label{expt}
\begin{figure}
\begin{center} 
\includegraphics[angle=0,scale=0.3]{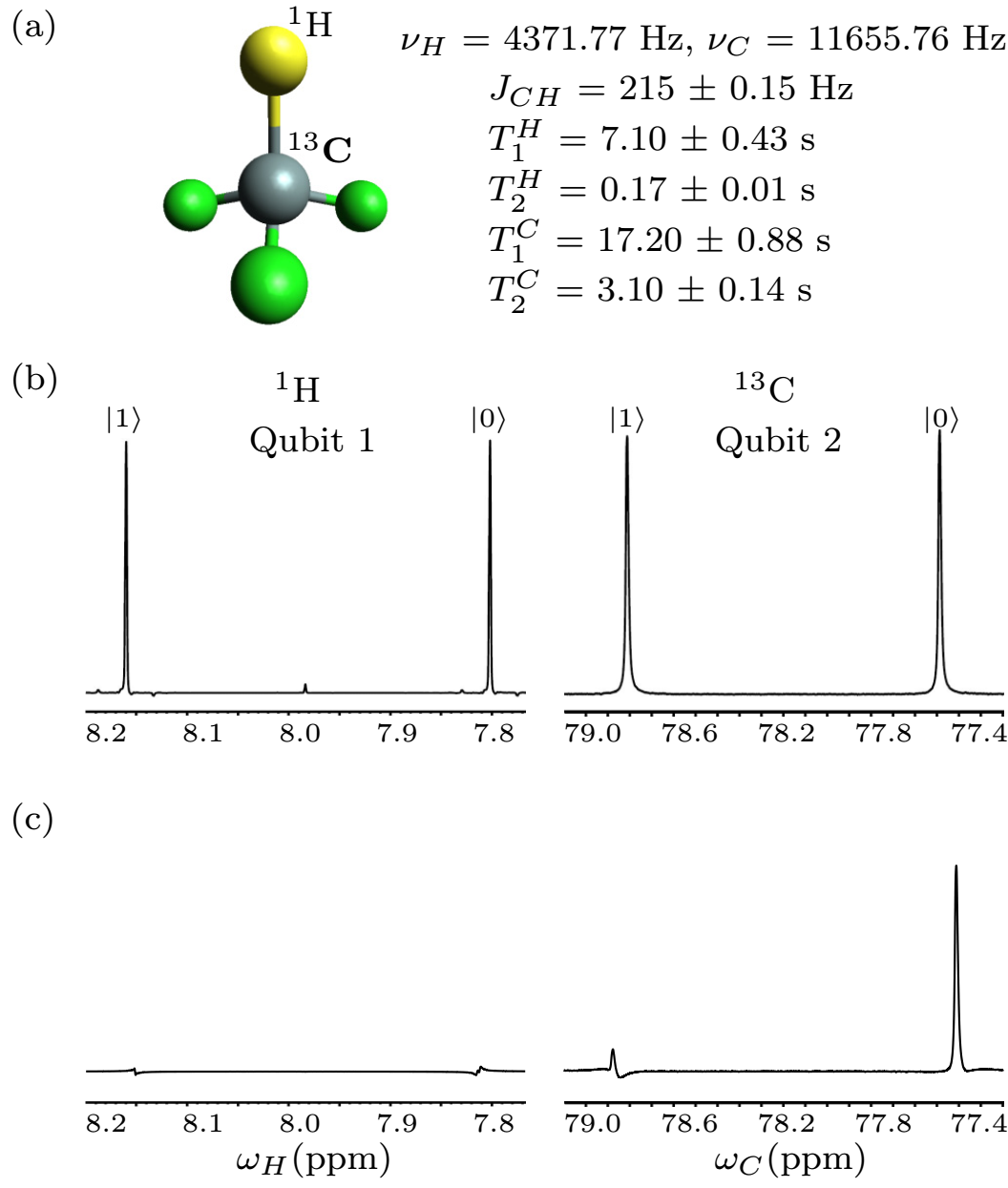}
\caption{(a) Molecular structure of $^{13}$C labeled chloroform with the two
qubits encoded as nuclear spins of ${}^{1}$H and ${}^{13}$C; system parameters
including chemical shifts $\nu_i$, scalar coupling strength $J$ (in Hz) and
relaxation times T$_{1}$ and T$_{2}$ (in seconds) are tabulated alongside. (b)
Thermal equilibrium NMR spectra of ${}^{1}$H (Qubit 1) and ${}^{13}$C (Qubit 2)
after a $\frac{\pi}{2}$ readout pulse. (c) NMR spectra of  ${}^{1}$H and
${}^{13}$C for the $\sigma$ NCC state. Each transition in the spectra is labeled
with the logical state ($\vert0\rangle$ or $\vert 1\rangle$) of the ``passive
qubit'' (not undergoing any transition).}
\label{molecule}
\end{center}
\end{figure}

\begin{figure}
\begin{center} 
\includegraphics[angle=0,scale=0.9]{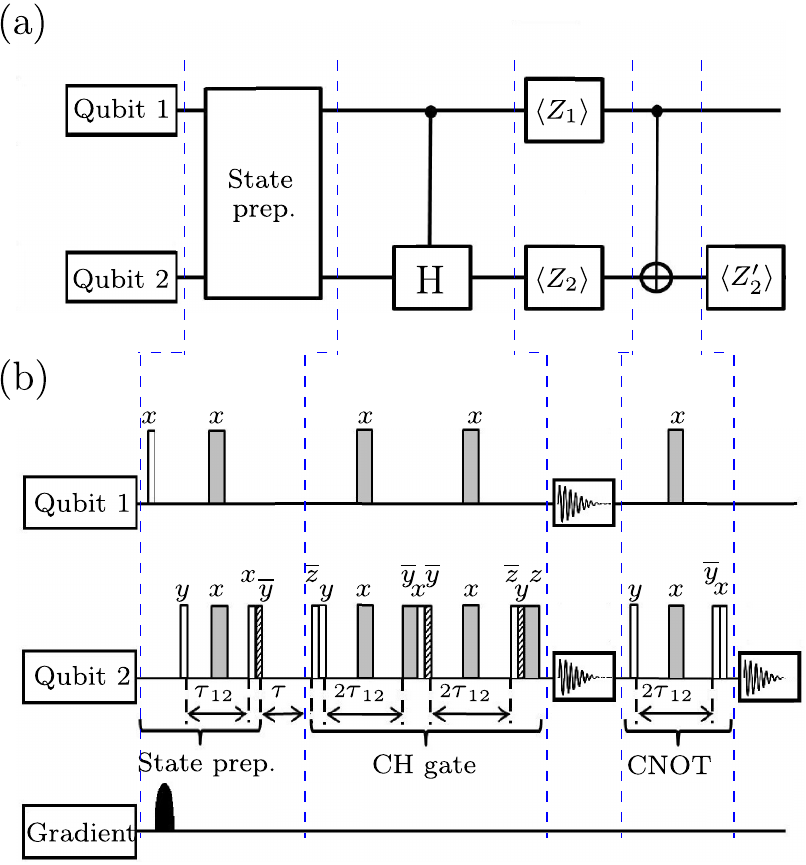}
\caption{(a) Quantum circuit  and (b) NMR pulse sequence to create and detect an
NCC state. Unfilled rectangles depict $\frac{\pi}{2}$ pulses, grey-shaded
rectangles depict $\pi$ pulses and filled rectangles depict $\frac{\pi}{4}$
pulses, respectively. Phases are written above each pulse, with a bar over a
phase indicating a negative phase.  The evolution period was set to $ \rm
\tau_{12}=\frac{1}{4J}$. The delay $\tau$ is the time for which the NCC state is
allowed to evolve before detection and the group of pulses and delays labeled as
CH gate implement a controlled-Hadamard operation. The measurements of $ \rm
\langle Z_1 \rangle$, $ \rm \langle Z_2 \rangle$ and $ \rm \langle Z_2^{\prime}
\rangle$ magnetizations in the circuit in (a) are represented by an FID
collection symbol at the corresponding points in the pulse sequence in (b).}
\label{ckt}
\end{center}
\end{figure}

In order to implemented the nonclassicality witness map $\mathcal{W}_\sigma$ on an NMR
sample of $ ^{13} $C-enriched chloroform dissolved in acetone-D6; the $ ^{1} $H
and $ ^{13} $C nuclear spins were used to encode the two qubits (see
Fig.~\ref{molecule} for experimental parameters).
Unitary operations were implemented by specially crafted transverse radio
frequency pulses of suitable amplitude, phase and duration. Since a
heteronuclear $ ^{1}$H-$^{13}$C spin system was used to encode the qubits,
standard pulse calibration methods available on the dedicated NMR spectrometer
software were used for pulse optimization and gave accurate results. A sequence
of spin-selective pulses interspersed with tailored free evolution periods were
used to prepare the system in an NCC state  as described below, written using
spin-angular momentum operators:
\begin{eqnarray*}
&& \rm I_{1z}+I_{2z} \stackrel{(\pi/2)^1_x}{\longrightarrow}-I_{1y}+I_{2z}\stackrel{Sp. Av.}{\longrightarrow} I_{2z} \stackrel{(\pi/2)^2_y}{\longrightarrow} I_{2x} \stackrel{\frac{1}{4J}}{\longrightarrow} \nonumber\\ && \quad\quad \rm \frac{I_{2x}+2I_{1z}I_{2y}}{\sqrt{2}}\stackrel{(\pi/2)^2_x}{\longrightarrow} \frac{I_{2x}+2I_{1z}I_{2z}}{\sqrt{2}} \stackrel{(-\pi/4)^2_y}{\longrightarrow} \nonumber \\ && \quad\quad\quad\quad \rm \frac{\left(I_{2z}+I_{2x}+2I_{1z}I_{2z}-2I_{1z}I_{2x}\right)}{2}
\end{eqnarray*}
One can begin with the system in thermal equilibrium and ignore the identity
part of the density matrix, which does not evolve under RF (radio frequency)
pulses. The RF pulses $( \rm \alpha)^{i}_{j}$ are written above each arrow, with
$\alpha $ denoting the pulse flip angle,  $ \rm i=1,2 $ denoting the qubit on
which the pulse is being applied and $ \rm j=x,y,z $ being the axis along 
which the pulse is applied. Spatial averaging (denoted by Sp. Av.) is achieved via a
dephasing z-gradient. The NMR spectra of the thermal state and the prepared NCC
state are shown in Fig.~\ref{molecule}(b), and the corresponding pulse sequence
is depicted in Fig.~\ref{ckt}(b).
The quantum circuit to implement the nonclassicality witness map is shown in
Fig.~\ref{ckt}(a).  The first module represents NCC state preparation using the
pulses as already described. The circuit to capture nonclassicality of the
prepared state consists of a controlled-Hadamard (CH)  gate, followed by
measurement on both qubits, a CNOT gate and finally detection
on `Qubit 2'. The CH gate is analogous to a CNOT gate, with a Hadamard gate
being implemented on the target qubit if the control qubit is in the state
$\vert 1 \rangle$ and a `no-operation' if the control qubit is in the state
$\vert 0 \rangle$.  The NMR pulse sequence corresponding to the quantum circuit
is depicted in Fig.~\ref{ckt}(b). The set of pulses grouped under the label
`State prep.' convert the thermal equilibrium state to the desired NCC state. A
dephasing z-gradient is applied on the gradient channel to kill undesired
coherences.  After a delay $\tau$ followed by the pulse sequence to implement
the CH gate, the magnetizations of both qubits were measured with
$\frac{\pi}{2}$ readout pulses (not shown in the figure).  In the last part of
detection circuit a CNOT gate is applied followed by a magnetization measurement
of `Qubit 2'; the scalar coupling time interval was set to $ \rm
\tau_{12}=\frac{1}{4J}$ where J is the strength of the scalar coupling between
the qubits. Refocusing pulses were used during all J-evolution to compensate for
unwanted chemical shift evolution during the selective pulses.  State fidelity
was computed using the Uhlmann-Jozsa
measure~\cite{uhlmann-rpmp-76,jozsa-jmo-94}(also see Eq.~(\ref{fidelity_eq})), and
the NCC state was prepared with a fidelity of 0.97 $\pm$ 0.02.

To detect the nonclassicality in the prepared NCC state via the map
$\mathcal{W}_\sigma $, the expectation values of the operators $\vert
00\rangle\langle 00\vert$ and $\vert 1+\rangle\langle 1+\vert$ are required.
Re-working the map brings it to the following form~\cite{rahimi-pra-10}
\begin{eqnarray}\label{SeqFID_eq} \rm \mathcal{W}_{\sigma}(\rho)  =
c_{opt}-\frac{1}{16}\left(1+\langle Z_1 \rangle+\langle Z_2\rangle+\langle
Z_2'\rangle\right)\times \left(1-\langle Z_1 \rangle+\langle Z_2\rangle-\langle
Z_2'\rangle\right) \end{eqnarray} where $ \rm \langle Z_1\rangle$ and $ \rm
\langle Z_2\rangle$ are the magnetizations of `Qubit 1' and `Qubit 2' after a CH
gate on the input state $\rho$, while $ \rm \langle Z_2'\rangle$ is the
magnetization of `Qubit 2' after a CNOT gate.  The theoretically expected
normalized values of $ \rm \langle Z_1\rangle$, $ \rm \langle Z_2\rangle$ and $
\rm \langle Z_2'\rangle$ for state $\rho\equiv\sigma$ are $0$, $1$ and $0$
respectively. Map value (MV) is $-0.067862<0$ and as desired this map does
indeed witness the presence of nonclassicality. The experimentally computed MV
for the prepared NCC state turns out to be $-0.0406 \pm 0.0056$, proving that
the map is indeed able to witness the nonclassicality present in the state.
\subsection{Map Value Dynamics} \label{mv-dyn}
\begin{figure}
\begin{center}
\includegraphics[angle=0,scale=0.65]{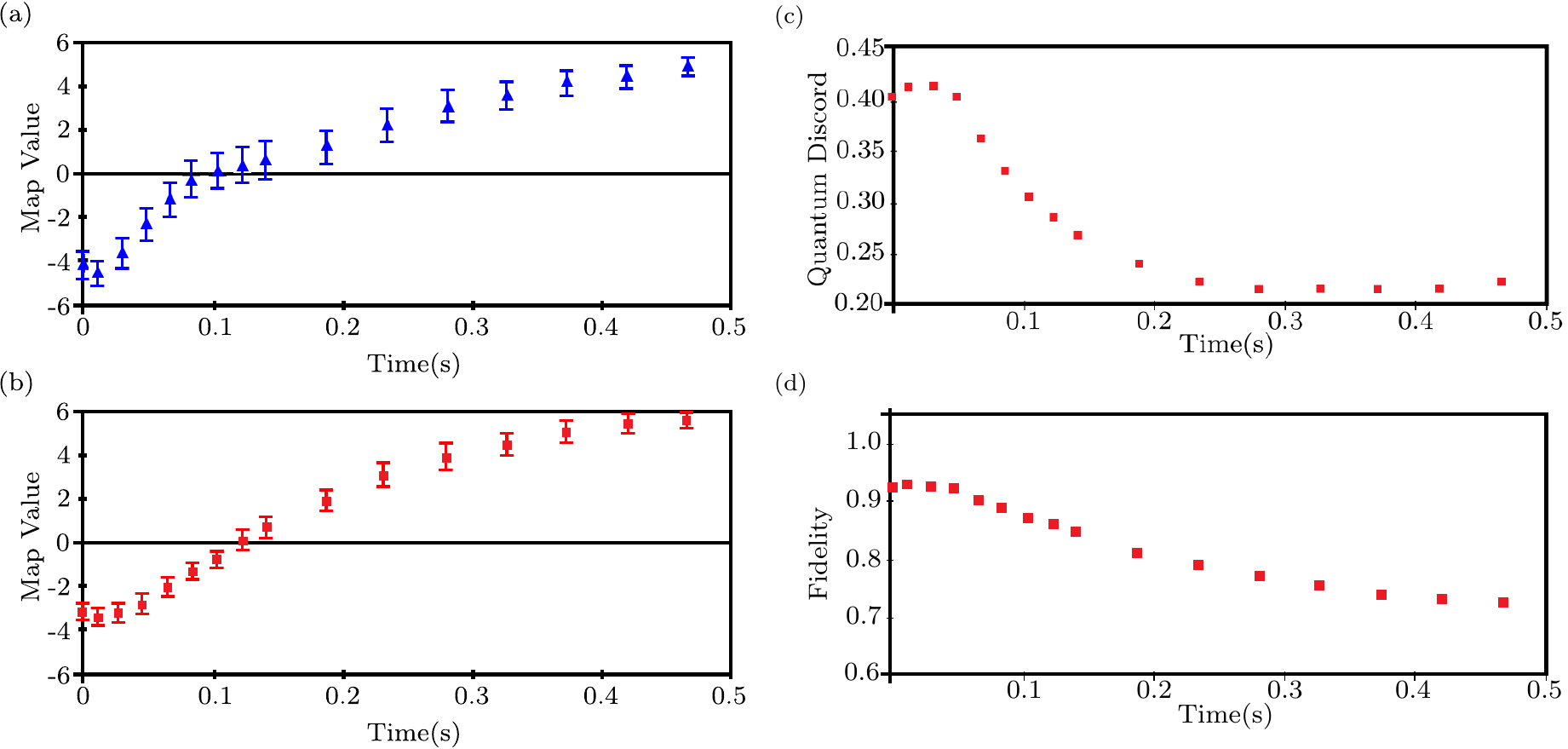} 
\caption{(a) Experimental
map value (in $\times10^{-2}$ units) plotted as a function of time. (b) Map
value (in $\times10^{-2}$ units) directly calculated from the tomographically
reconstructed state at each time instant (c) Time evolution of quantum discord
(characterizing total quantum correlations present in the state) for  the NCC
state (d) Time evolution of state fidelity. The red squares represent fidelity
of the experimentally prepared NCC state $\sigma_{{\rm exp}}(t)$ evolving in
time, w.r.t. the theoretical NCC state at time $ \rm t=0 $.}
\label{MV-NCC}
\end{center} 
\end{figure}

The prepared NCC state was allowed to evolve freely in time and the MV calculated at each time instant,  in
order to characterize the decoherence dynamics of the nonclassicality witness
map. As theoretically expected, one should get a negative MV for states which
are NCC. MV was measured at time instants which were integral multiples of $ \rm
\frac{2}{J} $ \ie $  \rm \frac{2n}{J} $ (with $ \rm n $ = 0, 1, 3, 5, 7, 9, 11,
13, 15, 20, 25, 30, 35, 40, 45 and 50), in order to avoid experimental errors
due to J-evolution. The results of experimental MV dynamics as a function of
time are shown in Fig.\ref{MV-NCC}(a). Experiments were repeated eight times to
estimate the errors as depicted in the figure. As seen from
Fig.~\ref{MV-NCC}(a), the MV remains negative (indicating the state is NCC) for
up-to 120 ms, which is approximately the $^{1}$H transverse relaxation time. The
standard NMR decoherence mechanisms are denoted by T$_2$ the spin-spin
relaxation time which causes dephasing among the energy eigenstates and T$_1$
the spin-lattice relaxation time, which causes energy exchange between the spins
and their environment. For comparison, the MV was also calculated directly using
Eq.~(\ref{sigmamap}) with $ \rm c = c_{opt}$, from the state which was
tomographically reconstructed at each time instant via full state
tomography~\cite{leskowitz-pra-04}. The results are shown in
Fig.~\ref{MV-NCC}(b), which are in good agreement with direct experimental MV
measurements. The state fidelity was also computed at the different time
instants and the results are shown in Fig.~\ref{MV-NCC}(c). The red squares in
Fig.~\ref{MV-NCC}(c) represent state fidelity of the experimental state $
\sigma_{{\rm exp}(t)}$ evolving in time, w.r.t. the theoretical NCC state
$\sigma_{{\rm theo}}(0) $ at time $ \rm t=0$ given in Eq.~(\ref{sigma}).

\subsection{Quantum Discord Dynamics}
\label{qd-dyn}
Map value evaluation of nonclassicality was also compared with the standard measure of nonclassicality, namely QD~\cite{ollivier-prl-02,luo-pra-08}. The state was reconstructed by
peRForming full quantum state tomography and the QD measure was computed from
the experimental data. Quantum mutual information can be quantified by the
equations: \begin{eqnarray} \rm I(\rho_{AB})&=& \rm
S(\rho_A)+S(\rho_B)-S(\rho_{AB}) \label{quantum-I} \nonumber \\ \rm
J_A(\rho_{AB})&=& \rm S(\rho_B)-S(\rho_B\vert \rho_A) \label{quantum-J}
\label{mutual} \end{eqnarray} where $  \rm  S(\rho_B\vert\rho_A)$ is the
conditional von Neumann entropy of subsystem $B$ when $A$ has already been
measured. QD is defined as the minimum difference between the two formulations
of mutual information in Eq.~(\ref{mutual}): \begin{equation} \rm
D_A(\rho_{AB})=S(\rho_A)-S(\rho_{AB})+S(\rho_B\vert\lbrace\Pi^A_j\rbrace)
\label{discord} \end{equation} QD hence depends on projectors $  \rm
\lbrace\Pi^A_j\rbrace$. The state of the system, after the outcome corresponding
to projector $ \rm \lbrace\Pi^A_j\rbrace$ has been detected, is \begin{equation}
\rm \tilde{\rho}_{AB}\vert \lbrace\Pi^A_j\rbrace=\frac{\left(\Pi^A_j\otimes
I_B\right)\rho_{AB}\left(\Pi^A_j\otimes I_B\right)}{p_j} \label{B}
\end{equation} with the probability $  \rm p_j=Tr\left((\Pi^A_j\otimes
I_B)\rho_{AB}(\Pi^A_j\otimes I_B)\right)$; $ \rm I_B$ is identity operator on
subsystem B. The state of the system B, after this measurement is
\begin{equation} \rm \rho_B\vert
\lbrace\Pi^A_j\rbrace=Tr_A\left(\tilde{\rho}_{AB}\vert
\lbrace\Pi^A_j\rbrace\right) \label{A} \end{equation} $  \rm
S\left(\rho_B\vert\lbrace\Pi^A_j\rbrace\right)$ is the missing information about
B before measurement $ \rm \lbrace\Pi^A_j\rbrace$. The expression
\begin{equation} \rm S(\rho_B\vert\lbrace\Pi^A_j\rbrace)=\sum_{j}{p_jS
\left(\rho_B\vert\lbrace\Pi^A_j\rbrace\right)} \label{cond-entropy}
\end{equation} is the conditional entropy appearing in Eq.~(\ref{discord}).  In
order to capture the true quantumness of the correlation one needs to peRForm an
optimization over all sets of von-Neumann type measurements represented by the
projectors $ \rm \lbrace\Pi^A_j\rbrace$. One can define two orthogonal vectors
(for spin half quantum subsystems), characterized by two real parameters
$\theta$ and $\phi$, on the Bloch sphere as: \begin{eqnarray} &&  \rm
\cos{\theta}\vert 0\rangle+e^{\textit{i}\phi}\sin{\theta}\vert 1\rangle
\nonumber \\ &&  \rm e^{-\textit{i}\phi}\sin{\theta}\vert
0\rangle-\cos{\theta}\vert 1\rangle \end{eqnarray} These vectors can be used to
construct the projectors $\rm \Pi^{A,B}_{1,2}$, which are then used to find out
the state of B after an arbitrary measurement was made on subsystem A. The
definition of conditional entropy (Eq.~(\ref{cond-entropy})) can be used to
obtain an expression which is parameterized by $\theta$ and $\phi$ for a given
state $\rm \rho_{AB}$. This expression is finally minimized by varying $\theta$
and $\phi$ and the results fed back into Eq.~(\ref{discord}), which yields a
measure of QD independent of the basis chosen for the measurement of the
subsystem.

To compare the detection via the positive map method  with the standard QD
measure,  the state was let evolve for a time $\tau$ and then reconstructed the
experimentally prepared via full quantum state tomography and calculated the QD
at all time instants where the MV was determined experimentally (the results are
shown in Fig.~\ref{MV-NCC}(d)). At $\tau$ = 0 s, a non-zero QD confirms the
presence of NCC and verifies the results given by MV. As the state evolves with
time, the QD parameter starts decreasing rapidly, in accordance with increasing
MV. Beyond 120 ms, while the MV becomes positive and hence fails to detect
nonclassicality, the QD parameter remains non-zero, indicating the presence of
some amount of nonclassicality (although by this time the state fidelity has
decreased to 0.7). However, value of QD is very close to zero and in fact cannot
be distinguished from contributions due to noise. One can hence conclude that
the positive map suffices to detect nonclassicality when decoherence processes
have not set in and while the fidelity of the prepared state is good. Once the
state has decohered however, a measure such as QD has to be used to verify if
the degraded state retains some amount of nonclassical correlations or not.
While the constructed nonclassicality witness is not optimal and hence cannot be
quantitatively compared with a stricter measure of nonclassicality such as a
measurement of the QD parameter, in most cases the witness suffices to detect
the presence of nonclassicality in a quantum state without having to resort to
more complicated experimental schemes.

\section{Conclusions} In the work  described in this
chapter, nonclassical correlations were detected experimentally in a separable
two-qubit quantum state, using a nonlinear positive map as a nonclassicality
witness.  The witness is able to detect nonclassicality  and its obvious
advantage lies in its using much fewer experimental resources as compared to
quantifying nonclassicality by measuring quantum discord via full quantum state
tomography.  It will be interesting to construct and utilize this map in
higher-dimensional quantum systems and for multi-qubits, where it is more challenging to distinguish between classical and quantum correlations. It has
been posited that quantum correlations, which can go beyond quantum entanglement
(and are captured by quantum discord and can thus be present even in separable
states), are responsible for achieving computational speedup in quantum
algorithms. It is important from the point of view of quantum information processing, to confirm the presence of such correlations in a quantum state,
without having to expend too much experimental effort.
The work described in this chapter is a step forward in this direction. Results of this chapter are contained in \href{https://journals.aps.org/pra/abstract/10.1103/PhysRevA.95.062318}{\rm Phys. Rev. A \textbf{95}, 062318 (2017)}. 

\chapter{ Experimental Classification of Entanglement in Arbitrary Three-Qubit States}\label{chapter_3QEntDet}

\section{Introduction} 
This chapter extends the goal of entanglement detection from a bipartite scenario, reported in Chapter \ref{chap2}, to multipartite case. Experimental characterization of arbitrary three-qubit pure states is undertaken. The three-qubit states can be classified into six inequivalent classes~\cite{dur-pra-00} under SLOCC~\citep{bennett-pra-00}. Protocols have been invented to carry out the classification of three-qubit states into the SLOCC classes~\cite{chi-pra-2010,zhao-pra-2013}. A recent proposal~\citep{datta-epd-18} aims to  classify any three-qubit pure entangled state into these six inequivalent classes by measuring only four observables.  Previously constructed scheme~\cite{dogra-pra-15} to experimentally realize a canonical form for general three-qubit states, is used to prepare arbitrary three-qubit states with an unknown amount of entanglement. Experimental implementation of the entanglement detection protocol is such that a single-shot (using only four experimental settings) is able to determine if a state belongs to the $\rm W $ class or to the ${\rm GHZ}$ class.  Schemes are devised to map the desired observables onto the $z$-magnetization of one of the subsystems, making it possible to experimentally measure its expectation value on NMR systems~\citep{singh-pra-16}. Mapping of the observables onto Pauli $z$-operators of a single qubit eases the experimental determination of the desired expectation value, since the NMR signal is proportional to the ensemble average of the Pauli $z$-operator.

The protocol has been tested on known three-qubit entangled states such as the ${\rm GHZ}$ state and the $W$ state as well as on randomly generated arbitrary three-qubit states with an unknown amount of entanglement. Seven representative states  belonging to the six SLOCC inequivalent classes  as well as twenty random states were prepared experimentally, with state fidelities ranging between 89\% to 99\%.  To decide the entanglement class of a state, the expectation values of four observables were experimentally measured in the state under investigation. All the seven representative states (namely, GHZ, W, $\rm W\overline{W}$, three bi-separable states and a separable state) were successfully detected within the experimental error limits.  Using this protocol, the experimentally randomly generated arbitrary three-qubit states were correctly identified  as belonging to either the GHZ, the W, the bi-separable or the separable class of states. Full quantum state tomography is performed to directly compute the observable value. Reconstructed density matrices were used to calculate the entanglement by computing negativity in each case, and the results compared well with those of the current protocol.

\section{Detecting Tripartite Entanglement} \label{Theory} There are six SLOCC inequivalent classes of entanglement in three-qubit systems, namely, the GHZ, W, three different biseparable classes and the separable class~\cite{dur-pra-00}.  A widely used measure of entanglement is the $n$-tangle~\cite{wong-pra-01,li-qip-12} and a non-vanishing three-tangle is a signature of the GHZ entangled class and can hence be used for their detection. For three parties A, B and C, the three-tangle $\tau$ is defined as
\begin{equation}
\label{3tangle}
\tau=C^{2}_{ \rm {A(BC)}}-C^{2}_{\rm{AB}}-C^{2}_{\rm{AC}}
\end{equation}  
with $C^{}_{\rm{AB}}$ and $C^{}_{\rm{AC}}$ being the concurrence that characterizes entanglement between A and B, and between A and C respectively; $C^{}_{\rm{A(BC)}}$ denotes the concurrence between A and the joint state of the subsystem comprising B and C~\cite{coffman-pra-00}. It was shown~\cite{coffman-pra-00} that for a pure state of form $ \vert \xi \rangle= \sum _{i,j,k=0}^{1} b_{ijk}\vert ijk \rangle $ with $ \sum _{i,j,k=0}^{1} \vert b_{ijk} \vert ^2 =1 $ the quadratic expression in concurrence can be written as 
\begin{eqnarray}\label{concu}
C^{2}_{\rm {A(BC)}}-C^{2}_{\rm{AB}}-C^{2}_{\rm {AC}}=4\vert d_1 -2d_2+4d_3\vert
\end{eqnarray}
with
\begin{eqnarray}\label{dls}
d_1 &=& b_{000}^2b_{111}^2 + b_{001}^2b_{110}^2 + b_{010}^2b{101}^2 + b_{100}^2b_{011}^2 \nonumber\\
d_2 &=& b_{000}b_{111}b_{011}b_{100} + b_{000}b_{111}b_{101}b_{010} + b_{000}b_{111}b_{110}b_{001}  \nonumber\\
&\textcolor[rgb]{1,1,1}{=}& + b_{011}b_{100}b_{101}b_{010} + b_{011}b_{100}b_{110}b_{001} + b_{101}b_{010}b_{110}b_{001} \nonumber\\
d_3 &=& b_{000}b_{110}b_{101}b_{011} + b_{111}b_{001}b_{010}b_{100}
\end{eqnarray}
The idea of using the three-tangle to investigate entanglement in three-qubit generic states is particularly interesting and general, as any three-qubit pure state can be written in the canonical form~\cite{acin-prl-01}\begin{equation}
\label{generic}
\vert\psi\rangle=a_0\vert 000 \rangle + a_1e^{\textit{i}
\theta}\vert 100 \rangle + a_2\vert 101 \rangle + a_3\vert
110 \rangle + a_4\vert 111 \rangle
\end{equation}
where $a_i\geq 0$, $\sum_i a^2_i=1$ and $\theta \in [0,\pi]$, and the class of states is written in the computational basis $\{ \vert 0 \rangle, \vert 1 \rangle \}$ of the qubits. On comparing the coefficients of general three-qubit pure state, $ \vert \xi \rangle $ with the generic state $ \vert \psi \rangle $ one can observe that $b_{000}=a_0$, $b_{100}=a_1e^{i\theta}$, $b_{101}=a_2$, $b_{110}=a_3$ and $b_{111}=a_4$ and hence can compute $d_l$ in Eq.~(\ref{dls}). On using resulting $d_l$ in Eq.~(\ref{concu}) the three-tangle for the generic state Eq.~(\ref{generic}) turns out to be~\citep{datta-epd-18} 
\begin{equation}
\tau_{\psi}=4a^2_0a^2_4
\end{equation}
Three-tangle can be measured experimentally by measuring the expectation value of the operator $O=2\sigma^{}_{1x}\sigma^{}_{2x}\sigma^{}_{3x}$, in the three-qubit state $\vert \psi \rangle$. Here $\sigma^{}_{x,y,z}$ are the Pauli matrices and $i=1,2,3$ denotes qubits label and the tensor product symbol between the Pauli operators has been omitted for brevity. One can compute
\begin{gather}\label{three-tang}
\langle \Psi \vert O \vert \Psi \rangle=
\begin{bmatrix}
   a_0 & 0 & 0 & 0 & a_1e^{-i\theta} & a_2 & a_3 & a_4 
\end{bmatrix}
\begin{bmatrix}
    0 & 0 & 0 & 0 & 0 & 0 & 0 & 1 \\
    0 & 0 & 0 & 0 & 0 & 0 & 1 & 0 \\
    0 & 0 & 0 & 0 & 0 & 1 & 0 & 0 \\
		0 & 0 & 0 & 0 & 1 & 0 & 0 & 0 \\
		0 & 0 & 0 & 1 & 0 & 0 & 0 & 0 \\
		0 & 0 & 1 & 0 & 0 & 0 & 0 & 0 \\
		0 & 1 & 0 & 0 & 0 & 0 & 0 & 0 \\
		1 & 0 & 0 & 0 & 0 & 0 & 0 & 0 
\end{bmatrix}
\begin{bmatrix}
    a_0 \\
    0 \\
    0 \\
    0 \\
    a_1e^{i\theta} \\
    a_2 \\
    a_3 \\
    a_4    
\end{bmatrix}
\end{gather} 
and this results in $\langle \psi \vert O \vert \psi \rangle^2= \langle O \rangle^{2}_{\psi}= 4\tau_\psi$. A non-zero expectation value of $O$ implies that the state under investigation is in the GHZ class~\citep{dur-pra-00}. In order to further categorize the classes of three-qubit generic states three more observables are defined as $O^{}_1=2\sigma^{}_{1x}\sigma^{}_{2x}\sigma^{}_{3z}$, $O^{}_2=2\sigma^{}_{1x}\sigma^{}_{2z}\sigma^{}_{3x}$, $O^{}_3=2\sigma^{}_{1z}\sigma^{}_{2x}\sigma^{}_{3x}$. Experimentally measuring the expectation values of the operators $O$, $O^{}_1$, $O^{}_2$ and $O^{}_3$ can reveal the entanglement class of every three-qubit pure state~\cite{zhao-pra-2013,datta-epd-18}. Table~\ref{classification table} summarizes the classification of the six SLOCC inequivalent classes of entangled states based on the expectation values of the observables $O$, $O^{}_1$, $O^{}_2$, $O^{}_3$. 
\begin{table}[h]
\begin{center}
\caption{\label{classification table}
Decision table for the classification of three-qubit pure
entangled states based on the expectation values of
operators $O$, $O^{}_1$, $O^{}_2$ and $O^{}_3$ in state $
\vert \psi \rangle $. Each class in the row is shown with
the expected values of the observables.}
\begin{tabular}{c | r r r r }
\hline
\textrm{Class} &
\textrm{$\langle O \rangle$}&
\textrm{$\langle O^{}_1 \rangle$} &
\textrm{$\langle O^{}_2 \rangle$} &
\textrm{$\langle O^{}_3 \rangle$} \\
\hline
\hline
GHZ    & $\neq 0$ & $*$ & $*$ & $*$  \\
W      & 0 & $\neq 0$ & $\neq 0$ & $\neq 0$ \\ 
BS$_1$ & 0 & 0 & 0 & $\neq 0$ \\ 
BS$_2$ & 0 & 0 & $\neq 0$ & 0 \\ 
BS$_3$ & 0 & $\neq 0$ & 0 & 0 \\ 
Separable & 0 & 0 & 0 & 0 \\
\hline
\hline
\end{tabular}
\end{center}
\begin{center}
$*$ May or may not be zero.
\end{center}
\end{table}
The six SLOCC inequivalent classes of three-qubit entangled states are GHZ, W, BS${}_1$, BS${}_2$, BS${}_3$ and separable. While GHZ and W classes are well known, BS$_1$ denotes a biseparable class having B and C subsystems entangled, the BS$_2$ class has subsystems A and C entangled, while the BS$_3$ class has subsystems A and B entangled. As has been summarized in Table~\ref{classification table} a non-zero value of $\langle O \rangle$ indicates that the state is in the GHZ class and this expectation value is zero for all other classes. For the W class of states all $\langle O_j\rangle$ are non-zero except $\langle O\rangle$. For the BS${}_1$ class only $\langle O_3\rangle$ is non-zero while only $\langle O_2\rangle$, and $\langle O_1\rangle$  are non-zero for the classes BS${}_2$ and BS${}_3$, respectively. For separable states all expectations are zero.

In order to experimentally realize the entanglement characterization protocol, one has to determine the expectation values $\langle O \rangle$, $\langle O_1 \rangle$, $\langle O_2 \rangle$ and $\langle O_3 \rangle$ for an experimentally prepared state $\vert \psi \rangle$. Next section describes the method to experimentally realize these expectation values based on subsystem measurement of the Pauli $z$-operator~\cite{singh-pra-16} and scheme for generating arbitrary three-qubit states~\cite{dogra-pra-15}.
\subsection{Mapping Pauli basis operators to single-qubit $z$-operators}
\label{Mapping}
A standard way to determine the expectation value of a desired observable in an experiment is to decompose the observable as a linear superposition of the observables accessible in the experiment~\cite{nielsen-book-02}. This task becomes particularly accessible while dealing with  the Pauli basis. Any observable for a three-qubit system, acting on an eight-dimensional Hilbert space can be decomposed as a linear superposition of 64 basis operators, and the Pauli basis is one possible basis for this decomposition. Let the set of Pauli basis operators be denoted as $\mathbb{B}=\{ \rm{B}_i; 0\leq i \leq 63\}$. For example, $O^{}_2$ has the form $\sigma^{}_{1x}\sigma^{}_{2z}\sigma^{}_{3x}$ and it is the element B$_{29}$ of the basis set $\mathbb{B}$. The four observables $O$, $O^{}_1$, $O^{}_2$ and $O^{}_3$ are represented by the elements B$_{21}$, B$_{23}$, B$_{29}$ and B$_{53}$ respectively of the Pauli basis set $\mathbb{B}$. Also by this convention the single-qubit $z$-operators for the first, second and third qubit \ie $\sigma^{}_{1z}$, $\sigma^{}_{2z}$ and $\sigma^{}_{3z}$ are the elements B$_{48}$, B$_{12}$ and B$_{3}$ respectively.  

\begin{figure} [ht]
\begin{center}
\includegraphics[angle=0,scale=1.2]{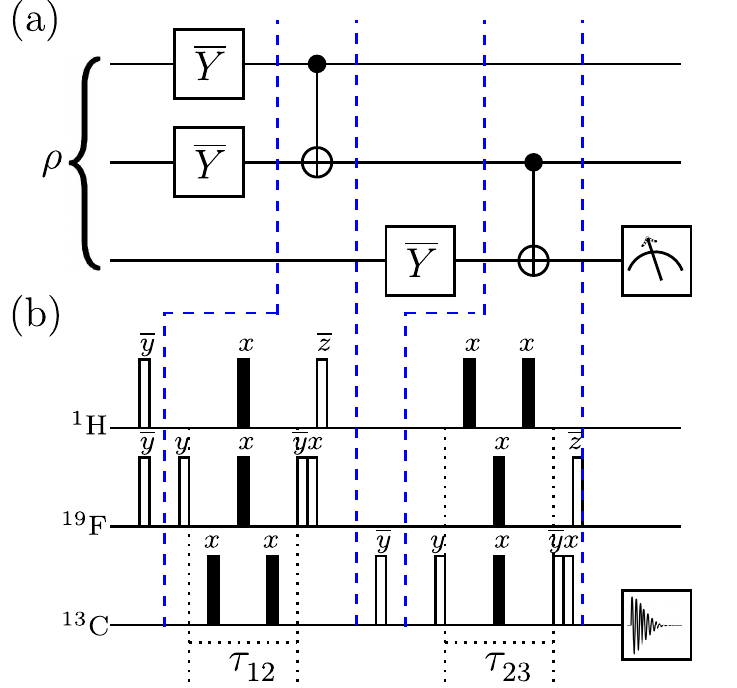}
\caption{(a) Quantum circuit to achieve mapping of the state $ \rho $ to either of the states $ \rho^{}_{21} $, $\rho^{}_{23} $, $ \rho^{}_{29} $ or $ \rho^{}_{53} $ followed by measurement of qubit 3 in the computational basis. (b) NMR pulse sequence of the quantum circuit given in (a). All the unfilled rectangles denote $ \frac{\pi}{2} $ spin-selective RF pulses while filled rectangles denote $ \pi $ pulses. Pulse phases are written above the respective pulse and a bar over a phase represents negative phase. Delays are given by $\tau^{}_{ij}=\frac{1}{8J_{ij}}$; $i,j$ label the qubit and $J$ is the coupling constant.}
\label{ckt+seq3Q}
\end{center}
\end{figure}
Table~\ref{Mapping Table} details the mapping of all 63 Pauli basis operators (excluding the 8$\otimes$8 identity operator) to the single-qubit Pauli $z$-operator. This mapping is particularly useful in an experimental setup where the expectation values of Pauli local $z$-operators are easily accessible. In NMR experiments, the $z$-magnetization of a nuclear spin in a state is proportional to the expectation value of Pauli $z$-operator of that spin in the state.

As an example of the mapping given in Table~\ref{Mapping Table}, the operator $O^{}_2$ has the form $\sigma^{}_{1x}\sigma^{}_{2z}\sigma^{}_{3x}$ and is the element B$_{29}$ of basis set $\mathbb{B}$. In order to determine $\langle O^{}_2 \rangle $ in the state $\rho=\vert \psi \rangle \langle \psi \vert$, one can map the state $\rho \rightarrow \rho^{}_{29}=U^{}_{29}.\rho.U^{\dagger}_{29} $ with $U^{}_{29}= {\rm CNOT}_{23}.\overline{Y}_3.$ ${\rm CNOT}_{12}.\overline{Y}_1 $. This is followed by finding $\langle \sigma^{}_{3z} \rangle $ in the state $\rho^{}_{29}$. The expectation value $\langle \sigma^{}_{3z} \rangle $ in the state $\rho^{}_{29}$ is equivalent to the expectation value of $\langle O^{}_2 \rangle $ in the state $\rho=\vert \psi \rangle \langle \psi \vert $(Table~\ref{Mapping Table}); the operation ${\rm CNOT}_{kl}$ is a controlled-NOT gate with $k$ as the control qubit and $l$ as the target qubit, and $X$, $\overline{X}$, $Y$ and $\overline{Y}$ represent local $\frac{\pi}{2}$ unitary rotations with phases $x$, $-x$, $y$ and $-y$ respectively.  The subscript on $\pi/2$ local unitary rotations denotes qubit number. The quantum circuit to achieve such a mapping is shown in Fig.~\ref{ckt+seq3Q}(a).

It should be noted that while measuring the expectation values of $O$, $O^{}_1$, $O^{}_2$ or $O^{}_3$, all the $\overline{Y}$ local rotations may not act in all these four cases. The mapping given in Table~\ref{Mapping Table} is used to decide which $\overline{Y}$ local rotation in the circuit~\ref{ckt+seq3Q}(a) will act.  All the basis operators in set $\mathbb{B}$ can be mapped to single-qubit $z$-operators in a similar fashion.  The mapping given in Table~\ref{Mapping Table} is not unique and there are several equivalent mappings which can be worked out as per the experimental requirements.

\section{NMR Implementation of Three-Qubit Entanglement Detection Protocol}
\label{NMR Implementation}
\begin{figure}
\begin{center}
\includegraphics[angle=0,scale=1.2]{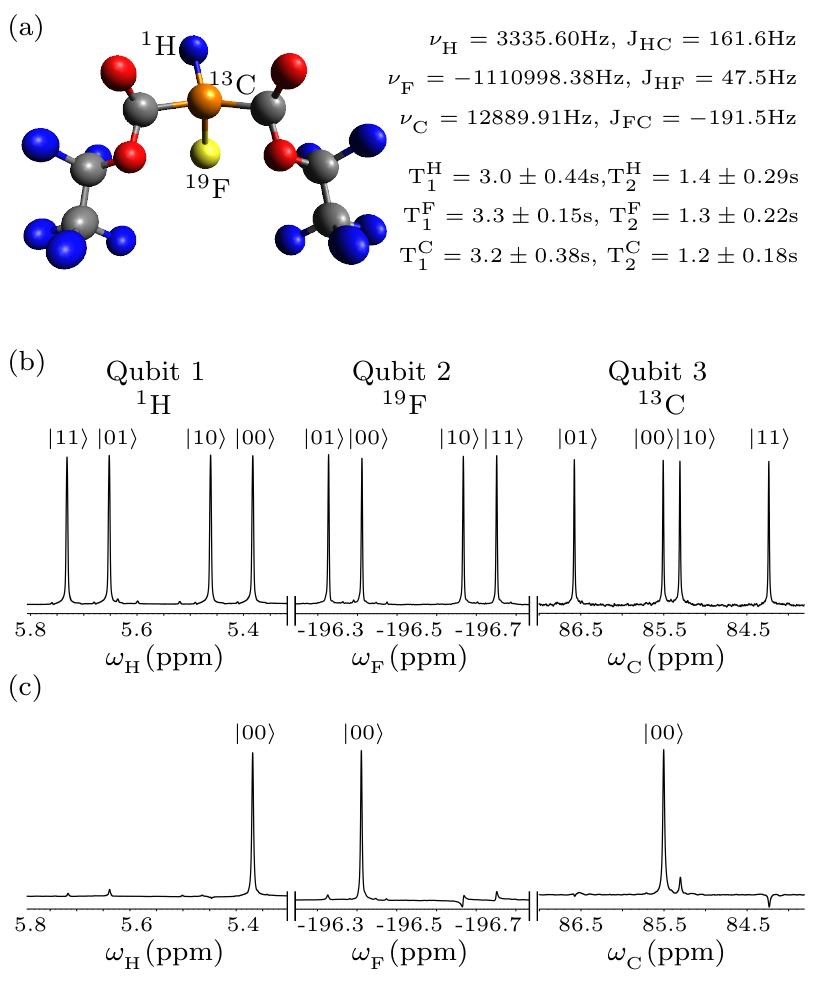}
\end{center}
\caption{(a) Molecular structure of $^{13}C$-labeled diethyl fluoromalonate and NMR parameters. NMR spectra of (b) thermal equilibrium state (c) pseudopure state. Each peak is labeled with the logical state of the qubit which is passive during the transition. Horizontal scale represents the chemical shifts in ppm.}
\label{molecule3Q}
\end{figure}

The Hamiltonian~\citep{ernst-book-90} for a three-qubit system in the rotating frame is given by
\begin{equation}\label{Hamiltonian}
\mathcal{H}= -\sum_{i=1}^{3} \nu_i I_{iz} + \sum_{i>j,i=1}^{3} J_{ij}I_{iz}I_{jz}
\end{equation}
where the indices $i,j=$ 1,2 or 3 represent the qubit number and $\nu_i$ is the respective chemical shift in rotating frame, $J_{ij}$ is the scalar coupling constant and $I_{iz}$ is the Pauli $z$-spin angular momentum operator of the $i^{\rm{th}}$ qubit. To implement the entanglement detection protocol experimentally, $^{13}$C labeled diethylfluoromalonate dissolved in acetone-D6 sample was used. $^{1}$H, $^{19}$F and $^{13}$C spin-half nuclei were encoded as qubit 1, qubit 2 and qubit 3 respectively. The system was initialized in the pseudopure (PPS) state \ie $\vert 000 \rangle$ using the spatial averaging~\cite{cory-physD-98,mitra-jmr-07} with the density operator being
\begin{equation}
\rho_{000}=\frac{1-\epsilon}{2^3}\mathbb{I}_8 +\epsilon
\vert 000 \rangle \langle 000 \vert 
\end{equation} 
where $\epsilon \approx 10^{-5}$ is the thermal polarization at room temperature and $ \mathbb{I}_8 $ is the 8 $ \times $ 8 identity operator. The experimentally determined NMR parameters (chemical shifts, T$_1$ and T$_2$ relaxation times and scalar couplings $\rm{J}_{ij}$) as well as the NMR spectra of the PPS state  are shown in Fig.~\ref{molecule3Q}. Each spectral transition is labeled with the logical states of the passive qubits (\ie qubits not undergoing any transition) in the computational basis. The state fidelity of the experimentally prepared PPS (Fig.~\ref{molecule3Q}(c)) was compute to be 0.98$\pm$0.01 and was calculated using the fidelity measure \citep{uhlmann-rpmp-76,jozsa-jmo-94} (also see Eqn.~\ref{fidelity_eq}). For the experimental reconstruction of density operator, full quantum state tomography (QST)\citep{leskowitz-pra-04,singh-pla-16} was performed using a preparatory pulse set $\left\lbrace III, XXX, IIY, XYX, YII, XXY, IYY \right\rbrace$, where $I$ implies ``no operation''. In NMR a $\frac{\pi}{2} $ local unitary rotation $X$($Y$) can be achieved using spin-selective transverse radio frequency (RF) pulses having phase $x$($y$). 

Experiments were performed at room temperature ($293$K)  on a Bruker Avance III 600-MHz FT-NMR spectrometer equipped with a QXI probe. Local unitary operations were achieved using highly accurate and calibrated spin selective transverse RF pulses of suitable amplitude, phase and duration. Non-local unitary operation were achieved by free evolution under the system Hamiltonian Eq.~(\ref{Hamiltonian}), of suitable duration under the desired scalar coupling with the help of embedded $\pi$ refocusing pulses. In the current study, the durations of $\frac{\pi}{2}$ pulses for $^{1}$H, $^{19}$F and $^{13}$C were 9.55 $\mu$s at 18.14 W power level, 22.80 $\mu$s at a power level of 42.27 W and 15.50 $\mu$s at a power level of  179.47 W, respectively.

\subsection{Measuring Observables by Mapping to Local $z$-Magnetization}\label{NMR Mapping}
As discussed in Sec.~\ref{Mapping}, the observables required to differentiate between six inequivalent classes of three-qubit pure entangled states can be mapped to the Pauli $z$-operator of one of the qubits. Further, in NMR the observed $z$-magnetization of a nuclear spin in a quantum state is proportional to the expectation value of $\sigma_{z} $-operator~\citep{ernst-book-90} of the spin in that state. The time-domain NMR signal \ie the free induction decay with appropriate phase gives Lorentzian peaks when Fourier transformed. These normalized experimental intensities give an estimate of the expectation value of $\sigma_{z}$  of the quantum state.
\begin{table}
\caption{\label{Mapping Table}
All sixty three product operators, for a three spin (half) system, mapped to the Pauli $z$-operators (of either spin 1, spin 2 or spin 3) by mapping initial state $ \rho \rightarrow \rho_i=U_i.\rho.U_i^{\dagger} $.} 
\scriptsize
\begin{tabular}{l l l l}
\hline
\textrm{Observable}&
\textrm{Initial State Mapped via}&
\textrm{Observable}&
\textrm{Initial State Mapped via}\\
\hline
\hline
$\langle B_{1} \rangle$ = Tr[$\rho_{1}.I_{3z}$] & $ U_{1}=\overline{Y}_3 $ & $\langle B_{33} \rangle$ = Tr[$\rho_{33}.I_{3z}$] & $ U_{33}={\rm CNOT}_{13}.\overline{Y}_3.X_1 $ \\

$\langle B_{2} \rangle$ = Tr[$\rho_{2}.I_{3z}$] & $ U_{2}=X_3 $ & $\langle B_{34} \rangle$ = Tr[$\rho_{34}.I_{3z}$] & $ U_{34}={\rm CNOT}_{13}.X_3.X_1 $ \\ 

$\langle B_{3} \rangle$ = Tr[$\rho_{3}.I_{3z}$] & $ U_{3}=\mathbb{I}_8 $ & $\langle B_{35} \rangle$ = Tr[$\rho_{35}.I_{3z}$] & $ U_{35}={\rm CNOT}_{13}.X_1 $ \\

$\langle B_{4} \rangle$ = Tr[$\rho_{4}.I_{2z}$] & $ U_{4}=\overline{Y}_2 $ & $\langle B_{36} \rangle$ = Tr[$\rho_{36}.I_{2z}$] & $ U_{36}={\rm CNOT}_{12}.\overline{Y}_2.X_1 $ \\

$\langle B_{5} \rangle$ = Tr[$\rho_{5}.I_{3z}$] & $ U_{5}={\rm CNOT}_{23}.\overline{Y}_3.\overline{Y}_2 $ & $\langle B_{37} \rangle$ = Tr[$\rho_{37}.I_{3z}$] & $ U_{37}={\rm CNOT}_{23}.\overline{Y}_3.{\rm CNOT}_{12}.\overline{Y}_2.X_1 $ \\

$\langle B_{6} \rangle$ = Tr[$\rho_{6}.I_{3z}$] & $ U_{6}={\rm CNOT}_{23}.X_3.\overline{Y}_2 $ & $\langle B_{38} \rangle$ = Tr[$\rho_{38}.I_{3z}$] & $ U_{38}={\rm CNOT}_{23}.X_3.{\rm CNOT}_{12}.\overline{Y}_2.X_1 $ \\

$\langle B_{7} \rangle$ = Tr[$\rho_{7}.I_{3z}$] & $ U_{7}={\rm CNOT}_{23}.\overline{Y}_2 $ & $\langle B_{39} \rangle$ = Tr[$\rho_{39}.I_{3z}$] & $ U_{39}={\rm CNOT}_{23}.{\rm CNOT}_{12}.\overline{Y}_2.X_1 $ \\

$\langle B_{8} \rangle$ = Tr[$\rho_{8}.I_{2z}$] & $ U_{8}=X_2 $ & $\langle B_{40} \rangle$ = Tr[$\rho_{40}.I_{2z}$] & $ U_{40}={\rm CNOT}_{12}.X_2.X_1 $ \\
$\langle B_{9} \rangle$ = Tr[$\rho_{9}.I_{3z}$] & $ U_{9}={\rm CNOT}_{23}.\overline{Y}_3.X_2 $ & $\langle B_{41} \rangle$ = Tr[$\rho_{41}.I_{3z}$] & $ U_{41}={\rm CNOT}_{23}.\overline{Y}_3.{\rm CNOT}_{12}.X_2.X_1 $ \\

$\langle B_{10} \rangle$ = Tr[$\rho_{10}.I_{3z}$] & $ U_{10}={\rm CNOT}_{23}.X_3.X_2 $ & $\langle B_{42} \rangle$ = Tr[$\rho_{42}.I_{3z}$] & $ U_{42}={\rm CNOT}_{23}.X_3.{\rm CNOT}_{12}.X_2.X_1 $ \\

$\langle B_{11} \rangle$ = Tr[$\rho_{11}.I_{3z}$] & $ U_{11}={\rm CNOT}_{23}.X_2 $ & $\langle B_{43} \rangle$ = Tr[$\rho_{43}.I_{3z}$] & $ U_{43}={\rm CNOT}_{23}.{\rm CNOT}_{12}.X_2.X_1 $ \\

$\langle B_{12} \rangle$ = Tr[$\rho_{12}.I_{3z}$] & $ U_{12}=\mathbb{I}_8 $ & $\langle B_{44} \rangle$ = Tr[$\rho_{44}.I_{2z}$] & $ U_{44}={\rm CNOT}_{12}.X_1 $ \\

$\langle B_{13} \rangle$ = Tr[$\rho_{13}.I_{3z}$] & $ U_{13}={\rm CNOT}_{23}.\overline{Y}_3 $ & $\langle B_{45} \rangle$ = Tr[$\rho_{45}.I_{3z}$] & $ U_{45}={\rm CNOT}_{23}.\overline{Y}_3.{\rm CNOT}_{12}.X_1 $ \\

$\langle B_{14} \rangle$ = Tr[$\rho_{14}.I_{3z}$] & $ U_{14}={\rm CNOT}_{23}.X_3 $ & $\langle B_{46} \rangle$ = Tr[$\rho_{46}.I_{3z}$] & $ U_{46}={\rm CNOT}_{23}.X_3.{\rm CNOT}_{12}.X_1 $ \\

$\langle B_{15} \rangle$ = Tr[$\rho_{15}.I_{3z}$] & $ U_{15}={\rm CNOT}_{23} $ & $\langle B_{47} \rangle$ = Tr[$\rho_{47}.I_{3z}$] & $ U_{47}={\rm CNOT}_{23}.{\rm CNOT}_{12}.X_1 $ \\

$\langle B_{16} \rangle$ = Tr[$\rho_{16}.I_{1z}$] & $ U_{16}=X_1 $ & $\langle B_{48} \rangle$ = Tr[$\rho_{48}.I_{1z}$] & $ U_{48}=\mathbb{I}_8 $ \\
$\langle B_{17} \rangle$ = Tr[$\rho_{17}.I_{3z}$] & $ U_{17}={\rm CNOT}_{13}.\overline{Y}_3.\overline{Y}_1 $ & $\langle
B_{49} \rangle$ = Tr[$\rho_{49}.I_{3z}$] & $ U_{49}={\rm CNOT}_{13}.\overline{Y}_3 $ \\

$\langle B_{18} \rangle$ = Tr[$\rho_{18}.I_{3z}$] & $ U_{18}={\rm CNOT}_{13}.X_3.\overline{Y}_1 $ & $\langle B_{50} \rangle$ = Tr[$\rho_{50}.I_{3z}$] & $ U_{50}={\rm CNOT}_{13}.X_3 $ \\

$\langle B_{19} \rangle$ = Tr[$\rho_{19}.I_{3z}$] & $ U_{19}={\rm CNOT}_{13}.\overline{Y}_1 $ & $\langle B_{51} \rangle$ = Tr[$\rho_{51}.I_{3z}$] & $ U_{51}={\rm CNOT}_{13} $ \\

$\langle B_{20} \rangle$ = Tr[$\rho_{20}.I_{2z}$] & $ U_{20}={\rm CNOT}_{12}.\overline{Y}_2.\overline{Y}_1 $ & $\langle B_{52} \rangle$ = Tr[$\rho_{52}.I_{2z}$] & $ U_{52}={\rm CNOT}_{12}.\overline{Y}_2 $ \\

$\langle B_{21} \rangle$ = Tr[$\rho_{21}.I_{3z}$] & $ U_{21}={\rm CNOT}_{23}.\overline{Y}_3.{\rm CNOT}_{12}.\overline{Y}_2.\overline{Y}_1 $ & $\langle B_{53} \rangle$ = Tr[$\rho_{53}.I_{3z}$] & $ U_{53}={\rm CNOT}_{23}.\overline{Y}_3.{\rm CNOT}_{12}.\overline{Y}_2 $ \\

$\langle B_{22} \rangle$ = Tr[$\rho_{22}.I_{3z}$] & $ U_{22}={\rm CNOT}_{23}.X_3.{\rm CNOT}_{12}.\overline{Y}_2.\overline{Y}_1 $ & $\langle B_{54} \rangle$ = Tr[$\rho_{54}.I_{3z}$] & $ U_{54}={\rm CNOT}_{23}.X_3.{\rm CNOT}_{12}.\overline{Y}_2 $ \\

$\langle B_{23} \rangle$ = Tr[$\rho_{23}.I_{3z}$] & $ U_{23}={\rm CNOT}_{23}.{\rm CNOT}_{12}.\overline{Y}_2.\overline{Y}_1 $ & $\langle B_{55} \rangle$ = Tr[$\rho_{55}.I_{3z}$] & $ U_{55}={\rm CNOT}_{23}.{\rm CNOT}_{12}.\overline{Y}_2 $ \\

$\langle B_{24} \rangle$ = Tr[$\rho_{24}.I_{2z}$] & $ U_{24}={\rm CNOT}_{12}.X_2.\overline{Y}_1 $ & $\langle B_{56} \rangle$ = Tr[$\rho_{56}.I_{2z}$] & $ U_{56}={\rm CNOT}_{12}.X_2 $ \\
$\langle B_{25} \rangle$ = Tr[$\rho_{25}.I_{3z}$] & $ U_{25}={\rm CNOT}_{23}.\overline{Y}_3.{\rm CNOT}_{12}.X_2.\overline{Y}_1 $ & $\langle B_{57} \rangle$ = Tr[$\rho_{57}.I_{3z}$] & $ U_{57}={\rm CNOT}_{23}.\overline{Y}_3.{\rm CNOT}_{12}.X_2 $ \\

$\langle B_{26} \rangle$ = Tr[$\rho_{26}.I_{3z}$] & $ U_{26}={\rm CNOT}_{23}.X_3.{\rm CNOT}_{12}.X_2.\overline{Y}_1 $ & $\langle B_{58} \rangle$ = Tr[$\rho_{58}.I_{3z}$] & $ U_{58}={\rm CNOT}_{23}.X_3.{\rm CNOT}_{12}.X_2 $ \\

$\langle B_{27} \rangle$ = Tr[$\rho_{27}.I_{3z}$] & $ U_{27}={\rm CNOT}_{23}.{\rm CNOT}_{12}.X_2.\overline{Y}_1 $ & $\langle B_{59} \rangle$ = Tr[$\rho_{59}.I_{3z}$] & $ U_{59}={\rm CNOT}_{23}.{\rm CNOT}_{12}.X_2 $ \\

$\langle B_{28} \rangle$ = Tr[$\rho_{28}.I_{2z}$] & $ U_{28}={\rm CNOT}_{12}.\overline{Y}_1 $ & $\langle B_{60} \rangle$ = Tr[$\rho_{60}.I_{2z}$] & $ U_{60}={\rm CNOT}_{12} $ \\

$\langle B_{29} \rangle$ = Tr[$\rho_{29}.I_{3z}$] & $ U_{29}={\rm CNOT}_{23}.\overline{Y}_3.{\rm CNOT}_{12}.\overline{Y}_1 $ & $\langle B_{61} \rangle$ = Tr[$\rho_{61}.I_{3z}$] & $ U_{61}={\rm CNOT}_{23}.\overline{Y}_3.{\rm CNOT}_{12} $ \\

$\langle B_{30} \rangle$ = Tr[$\rho_{30}.I_{3z}$] & $ U_{30}={\rm CNOT}_{23}.X_3.{\rm CNOT}_{12}.\overline{Y}_1 $ & $\langle B_{62} \rangle$ = Tr[$\rho_{62}.I_{3z}$] & $ U_{62}={\rm CNOT}_{23}.X_3.{\rm CNOT}_{12} $ \\

$\langle B_{31} \rangle$ = Tr[$\rho_{31}.I_{3z}$] & $ U_{31}={\rm CNOT}_{12}.{\rm CNOT}_{23}.\overline{Y}_1 $ & $\langle B_{63} \rangle$ = Tr[$\rho_{63}.I_{3z}$] & $ U_{63}={\rm CNOT}_{23}.{\rm CNOT}_{12} $ \\

$\langle B_{32} \rangle$ = Tr[$\rho_{32}.I_{1z}$] & $ U_{32}=X_1 $ &  &  \\
\hline
\hline
\end{tabular}
\end{table}
Let $\mathcal{\hat{O}}$ be the observable whose expectation value is to be measured in a state $ \rho =\vert \psi \rangle \langle \psi \vert $. Instead of measuring $\langle \mathcal{\hat{O}}\rangle _{\rho}$, the state $ \rho $ can be mapped to $ \rho_i $ using $ \rho_i=U_i.\rho.U^{\dagger}_i $ followed by $ z $-magnetization measurement of one of the qubits. Table~\ref{Mapping Table} lists the explicit forms of $U_i$ for all the basis elements of the Pauli basis set $\mathbb{B}$. In the present study, the observables of interest are $O$, $O^{}_1$, $O^{}_2$ and $O^{}_3$  as described in Sec.~\ref{Mapping} and Table~\ref{classification table}. The quantum circuit to achieve the required mapping is shown in Fig.~\ref{ckt+seq3Q}(a). The circuit is designed to map the state $ \rho $ to either of the states $ \rho^{}_{21} $, $ \rho^{}_{23} $, $ \rho^{}_{29} $ or $ \rho^{}_{53} $ followed by a $\sigma_z$ measurement on the third qubit in the mapped state. Depending upon the experimental settings, $ \langle B_3 \rangle $ in the mapped states is indeed the expectation values of $O$, $O^{}_1$, $O^{}_2$ or $O^{}_3$ in the initial state $ \rho $. The NMR pulse sequence to achieve the quantum mapping of circuit in Fig.~\ref{ckt+seq3Q}(a) is shown in Fig.~\ref{ckt+seq3Q}(b). The unfilled rectangles represent $\frac{\pi}{2}$ spin-selective pulses while the filled rectangles represent $\pi$ pulses. Evolution under chemical shifts has been refocused during all the free evolution periods (denoted by $ \tau_{ij}=\frac{1}{8J_{ij}}$)  and $\pi$ pulses are embedded in between the free evolution periods in such a way that the system evolves only under the desired scalar coupling $J_{ij}$.

\subsection{Implementing the Entanglement Detection Protocol}
\label{demo states}
\begin{figure} 
\begin{center}
\includegraphics[angle=0,scale=1.2]{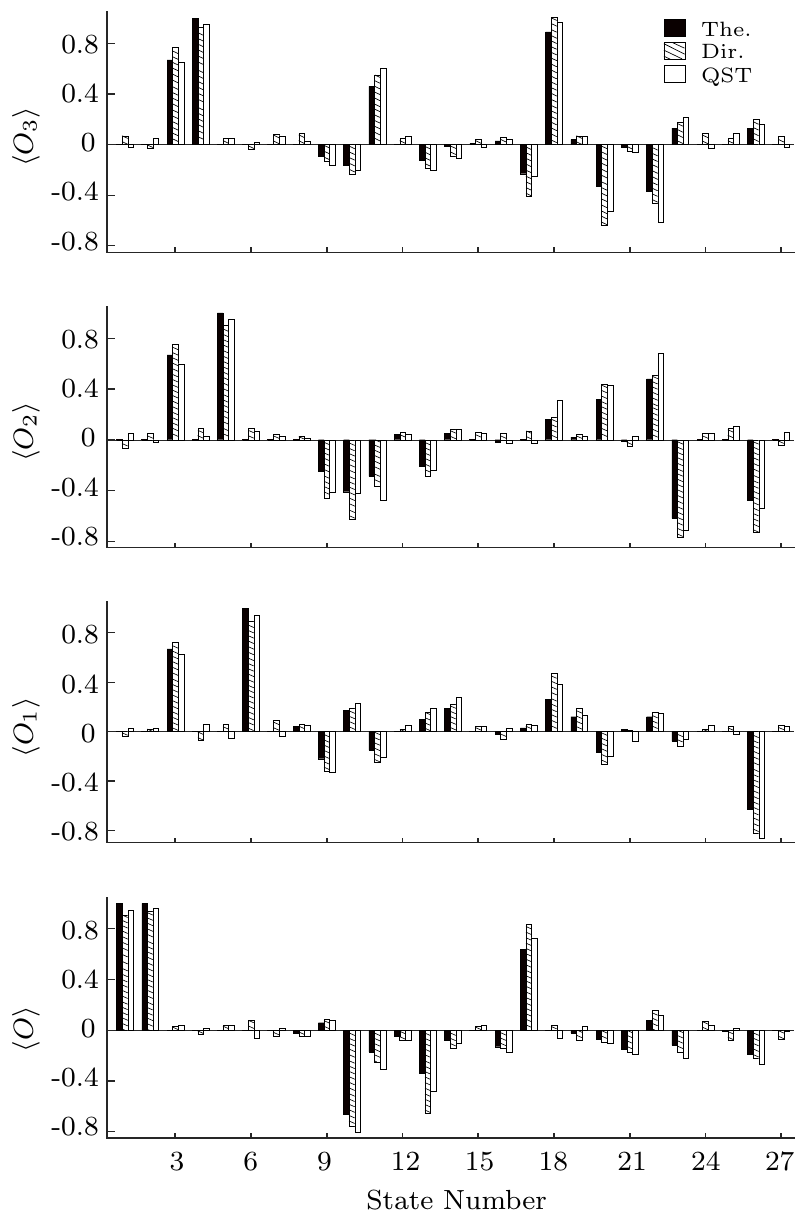}
\caption{Bar plots of the expectation values of the observables $O$, $O_1$, $O_2$ and $O_3$ for states numbered from 1-27 (Table~\ref{result table}). The horizontal axes denote the state number while the vertical axes represent the values of the respective observable. Black, cross-hatched and unfilled bars represent the theoretical (The.), directly (Dir.) measured from experiment, and QST-derived expectation values, respectively.}
\label{ResultPlot}
\end{center}
\end{figure}

The three-qubit system  was prepared in twenty seven different states in order to experimentally demonstrate the efficacy of the entanglement detection protocol. Seven representative states were prepared from the six inequivalent entanglement classes \ie  GHZ (GHZ and $\rm W\overline{W}$ states), W, three bi-separable and a separable class of states. In addition, twenty generic states were randomly generated (labeled as R$_1$, R$_2$, R$_3$,......., R$_{20}$). To prepare the random states the MATLAB\textsuperscript{\textregistered}-2016a random number generator was used. A recent experimental scheme~\citep{dogra-pra-15} was utilized to prepare the generic three-qubit states. For the details of quantum circuits as well as NMR pulse sequences used for state preparation see \citep{dogra-pra-15}. All the prepared states had state fidelities ranging between 0.89 to 0.99. Each prepared state $\rho$ was passed through the detection circuit \ref{ckt+seq3Q}(a) to yield the expectation values of the observables $O$,$O^{}_1$, $O^{}_2$ and $O^{}_3$ as described in Sec.~\ref{NMR Mapping}. Further, full QST \citep{cory-physD-98} was performed  to directly estimate the expectation value of $O$, $O^{}_1$, $O^{}_2$ and $O^{}_3$ for all the twenty seven states. The results of the experimental implementation of the three-qubit entanglement detection protocol are tabulated in Table~\ref{result table}. For a visual representation of the data in Table~\ref{result table}, bar charts have been plotted  and are shown in Fig.\ref{ResultPlot}. The seven known states were numbered as 1-7 while twenty random states were numbered as 8-27 in accordance with Table~\ref{result table}. Horizontal axes in plots of Fig.~\ref{ResultPlot}denote the state number while vertical axes represent the value of the respective observable. Black, cross-hatched and unfilled bars represent theoretical (The.), direct (Dir.) experimental and QST based expectation values, respectively. To further quantify the entanglement quotient, the entanglement measure, negativity~\citep{weinstein-pra-10,vidal-pra-02} was also computed theoretically as well as experimentally in all the cases (Table~\ref{negativity table}). Experiments were repeated several times for error estimation and to validate the reproducibility of the experimental results. All the seven representative states belonging to the six inequivalent entanglement classes were detected successfully within the experimental error limits, as suggested by the experimental results in first seven rows of Table~\ref{result table} in comparison with Table~\ref{classification table}.  The errors in the experimental expectation values reported in the Table~\ref{result table} were in the range 3.1\%-8.5\%.  The entanglement detection protocol with only four observables is further supported by negativity measurements (Table~\ref{negativity table}). It is to be noted here that one will never be able to conclude that the result of an experimental observation is exactly zero. However it can be established that the result is non-zero. This has to be kept in mind while interpreting the experimentally obtained values of the operators involved via the decision  Table~\ref{classification table}.

\begin{table}
\begin{center}
\caption{\label{negativity table} Theoretically calculated and experimentally measured values of negativity.}
\scriptsize
\begin{tabular}{c | c  c|| c| c c}
\hline
Negativity $\rightarrow$ &  Theoretical  & Experimental &Negativity $\rightarrow$ &  Theoretical  & Experimental \\

State $\downarrow$ &  & & State $\downarrow$ &  &   \\
\hline
\hline
GHZ & 0.5 & 0.46 $\pm$ 0.03 & R$_{8}$ & 0   &0.02 $\pm$ 0.02 \\
$\rm W\overline{W}$ & 0.37 & 0.35 $\pm$ 0.03 & R$_{9}$ & 0.07&0.06 $\pm$ 0.03 \\
W & 0.47 & 0.41 $\pm$ 0.02 & R$_{10}$ &0.38&0.35$\pm$ 0.08 \\
BS$_1$  & 0 & 0.03 $\pm$ 0.02 & R$_{11}$ &0.32&0.28$\pm$ 0.06 \\
BS$_2$  & 0 & 0.05 $\pm$ 0.02 & R$_{12}$ &0.05&0.04$\pm$ 0.02 \\
BS$_3$  & 0 & 0.03 $\pm$ 0.03 & R$_{13}$ &0.18&0.15$\pm$ 0.03 \\
Sep     & 0 & 0.02 $\pm$ 0.01 & R$_{14}$ &0.08&0.07$\pm$ 0.02 \\
R$_{1}$ & 0.02&0.04 $\pm$ 0.02 & R$_{15}$ &0.34&0.32$\pm$ 0.06 \\
R$_{2}$ & 0.16&0.12 $\pm$ 0.04 & R$_{16}$ &0.30&0.28$\pm$ 0.06 \\
R$_{3}$ & 0.38&0.35 $\pm$ 0.07 & R$_{17}$ &0 & 0.03$\pm$ 0.02 \\
R$_{4}$ & 0.38&0.34 $\pm$ 0.06 & R$_{18}$ &0 & 0.02$\pm$ 0.02 \\
R$_{5}$ & 0.03&0.04 $\pm$ 0.02 & R$_{19}$ &0.39&0.36$\pm$ 0.09 \\
R$_{6}$ & 0.21&0.18 $\pm$ 0.04 & R$_{20}$ &0 & 0.02$\pm$ 0.02 \\
R$_{7}$ & 0.09&0.08 $\pm$ 0.03 & & & \\
\hline
\hline
\end{tabular}
\end{center}
\end{table}
\begin{table} 
\caption{\label{result table}
Results of the three-qubit entanglement detection protocol for twenty seven states. Label BS is for biseparable states while R is for random states. First column depicts the state label, top row lists the observable (Obs.) while second row specify if the observable value is theoretical (The.), direct experimental (Dir.) or from QST.}
\scriptsize
\begin{tabular}{c | c c c | c c c | c c c | c c c}
\hline
Obs. $\rightarrow$ &  \multicolumn{3}{c}{$\langle O
\rangle$} & \multicolumn{3}{c}{$\langle O^{}_{1} \rangle$} &
\multicolumn{3}{c}{$\langle O^{}_{2} \rangle$} &
\multicolumn{3}{c}{$\langle O^{}_{3} \rangle$} \\

State($F$) $\downarrow$ & The. & Dir. & QST & The. & Dir. &
QST & The. & Dir. & QST & The. & Dir. & QST  \\
\hline
\hline
GHZ(0.95 $\pm$ 0.03) & 1.00 & 0.91 & 0.95 & 0 & -0.04 & 0.03 & 0 & -0.07 & 0.05 & 0 & 0.07 & -0.02 \\
$\rm W\overline{W}$(0.98 $\pm$ 0.01) & 1.00 & 0.94 & 0.96 & 0 & 0.02 & 0.03 & 0 & 0.05 & -0.02 & 0 & -0.03 & 0.05 \\
W(0.96 $\pm$ 0.02) & 0 & 0.05 & 0.04 & 0.67 & 0.60 & 0.62 & 0.67 & 0.61 & 0.69 & 0.67 & 0.59 & 0.63 \\
BS$_1$(0.95 $\pm$ 0.02) & 0 & -0.03 & 0.02 & 0 & -0.07 & 0.06 & 0 & 0.09 & 0.03 & 1.00 & 0.93 & 0.95 \\
BS$_2$(0.96 $\pm$ 0.03) & 0 & 0.04 & 0.04 & 0 & 0.06 & -0.05 & 1.00 & 0.90 & 0.95 & 0 & 0.05 & 0.05 \\
BS$_3$(0.95 $\pm$ 0.04) & 0 & 0.08 & -0.06 & 1.00 & 0.89 & 0.94 & 0 & 0.09 & 0.07 & 0 & -0.04 & 0.02 \\
Sep(0.98 $\pm$ 0.01) & 0 & -0.05 & 0.02 & 0 & 0.09 & -0.04 & 0 & 0.04 & 0.03 & 0 & 0.08 & 0.07 \\
R$_{1}$	(	0.91 $\pm$ 0.02	)	& -0.02	& -0.05	&	-0.05	&	0.04	&	0.06	& 0.05	&	0.00	&	0.03	&	0.01	& 0.00	&	0.09	&	0.03	\\
R$_{2}$	(	0.94 $\pm$ 0.03	)	&	0.06	& 0.09	&	0.08	&	-0.22	&	-0.32	& -0.33	&	-0.25	&	-0.46	&	-0.41	& -0.09	&	-0.13	&	-0.16	\\
R$_{3}$	(	0.93 $\pm$ 0.03	)	&	-0.66	& -0.76	&	-0.80	&	0.17	&	0.19	& 0.23	&	-0.41	&	-0.63	&	-0.42	& -0.16	&	-0.23	&	-0.20	\\
R$_{4}$	(	0.91 $\pm$ 0.01	)	&	-0.17	& -0.25	&	-0.31	&	-0.15	&	-0.25	& -0.21	&	-0.29	&	-0.37	&	-0.48	& 0.46	&	0.55	&	0.60	\\
R$_{5}$	(	0.94 $\pm$ 0.03	)	&	-0.05	& -0.08	&	-0.08	&	0.00	&	0.02	& 0.05	&	0.04	&	0.06	&	0.04	& 0.00	&	0.05	&	0.07	\\
R$_{6}$	(	0.90 $\pm$ 0.02	)	&	-0.34	& -0.65	&	-0.48	&	0.10	&	0.16	& 0.19	&	-0.21	&	-0.29	&	-0.24	& -0.12	&	-0.19	&	-0.20	\\
R$_{7}$	(	0.93 $\pm$ 0.03	)	&	-0.08	& -0.14	&	-0.10	&	0.19	&	0.22	& 0.28	&	0.05	&	0.08	&	0.08	& -0.01	&	-0.09	&	-0.11	\\
R$_{8}$	(	0.94 $\pm$ 0.01	)	&	0.00	& 0.03	&	0.04	&	0.00	&	0.04	& 0.04	&	0.00	&	0.06	&	0.05	& 0.01	&	0.04	&	-0.02	\\
R$_{9}$	(	0.95 $\pm$ 0.02	)	&	-0.13	& -0.14	&	-0.17	&	-0.02	&	-0.06	& 0.03	&	-0.02	&	0.05	&	-0.03	& 0.03	&	0.06	&	0.04	\\
R$_{10}$	(	0.92 $\pm$ 0.03	)	&	0.64 &	0.84	&	0.73	&	0.03	&	0.06 &	0.05	&	0.00	&	0.07	& -0.03	&	-0.23	&	-0.41	&	-0.25	\\
R$_{11}$	(	0.93 $\pm$ 0.03	)	&	0.00 &	0.04	&	-0.06	&	0.26	&	0.47 &	0.38	&	0.16	&	0.18	&	0.31 &	0.89	&	1.01	&	0.97	\\
R$_{12}$	(	0.89 $\pm$ 0.02	)	&-0.02	&	-0.08	&	0.03	&	0.12	& 0.19	&	0.13	&	0.02	&	0.04	& 0.03	&	0.04	&	0.07	&	0.07	\\
R$_{13}$	(	0.92 $\pm$ 0.03	)	& -0.07	&	-0.09	&	-0.10	&	-0.17	& -0.26	&	-0.20	&	0.32	&	0.44	& 0.43	&	-0.33	&	-0.64	&	-0.53	\\
R$_{14}$	(	0.94$\pm$ 0.04	)	& -0.15	&	-0.17	&	-0.19	&	0.02	& 0.01	&	-0.08	&	-0.01	&	-0.05	& 0.03	&	-0.02	&	-0.05	&	-0.06	\\
R$_{15}$	(	0.94 $\pm$ 0.03	)	&	0.08 &	0.16	&	0.12	&	0.12	&	0.16 &	0.15	&	0.48	&	0.51	&	0.68 &	-0.37	&	-0.46	&	-0.61	\\
R$_{16}$	(	0.93 $\pm$ 0.02	)	& -0.12	&	-0.17	&	-0.22	&	-0.08	& -0.12	&	-0.06	&	-0.62	&	-0.77	& -0.71	&	0.13	&	0.18	&	0.22	\\
R$_{17}$	(	0.93 $\pm$ 0.04	)	&	0.00 &	0.07	&	0.04	&	0.00	&	0.02 &	0.05	&	0.00	&	0.05	&	0.05 &	0.00	&	0.09	&	-0.03	\\
R$_{18}$	(	0.90 $\pm$ 0.02	)	& -0.01	&	-0.08	&	0.02	&	0.00	& 0.04	&	-0.02	&	0.00	&	0.09	& 0.11	&	0.00	&	0.05	&	0.09	\\
R$_{19}$	(	0.94 $\pm$ 0.02	)	& -0.19	&	-0.22	&	-0.27	&	-0.63	& -0.82	&	-0.86	&	-0.48	&	-0.73	& -0.54	&	0.13	&	0.20	&	0.16	\\
R$_{20}$	(	0.93 $\pm$ 0.03	)	&	0.00 &	-0.07	&	-0.01	&	0.00	&	0.05 &	0.04	&	0.00	&	-0.04	&	0.06 &	0.00	&	0.07	&	-0.02	\\
\hline
\hline
\end{tabular}
\end{table}

The results for the twenty randomly generated generic states, numbered from 8-27 (R$_1$-R$_{20}$), are interesting. For instance, states R$_{10}$ and R$_{11}$ have a negativity of approximately 0.35 which implies that these states have genuine tripartite entanglement.  On the other hand the experimental results of current detection protocol (Table~\ref{result table}) suggest that R$_{10}$ has a nonzero 3-tangle, which is a signature of the GHZ class. The states R$_{3}$, R$_{4}$, R$_{6}$, R$_{7}$, R$_{14}$, R$_{16}$ and R$_{19}$ also belong to the GHZ class as they all have non-zero 3-tangle as well as finite negativity. On the other hand, the state R$_{11}$ has a vanishing 3-tangle with non-vanishing expectation values of $O_1$, $O_2$ and $O_3$ which indicates that this state belongs to the W class. The states  R$_{2}$, R$_{13}$ and R$_{15}$ were also identified as members of the W class using the detection protocol.  These results demonstrate the fine-grained state discrimination power of the entanglement detection protocol as compared to procedures that rely on QST.
Furthermore, all vanishing expectation values as well as a near-zero negativity, in the case of R$_8$ state, imply that it belongs to the separable class. The randomly generated states R$_{1}$, R$_{5}$, R$_{17}$, R$_{18}$ and R$_{20}$ have also been identified as belonging to the separable class of states. Interestingly, R$_{12}$ has vanishing values of 3-tangle, negativity, $ \langle O_2 \rangle $ and $ \langle O_3 \rangle $ but has a finite value of $ \langle O_1 \rangle $, from which one can conclude that this state belongs to the  bi-separable BS$_3$ class.

\section{Generalized Three-Qubit Pure State Entanglement Classification}\label{Theory3Q New}

Above experimental demonstration \cite{singh-pra-18} of the three-qubit entanglement classification scheme~\citep{datta-epd-18} is limited to five-parameter generic state, Eq.~(\ref{generic})~\citep{acin-prl-01}. As two different three-qubit pure states of form $ \vert \xi \rangle= \sum _{i,j,k=0}^{1} b_{ijk}\vert ijk \rangle $ can have same generic state representation~\citep{acin-prl-01}. In order to detect and classify the entanglement in generalized three-qubit pure states the proposal in Ref.~\citep{zhao-pra-13} is explored. For the experimental demonstration concurrence~\citep{wootters-prl-98, wootters-qic-01,rungta-pra-01} based entanglement classification protocol~\citep{zhao-pra-13} is investigated. Core of the experimental procedure followed here depends upon finding the expectation value of a desired Pauli operators efficiently and to achieve this, the schemes described in Refs.~\citep{singh-pra-16, singh-arxiv-18,singh-pra-18} were utilized. Further, experimental procedures~\citep{dogra-pra-15} were developed to prepare any desired three-qubit generic states and used it in the current study to prepare states with arbitrary entanglement quotient.

Consider a three-qubit pure state $ \vert \Psi \rangle $. The state is fully separable if one can write $ \vert \Psi \rangle = \vert \psi_1 \rangle \otimes \vert \psi_2 \rangle \otimes \vert \psi_3 \rangle $. In case $ \vert \Psi \rangle $ is biseparable under bipartition $ 1 \vert 23 $ then it is always possible to write $ \vert \Psi \rangle = \vert \psi_1 \rangle \otimes \vert \psi_{23} \rangle $ where second and third qubits are in an entangled state $ \vert \psi_{23} \rangle $. Similarly one many have other two bipartitions as $ 2 \vert 13 $ and $ 3 \vert 12 $. In case $ \vert \Psi \rangle  $ can not be written as a fully separable or biseparable then the state is genuinely entangled. There are two SLOCC inequivalent classes of genuine three-qubit entanglement \citep{dur-pra-00} namely, GHZ and W class. Hence any three-qubit pure state can belong to either of the six SLOCC inequivalent classes \ie GHZ, W, three different biseparable classes or separable \citep{dur-pra-00}.

Following is the outline of the procedure \cite{zhao-pra-13} for generalized three-qubit pure state entanglement classification. Entanglement measure concurrence \citep{wootters-prl-98, wootters-qic-01,rungta-pra-01} is utilized to identify the above mentioned biseparable states in three-qubit pure states. The most general three-qubit pure state can be written as $ \vert \Psi \rangle= \sum _{i,j,k=0}^{1} a_{ijk}\vert ijk \rangle $ with $ \sum _{i,j,k=0}^{1} \vert a_{ijk} \vert ^2 =1 $. Concurrence for state $ \rho=\vert \Psi \rangle \langle \Psi \vert $ is given by $ C(\rho)=\sqrt{1-(tr\rho_1)^2} $ where $ \rho_1=tr_2(\rho) $ being the reduced density operator of first party. Squared concurrence for a three-qubit pure state under bipartiton $ 1 \vert 23 $ is given by
\begin{eqnarray}\label{conr}
C^2_{1 \vert 23}(\rho)=\left( \sum \limits _{j,k=0}^{1} \vert a_{0jk} \vert^2 \right)\left( \sum \limits _{j,k=0}^{1} \vert a_{1jk} \vert^2 \right) -\Big\vert \sum \limits _{j,k=0}^{1} a_{0jk}a_{1jk}^* \Big\vert^2
\end{eqnarray}
Further it was shown in \citep{zhao-pra-13} that after a lengthy calculation the squared concurrence, Eq.~(\ref{conr}), can be written as a quadratic polynomial of the expectation values of Pauli operators for three spin system. Let us symbolize $ C^2_{1 \vert 23}(\rho) $ as $ G_1(\rho) $ and it takes the form
\begin{eqnarray}
\label{G1}
G_1(\rho)=&\frac{1}{16}&(3 - \langle\sigma_0\sigma_0\sigma_3 \rangle^2 
- \langle\sigma_0\sigma_3\sigma_0 \rangle^2 
+ \langle\sigma_3\sigma_3\sigma_0 \rangle^2 \nonumber \\
&-3& \langle\sigma_3\sigma_0 \sigma_0  \rangle^2  + \langle\sigma_3\sigma_0\sigma_3 \rangle^2 - \langle\sigma_0\sigma_3\sigma_3 \rangle^2 + \langle\sigma_3\sigma_3\sigma_3 \rangle^2 \nonumber\\
&-3&\langle\sigma_1\sigma_0 \sigma_0  \rangle^2 + \langle\sigma_1\sigma_0\sigma_3 \rangle^2 + \langle\sigma_1\sigma_3\sigma_0 \rangle^2 + \langle\sigma_1\sigma_3\sigma_3 \rangle^2 \nonumber\\
&-3&\langle\sigma_2\sigma_0 \sigma_0  \rangle^2 + \langle\sigma_2\sigma_0\sigma_3 \rangle^2 + \langle\sigma_2\sigma_3\sigma_0 \rangle^2 + \langle\sigma_2\sigma_3\sigma_3 \rangle^2)
\end{eqnarray}
with $ \sigma_0=\vert 0 \rangle\langle 0 \vert+\vert 1 \rangle\langle 1 \vert  $, $ \sigma_1=\vert 0 \rangle\langle 1 \vert+\vert 1 \rangle\langle 0 \vert  $, $ \sigma_2=i(\vert 1 \rangle\langle 0 \vert-\vert 0 \rangle\langle 1 \vert)  $ and $ \sigma_3=\vert 0 \rangle\langle 0 \vert-\vert 1 \rangle\langle 1 \vert  $ being Pauli spin matrices in computational basis. Similar expressions for squared concurrences under other two bipartitions \ie $ C^2_{2 \vert 13}(\rho) $ and $ C^2_{3 \vert 12}(\rho) $ can also be written by permutation and symbolized by $ G_2(\rho) $ and $ G_3(\rho) $ respectively \citep{zhao-pra-13}.

\label{th1}
As described in \textit{Theorem 1} of \citep{zhao-pra-13}, for any three-qubit pure state $ \rho= \vert \Psi \rangle \langle \Psi \vert $,

(i) $ \vert \Psi \rangle $ is fully separable iff $ G_l(\rho)=0 $, for $ l=2,3 $ or $ l=1,2 $ or $ l=1,3 $.

(ii) $ \vert \Psi \rangle $ is separable between \textit{l}$\rm ^{th}$ qubit and rest  iff $ G_l(\rho)=0 $ and $ G_m(\rho)>0$ with $ l,m\in \{ 1,2,3 \} $ and $ l\neq m $.

(iii) $ \vert \Psi \rangle $ is genuinely entangled iff $ G_l(\rho)>0 $, for $ l=2,3 $ or $ l=1,2 $ or $ l=1,3 $.\\

Hence computing the nonlinear entanglement witnesses $ G_l(\rho) $, through experimentally measured expectation values of Pauli operators in an arbitrary three-qubit pure state $ \rho= \vert \Psi \rangle \langle \Psi \vert $, can immediately reveal the entanglement class of the state.

\subsection{Framework for Experimental Implementation}
Experimental creation of arbitrary general three-qubit pure states is a non-trivial task but one can resort to generic state \citep{acin-prl-01} for the demonstration of above discussed entanglement classification protocol.

It has been established \citep{acin-prl-01} that any three-qubit pure state can be transformed to generic state of canonical form \ref{generic}. The idea of five parameter generic state representation of three-qubit states was motivated from two-qubit generic states utilizing Schmidt decomposition \cite{schmidt-ma-1907, ekert-ajp-95} where any two-qubit state was shown to have form $\vert \Psi \rangle = cos\theta \vert 00 \rangle + sin\theta \vert 11 \rangle $ with $0 \leq \theta \leq \pi/2$, the relative phase has been absorbed into any of the local bases. Although the three-qubit generic representation doesn't follow from Schmidt decomposition but it was shown that combining adequately the local changes of bases corresponding to U(1)$\times$U(3)$\times$SU(2)$\times$SU(2)$\times$SU(2) transformations, one can always do with five terms, which precisely can carry only five entanglement parameters, leading thus to a non-superfluous unique representation.

It should be noted that the entanglement classification procedure outlined in Sec~\ref{Theory} works for any three-qubit pure state but arbitrary generic states were chosen for experimental demonstration. Two or more different states may have same generic canonical representation \citep{acin-prl-01}. Entanglement properties for the class of all such states can be fully characterized resorting only to the SLOCC equivalent generic state representative of that class. Such choice of states further ease the experimental efforts as nearly 40\% of the expectation values of Pauli operators appearing in the expressions of $ G_l(\rho) $ (\textit{e.g.} Eq.~(\ref{G1})) vanish in the case of generic states (Eq.~(\ref{generic})). In the recent work \citep{singh-arxiv-18} discussed in the previous sections, the entanglement classification of arbitrary three-qubit pure states was experimentally demonstrated. To do so only four observables suffice to classify the entanglement class. In contrast, the current classification works not only for generic states but also for any arbitrary three-qubit pure state of form $ \vert \Psi \rangle= \sum _{i,j,k=0}^{1} a_{ijk}\vert ijk \rangle $ and not limited to the canonical form \ref{generic}.

Further, as per \textit{Theorem 1}-(iii) the current entanglement classification protocol enables us to decide if a given pure state has genuine three-qubit entanglement or not but doesn't say anything if state belongs to GHZ or W class. To overcome this an observable is defined as $O=2\sigma^{}_{1}\sigma^{}_{1}\sigma^{}_{1}$ and use $n$-tangle \cite{wong-pra-01,li-qip-12} as an entanglement measure. For a three-qubit system, a non-vanishing 3-tangle, $ \tau $, implies it belongs to GHZ class. One may easily verify that for a given generic state $ \vert \Psi \rangle $, the 3-tangle \ie\; $ \tau_{\Psi}=\langle \Psi \vert O \vert \Psi \rangle ^2/4$, Eq.~(\ref{three-tang}). Having defined $ O $ in addition to $ G_l(\rho) $, the protocol is equipped to experimentally classify any three-qubit pure state.

\section{NMR Implementation of Generalized Three-Qubit Entanglement Classification Protocol}
To experimentally implement the entanglement classification protocol discussed in Sec.~\ref{Theory3Q New} the experimental setup discussed in Sec.~\ref{Mapping}-\ref{NMR Implementation} is used. Three-qubit system is prepared in twenty seven different states. Seven states were prepared from the six SLOCC inequivalent entanglement classes \ie  GHZ (GHZ and $\rm W\overline{W}$ states), W, three bi-separable and a separable class of states. Three biseparable class states under partitions $ 1\vert 23 $, $ 2\vert 13 $ and $ 3\vert 12 $ are labeled as BS$ _1 $, BS$ _2 $ and BS$ _3 $ respectively. Additionally, twenty random generic states were prepared and labeled as R$_1$, R$_2$, R$_3$,......., R$_{20}$. To prepare the random states the random number generator available at \textit{www.random.org} was used. To experimentally prepare the desired three-qubit generic state, procedure outlined in \citep{dogra-pra-15} is followed. Ref.~\citep{dogra-pra-15} details the quantum circuits as well as NMR pulse sequences required to prepare all the desired quantum states in the current study. All such prepared states were found to have the fidelity (F) in the range 0.88 to 0.99. For each such prepared state the expectation values of the Pauli operators were found as described in Sec.~\ref{NMR Mapping} which in turn were used to compute $ G_l(\rho) $ using Eq.~(\ref{G1}). $ \langle O \rangle $ was also found in all the cases as it serves as an entanglement witness of the GHZ class.
\begin{table} 
\caption{\label{result table 1} Results of the three-qubit entanglement classification protocol for twenty seven states. Label BS is for biseparable states while R is for random states. First column depicts the state label, top row lists the observable (Obs.) while the second row specifies if the observable value obtained is theoretical (The.), from QST or direct experimental (Dir.).}
\scriptsize
\begin{tabular}{c | c c c | c c c | c c c | c c c}
\hline
Obs. $\rightarrow$ &  \multicolumn{3}{c}{$ \langle O \rangle $} & \multicolumn{3}{c}{$G_1$} & \multicolumn{3}{c}{$G_2$} &
\multicolumn{3}{c}{$G_3$} \\

State (F) $\downarrow$ & The. & QST & Dir. & The. & QST & Dir. & The. & QST & Dir. & The. & QST & Dir.  \\
\hline
\hline

GHZ(0.96$ \pm $0.01) & 1.00 & 0.96 & 0.91 & 0.25 & 0.23 & 0.22 & 0.25 & 0.24 & 0.21 & 0.25 & 0.22 & 0.24 \\
$\rm W\overline{W}$(0.95$ \pm $0.02)  & 1.00 & 0.95 & 0.94 & 0.14 & 0.11 & 0.13 & 0.14 & 0.13 & 0.12 & 0.14 & 0.15 & 0.13 \\
W(0.96$ \pm $0.02) & 0 & 0.03 & 0.02 & 0.22 & 0.19 & 0.21 & 0.22 & 0.24 & 0.25 & 0.22 & 0.25 & 0.23 \\
BS$_1$(0.98$ \pm $0.01) & 0 & 0.04 & 0.02 & 0 & 0.04 & 0.03 & 0.25 & 0.22 & 0.20 & 0.25 & 0.21 & 0.23 \\
BS$_2$(0.94$ \pm $0.03) & 0 & 0.04 & 0.03 & 0.25 & 0.21 & 0.24 & 0 & 0.02 & 0.03 & 0.25 & 0.24 & 0.27 \\
BS$_3$(0.95$ \pm $0.02) & 0 & 0.01 & 0.02 & 0.25 & 0.27 & 0.21 & 0.25 & 0.26 & 0.22 & 0 & 0.02 & 0.03 \\
Sep(0.98$ \pm $0.01) & 0 & 0.01 & 0.02 & 0 & 0.03 & 0.01 & 0 & 0.02 & 0.02 & 0 & 0.03 & 0.01 \\
R$_{ 1}$(0.92$ \pm $0.03) & 0 & 0.02 & 0.02 & 0 & 0.01 & 0.02 & 0 & 0.01 & 0.03 & 0 & 0.01 & 0.02 \\
R$_{ 2}$(0.93$ \pm $0.02) & -0.43 & -0.45 & -0.40 & 0.17 & 0.15 & 0.18 & 0.23 & 0.25 & 0.22 & 0.24 & 0.26 & 0.27 \\
R$_{ 3}$(0.96$ \pm $0.02) & -0.27 & -0.25 & -0.25 & 0.08 & 0.07 & 0.09 & 0.07 & 0.08 & 0.09 & 0.06 & 0.08 & 0.09 \\
R$_{ 4}$(0.94$ \pm $0.03) & -0.13 & -0.15 & -0.17 & 0.12 & 0.11 & 0.15 & 0.13 & 0.13 & 0.15 & 0.14 & 0.16 & 0.12 \\
R$_{ 5}$(0.93$ \pm $0.02) & 0.56 & 0.60 & 0.55 & 0.15 & 0.15 & 0.18 & 0.15 & 0.14 & 0.17 & 0.14 & 0.17 & 0.16 \\
R$_{ 6}$(0.89$ \pm $0.01) & 0 & 0.03 & 0.02 & 0.22 & 0.28 & 0.025 & 0.17 & 0.19 & 0.20 & 0.11 & 0.10 & 0.13 \\
R$_{ 7}$(0.96$ \pm $0.02) & -0.16 & -0.19 & -0.18 & 0.09 & 0.10 & 0.12 & 0.08 & 0.10 & 0.10 & 0.09 & 0.11 & 0.12 \\
R$_{ 8}$(0.93$ \pm $0.02) & 0 & 0.02 & 0.03 & 0.08 & 0.10 & 0.11 & 0.02 & 0.01 & 0.03 & 0.08 & 0.10 & 0.11 \\
R$_{ 9}$(0.97$ \pm $0.03) & 0.23 & 0.20 & 0.25 & 0.16 & 0.14 & 0.19 & 0.16 & 0.15 & 0.16 & 0.14 & 0.13 & 0.13 \\
R$_{10}$(0.93$ \pm $0.02) & -0.01 & 0.01 & 0.02 & 0 & 0.01 & 0.02 & 0.12 & 0.14 & 0.10 & 0.10 & 0.12 & 0.13 \\
R$_{11}$(0.94$ \pm $0.01) & 0.18 & 0.20 & 0.21 & 0.02 & 0.01 & 0.01 & 0.04 & 0.02 & 0.02 & 0.03 & 0.01 & 0.02 \\
R$_{12}$(0.95$ \pm $0.02) & 0.41 & 0.50 & 0.48 & 0.08 & 0.11 & 0.10 & 0.07 & 0.10 & 0.10 & 0.08 & 0.11 & 0.10 \\
R$_{13}$(0.93$ \pm $0.01) & 0.09 & 0.12 & 0.13 & 0.13 & 0.10 & 0.10 & 0.14 & 0.17 & 0.15 & 0.12 & 0.13 & 0.11 \\
R$_{14}$(0.94$ \pm $0.02) & 0.05 & 0.03 & 0.02 & 0.15 & 0.17 & 0.18 & 0.20 & 0.22 & 0.21 & 0.20 &  0.19 & 0.17 \\
R$_{15}$(0.98$ \pm $0.01) & 0.04 & 0.02 & 0.02 & 0.02 & 0.02 & 0.01 & 0.04 & 0.03 & 0.03 & 0.03 & 0.01 & 0.02 \\
R$_{16}$(0.96$ \pm $0.01) & 0 & -0.02 & -0.02 & 0 & 0.01 & 0.02 & 0 & 0.01 & 0.03 & 0 & 0.01 & 0.01 \\
R$_{17}$(0.95$ \pm $0.02) & 0 & 0.01 & 0.02 & 0.05 & 0.08 & 0.08 & 0.10 & 0.08 & 0.11 & 0.08 & 0.10 & 0.09 \\
R$_{18}$(0.90$ \pm $0.02) & -0.18 & -0.20 & -0.21 & 0.22 & 0.25 & 0.25 & 0.22 & 0.20 & 0.21 & 0.23 & 0.22 & 0.25 \\
R$_{19}$(0.94$ \pm $0.02) & 0 & 0.01 & 0.01 & 0 & 0.03 & 0.01 & 0.23 & 0.25 & 0.22 & 0.23 & 0.21 & 0.20 \\
R$_{20}$(0.96$ \pm $0.02) & 0 & 0.02 & 0 & 0 & 0.01 & 0.02 & 0 & 0.02 & 0.02 & 0 & 0.01 & 0.02 \\
\hline
\hline
\end{tabular}
\end{table}

\begin{figure}
\begin{center}
\includegraphics[angle=0,scale=1.2]{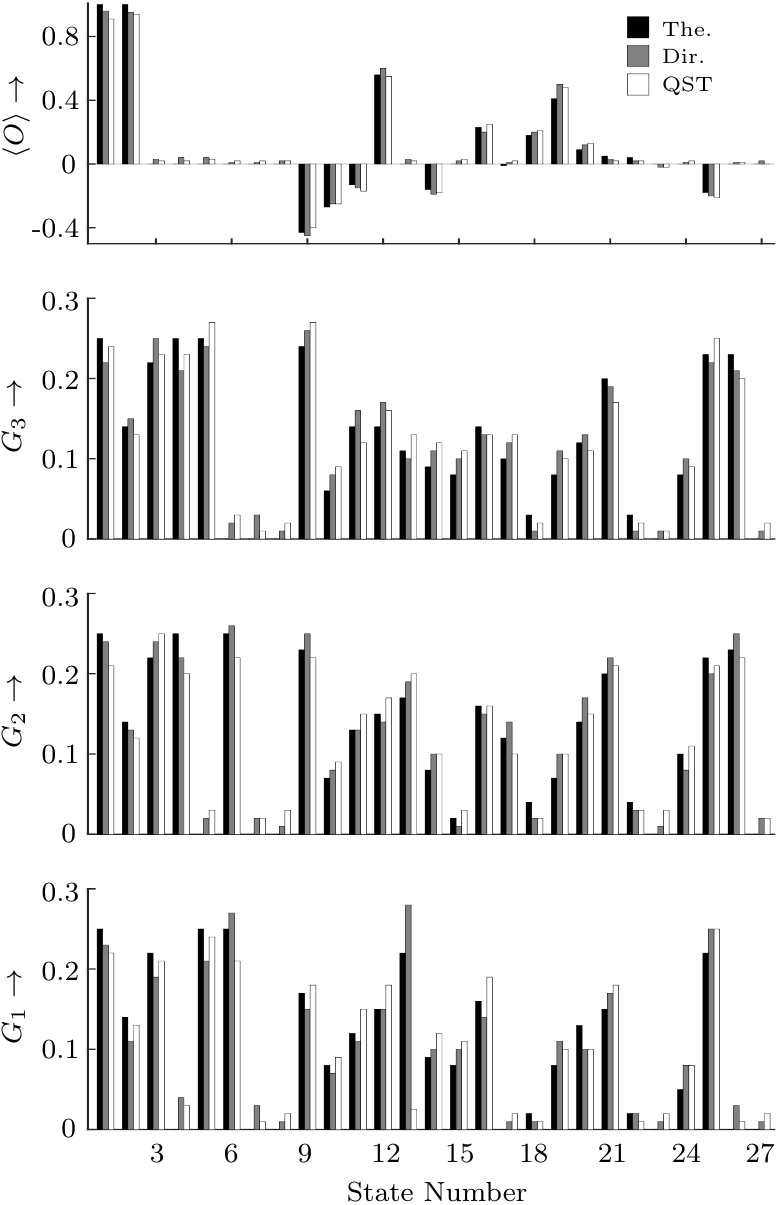}
\end{center}
\caption{Bar plots of the expectation value of the observable $O$ and the squared concurrences $G_1$, $G_2$ and $G_3$ for states numbered from 1-27 (Table~\ref{result table}). The horizontal axes denote the state number while the vertical axes represent the values of the respective observable. Black, gray and unfilled bars represent the theoretical (The.), directly (Dir.) measured from experiment, and QST-derived values, respectively.} 
\label{ResultPlot1}

\end{figure}

Experimental results of the three-qubit entanglement classification and detection protocol are shown in Table~\ref{result table 1}. A bar chart has been plotted in Fig.~\ref{ResultPlot1} for a visual representation of the experimental results of Table~\ref{result table 1}. To obtain the bar plots of Fig.~\ref{ResultPlot1}, the experimentally prepared states were numbered from 1 to 27 as per the order shown in Table~\ref{result table 1}. As detailed in Sec.~\ref{Theory}, the concurrence $ G_l(\rho) $ acts as the entanglement witness and the additional observable $ O $ helps in the experimental discrimination of GHZ class states from rest of the states. In order to further validate the results negativity \citep{weinstein-pra-10,vidal-pra-02}  has been computed from experimentally reconstructed state via QST \citep{leskowitz-pra-04} and the results are shown in the Table~\ref{negativity table 1}. In each case experiments were repeated several times for experimental error estimates. Experimental errors were in the range of 2.2\% - 5.7\% for the values reported in the Table~\ref{result table 1}.  
\begin{table}
\caption{\label{negativity table 1}
Theoretically calculated and experimentally measured negativity values for all twenty seven states under investigation.
}
\begin{center}
\scriptsize
\begin{tabular}{c | c  c || c | c c}
\hline
Negativity $\rightarrow$ &  Theoretical  & Experimental & Negativity $\rightarrow$ &  Theoretical  & Experimental \\

State $\downarrow$ &  &  & State $\downarrow$ &  &   \\
\hline
\hline
GHZ & 0.5 & 0.47 $\pm$ 0.02 & R$_{8}$ & 0.22 & 0.21 $\pm$ 0.02 \\
$\rm W\overline{W}$ & 0.37 & 0.39 $\pm$ 0.02 & R$_{9}$ & 0.39 & 0.37 $\pm$ 0.04 \\
W       & 0.47 & 0.44 $\pm$ 0.01 & R$_{10}$ &0.03 & 0.01 $\pm$ 0.01 \\
BS$_1$  & 0    & 0.02 $\pm$ 0.02 & R$_{11}$ &0.17 & 0.14 $\pm$ 0.02 \\
BS$_2$  & 0    & 0.03 $\pm$ 0.01 & R$_{12}$ &0.27 & 0.30 $\pm$ 0.03 \\
BS$_3$  & 0    & 0.02 $\pm$ 0.02 & R$_{13}$ &0.16 & 0.12 $\pm$ 0.04 \\
Sep     & 0    & 0.02 $\pm$ 0.02 & R$_{14}$ &0.42 & 0.37 $\pm$ 0.04 \\
R$_{1}$ & 0    & 0.01 $\pm$ 0.01 & R$_{15}$ &0.02 & 0.03 $\pm$ 0.01 \\
R$_{2}$ & 0.46 & 0.43 $\pm$ 0.04 & R$_{16}$ &0    & 0.01 $\pm$ 0.01 \\
R$_{3}$ & 0.26 & 0.24 $\pm$ 0.03 & R$_{17}$ &0.26 & 0.22 $\pm$ 0.03 \\
R$_{4}$ & 0.18 & 0.17 $\pm$ 0.03 & R$_{18}$ &0.47 & 0.41 $\pm$ 0.04 \\
R$_{5}$ & 0.38 & 0.35 $\pm$ 0.02 & R$_{19}$ &0    & 0.02 $\pm$ 0.02 \\
R$_{6}$ & 0.40 & 0.37 $\pm$ 0.04 & R$_{20}$ &0    & 0.03 $\pm$ 0.02\\
R$_{7}$ & 0.29 & 0.31 $\pm$ 0.03 & \\
%
\hline
\hline
\end{tabular}
\end{center}
\end{table}
One may observe from Table~\ref{result table 1} that the
seven states, from six SLOCC inequivalent classes, were
prepared with experimental fidelity $ \geq $ 0.95. The
entanglement classes of all these seven states were
correctly identified with the current protocol. It may
further be noted that the states R$ _2 $, R$ _3 $, R$ _4 $,
R$ _5 $, R$ _6 $, R$ _7 $, R$ _8 $, R$ _9 $, R$ _{11} $, R$
_{12} $, R$ _{13} $, R$ _{14} $, R$ _{17} $ and R$ _{18} $
have two non-zero concurrences and hence are all genuinely
entangled states. This fact is further supported by
negativity of these states reported in Table~\ref{negativity
table 1}. As discussed earlier, in order to discriminate GHZ
class from the rest one can resort to the observable $ O $.
Non-vanishing values of $ \langle O \rangle $ in
Table~\ref{result table 1} imply that the states R$ _2 $, R$
_3 $, R$ _4 $, R$ _5 $, R$ _7 $, R$ _9 $, R$ _{11} $, R$
_{12} $, R$ _{13} $ and R$ _{18} $ belong to GHZ class. In
contrast genuinely entangled states R$ _6 $, R$ _8 $, R$
_{14} $ and R$ _{17} $ have vanishing values of $ \langle O
\rangle $ and hence have vanishing 3-tangle as well so they
were identified as W class members. States R$ _{10} $ and R$
_{19} $ have the vanishing concurrence $ G_1 $ implying that
state belong to BS$ _1 $ class. Also states R$ _{1} $, R$
_{15} $, R$ _{16} $ and R$ _{20} $ were identified as
separable as all the observables have near zero values as
well as zero negativity.

\section{Effect of Mixedness in the Prepared States}
While the  proposed entanglement classification protocol assumes the state under investigation to be pure, the experimentally prepared states are invariably mixed. The experimentally prepared density operator $\rho_e $ can be expanded in terms of its eigenvalues $ \lambda_j $ and corresponding eigenvectors $ \vert \lambda_j \rangle $ as $\rho_e=\sum_{j=1}^8 \lambda_j \vert \lambda_j \rangle \langle \lambda_j \vert $, obeying the normalization condition $\sum_{j=1}^8 \lambda_j=1$.  For a pure state $\rho_p$, only one of the eigenvalue can be non-zero, hence one can take $\lambda_1^p=1 $ and other eigenvalues to be zero. The expectation value of the observable $ \hat{\mathcal{O}} $ can then be written as
\begin{equation}\label{idealexpectation}
\langle \hat{\mathcal{O}} \rangle _p=\langle \lambda_1^p \vert \hat{\mathcal{O}} \vert  \lambda_1^p \rangle = Tr[\rho_p.\hat{\mathcal{O}}] 
\end{equation} 
In an actual experiment the situation is different and several eigenvalues of the density operator may be non-zero. The errors can arise either from the mixedness present in the experimentally prepared state $ \rho_e $ or in the experimental measurement of $ \langle \hat{\mathcal{O}} \rangle $.  These errors are dominantly caused by imperfections in the unitary rotations used in state
preparation, rf inhomogeneity of the applied magnetic field, as well as T$ _2 $ and T$ _1 $ decoherence processes.

Let $\lambda_1$ be the maximum eigenvalue of the experimentally prepared state $\rho_e$. Mixedness is indicated by non-zero eigenvalues  $\lambda_j$ for $j\neq 1$. The expectation  value of  $\hat{\mathcal{O}}$ can  be written as an equation similar to Eq.~(\ref{idealexpectation}).
\begin{equation}
\langle \hat{\mathcal{O}} \rangle e=Tr[\rho_e.\hat{\mathcal{O}}]=\sum_{j=1}^8 \lambda_i Tr[P_i .\hat{\mathcal{O}}]=\sum_{j=1}^8 \lambda_i o_i
\end{equation}
The question is that if one approximate the state to be a pure state corresponding to the largest eigenvalue
$\lambda_1$ and take $ \langle \hat{\mathcal{O}} \rangle_p=\langle \lambda_1 \vert \hat{\mathcal{O}} \vert  \lambda_1 \rangle
$, how much error is introduced and how do these errors affect the results.

In order to estimate the error in the value of $\langle\hat{\mathcal{O}} \rangle $ due to the mixedness, one can define the fractional error as \begin{equation}
\Delta=\frac{ \langle\hat{\mathcal{O}} \rangle_p -\langle\hat{\mathcal{O}} \rangle _e }{\langle\hat{\mathcal{O}} \rangle _p} \cong
(1-\lambda_1)-\frac{\sum_{j=2}^8 \lambda_j o_j}{o_1}
\end{equation}
where $o_j$s depend upon the operator involved.  The experimental states have a minimum $ \lambda_1=0.88$ while $ \lambda_1 \geq 0.92 $ in other cases. In case of all the four observables $ O $, $ O_1 $, $ O_2 $ and $ O_3 $, $ \Delta $ was computed for all the 27 experimentally prepared states and the obtained values as percentage error were in the range 1.1\% $ \leq \Delta \leq $ 9.3\%.

In the light of the errors introduced by the mixedness
present in the experimentally prepared states the detection
protocol has to take $ \Delta $ error values into
consideration in addition to the experimental errors
reported in the Table~\ref{result table} for deciding the
class of three-qubit entanglement. As is evident from the
above analysis, in the worst-case scenario the protocol
works 90\% of the time. To further increase the fidelity of
the protocol, one can repeat the entire scheme on the same
prepared state, a number of times.

\section{Conclusions}
This chapter is aimed at the experimental classification of
arbitrary three-qubit pure states. To accomplish the goal
two different strategies were followed. In the first
classification protocol only four observable were defined
and measured experimentally for the classification. In the
second case, the entanglement measure concurrence was
measured experimentally to detect the class of three-qubit
pure states. Former protocol has the limitation that it can
only classify the entanglement class of the states in
generic form while the later is capable of detecting the
entanglement class of any arbitrary three-qubit pure state.
All the representative states from six SLOCC inequivalent
classes were detected by both the protocols.
Detection/classification protocols were further used to detect the entanglement class of twenty randomly generated
three-qubit pure states and both the protocols correctly
identified the  entanglement class within the experimental
error limits. Results were further verified and
substantiated by full quantum state tomography as well as
negativity calculations which indeed confirms the results
obtained by both the protocols. Results of this chapter are
contained in
\href{https://journals.aps.org/pra/abstract/10.1103/PhysRevA.98.032301}{\rm
Phys. Rev. A \textbf{98}, 032301 (2018)} and
\href{https://doi.org/10.1007/s11128-018-2105-5}{\rm Quant.
Info. Proc. \textbf{17}, 334 (2018)}.

\chapter{Detection of Qubit-Ququart Pseudo-Bound Entanglement}\label{BoundEnt}

\section{Introduction}
This chapter details the experimental investigations of a special type of entanglement which is fundamentally different from the entanglement we usually encounter in QIP. Entanglement exists in two fundamentally different forms~\citep{horodecki-pla-97,horodecki-prl-98}: ``free'' entanglement which can be distilled into EPR pairs using local operations and classical communications (LOCC)~\citep{nielsen-prl-99} and ``bound'' entanglement which cannot be distilled into EPR pairs via LOCC, as discussed in Sec-(\ref{boundEnt}) . The term ``bound'' essentially implies that although
correlations were established during the state preparation, they cannot brought into a ``free'' form in terms of EPR pairs by a distillation process, and used wherever EPR pairs can be used as a resource. PPT entangled states are a prime example of bound entangled states and have been shown to be useful to establish a secret key~\citep{horodecki-prl-05}, in the conversion of pure entangled states~\citep{ishizaka-prl-04} and for quantum secure communication~\citep{zhou-prl-18}.  While the existence of ``bound'' entangled states has been proved beyond doubt, there are still only a few known classes of such states~\cite{horodecki-pla-97,smolin-pra-01,acin-prl-01,sengupta-pra-11}. The problem of finding all such PPT entangled states is
still unsolved at the theoretical level.\\

Experimentally, bound entanglement has been created using four-qubit polarization states~\citep{amselem-natureP-09,kaneda-prl-12} and entanglement unlocking of a four-qubit bound entangled state was also demonstrated~\citep{lavoie-prl-10}.  Entanglement was characterized in bit-flip and phase-flip lossless quantum channel and the experiments were able to differentiate between free entangled, bound-entangled and separable states~\citep{amselem-sr-13}.  Continuous variable photonic bound-state entanglement has been created and detected in various experiments~\citep{dobek-lp-13,diGuglielmo-prl-11,steinhoff-pra-14}. Two photon qutrit Bound-entangled states of two qutrits were investigated utilizing orbital angular momentum degrees of freedom~\citep{hiesmayr-njp-13}. In NMR, a three-qubit system was used to prepare a three-parameter pseudo-bound entangled state~\citep{suter-pra-10}.\\

In the work described in this chapter, experiments were
performed to create and characterize a one-parameter family of qubit-ququart PPT entangled states using three nuclear spins on an NMR quantum information processor. There are a few proposals to detect PPT entanglement in the class of states introduced in Reference~\citep{horodecki-pla-97} by exploring local sum uncertainties~\citep{zhao-pra-13} and by measuring individual spin magnetisation along different directions~\citep{akbari-ijqi-17}. The proposal of Ref.~\citep{akbari-ijqi-17} is implemented to experimentally detect PPT entanglement in states prepared on an NMR quantum information processor.  The family of states considered in the current study is an incoherent mixture of five pure states and the relative strengths of the components of the mixture is controlled by a real parameter. Different PPT-entangled states were prepared experimentally, parameterized by a real parameter. These states represent five points on the one-parameter family of states. Discrete values of the real parameter were used which were uniformly distributed over the range for which the current detection protocol detects the entanglement.  In order to experimentally detect entanglement in these states, three Pauli operators need to be measured in each case. Previously developed schemes~\citep{singh-pra-16,singh-pra-18,singh-qip-18} were utilized to measure the required observables, which unitarily map the desired state followed by  NMR ensemble average measurements. In each case full quantum state
tomography (QST)~\citep{leskowitz-pra-04,singh-pla-16} was
also performed to verify the success of the detection
protocol as well as to establish that the experimentally
created states are indeed PPT entangled.  This work is
important both in the context of preparing and
characterizing bound entangled states and in devising new
experimental schemes to detect PPT entangled states which
use much fewer resources than are required by full quantum
state tomography schemes.  It should be noted here that we
prepare the PPT entangled states using NMR in the sense that
the total density operator for the spin ensemble always
remains close to the maximally mixed state and at any given
instance one is dealing with pseudo-entangled
states~\citep{laflemme-ptsca-98}. 

\section{Bound Entanglement in a Qubit-Ququart System}
\label{theory}
Consider a 3-qubit quantum system with an 8-dimensional
Hilbert space ${\mathcal H}={\mathcal H}_1\otimes{\mathcal
H}_2{}\otimes{\mathcal H}_3$, where ${\mathcal H}_i$
represent qubit Hilbert spaces.  If we choose to club the
last two qubits into a single system with a four-dimensional
Hilbert space ${\mathcal H}_q={\mathcal
H}_2{}\otimes{\mathcal H}_3$, the three-qubit system can be
reinterpreted as a qubit-ququart bipartite system with
Hilbert space ${\mathcal H}={\mathcal H}_1\otimes{\mathcal
H}_q$.\\
 
Formally one can say that the four ququart basis vectors
$\vert e_i \rangle$ are mapped to the logical state vectors
of the second and third qubits as $ \vert e_1 \rangle
\leftrightarrow \vert 00 \rangle ,\; \vert e_2
\rangle\leftrightarrow \vert 01 \rangle ,\; \vert e_3
\rangle\leftrightarrow \vert 10 \rangle$ and $\vert e_4
\rangle\leftrightarrow \vert 11 \rangle $ in the
computational basis. With this understanding, we will freely
use the three-qubit computational basis for this
qubit-ququart system, where all along it is understood that
the last two qubits form a ququart. For this system consider
a family of PPT bound entangled states parametrized by a
real parameter  $ b \in (0,1)$ introduced by
Horodecki~\citep{horodecki-pla-97}.  \begin{eqnarray}
\label{BE} \sigma_b=\frac{7b}{7b+1}\sigma_{\rm insep}+
\frac{1}{7b+1}\vert \phi_b \rangle \langle \phi_b\vert
\end{eqnarray} with \begin{eqnarray}\label{BE-1} \sigma_{\rm
insep}&=&\frac{2}{7}\sum_{i=1}^3 \vert \psi_i\rangle\langle
\psi_i \vert+\frac{1}{7}\vert 011 \rangle\langle 011 \vert,
\nonumber\\ \vert \phi_b\rangle &=& \vert 1 \rangle
\otimes\frac{1}{\sqrt{2}}\left(\sqrt{1+b}\vert 00
\rangle+\sqrt{1-b}\vert 11 \rangle \right), \nonumber\\
\vert \psi_1 \rangle &=& \frac{1}{\sqrt{2}}(\vert 000
\rangle + \vert 101 \rangle),\nonumber\\ \vert \psi_2
\rangle &=& \frac{1}{\sqrt{2}}(\vert 001 \rangle + \vert 110
\rangle),\nonumber\\ \vert \psi_3 \rangle &=&
\frac{1}{\sqrt{2}}(\vert 010 \rangle + \vert 111 \rangle)
\end{eqnarray} It has been shown in~\citep{horodecki-pla-97}
that the states in the family $\sigma_b$  defined above  are
entangled for $ 0<b<1 $ and is separable in the limiting
cases $ b=0 \;\; \rm or \;\; 1 $. One can explicitly write
the density operator for the mixed PPT entangled states
defined in Equation~(\ref{BE}) in the computational basis as
\begin{equation} \label{BE-mtrx}
\sigma_b=\frac{1}{1+7b}\left[ \begin{array}{cccccccc}
b&0&0&0&0&b&0&0\\ 0&b&0&0&0&0&b&0\\ 0&0&b&0&0&0&0&b\\
0&0&0&b&0&0&0&0\\
0&0&0&0&\frac{(1+b)}{2}&0&0&{\frac{\sqrt{1-b^2}}{2}}\\
b&0&0&0&0&b&0&0\\ 0&b&0&0&0&0&b&0\\
0&0&b&0&{\frac{\sqrt{1-b^2}}{2}}&0&0&\frac{(1+b)}{2}\\
\end{array} \right] \end{equation} It is interesting to
observe that for $ b=0 $, this  family of states reduce to a
separable state in 2$ \otimes $4 dimensions while it is
still entangled in the three-qubit space and the
entanglement  is restricted to the two qubits forming the
ququart.\\

Having defined the family of PPT entangled  states in
Eq.(\ref{BE-mtrx}) parameterized by `$ b $', the method that
used to experimentally detect their entanglement using a
protocol proposed in Reference~\citep{akbari-ijqi-17} is
described as follows.  Although the family of states in Eq.~(\ref{BE-mtrx}) is PPT entangled in $2 \otimes
4$-dimensional Hilbert space, it is useful to exploit the
underlying three-qubit structure.  For the detection
protocol, we define three observables $B_i$, with $i=1,2,3$
(here $B_1$ acting on the qubit space and $B_2$ and $B_3$
act in the state  space of qubits 2 and 3 forming the
ququart)

\begin{equation}\label{BE-obs}
B_1=\mathbb{I}_2 \otimes \sigma_x \otimes \sigma_x,  \;\; B_2=\mathbb{I}_2 \otimes \sigma_y \otimes \sigma_y,  \;\; B_3=\sigma_z \otimes \sigma_z \otimes \sigma_z
\end{equation}
where $ \sigma_{x,y,z} $ are the Pauli operators and $
\mathbb{I}_2 $ is the 2$ \times $2 identity operator.
Although the observables $B_j$ defined above are written in
the three qubit notation, they are bonafide observables of
the qubit-ququart system. The main result of
Reference~\citep{akbari-ijqi-17} is that any three-qubit
separable state, $ \rho_s $, obeys the four inequalities
given by 
\begin{equation}\label{BE-inequality}
\vert \langle B_1 \rangle_{\rho_s} \pm \langle B_2 \rangle_{\rho_s} \pm \langle B_3 \rangle_{\rho_s} \vert \leq 1 
\end{equation}
Therefore, if a states violates even one of the four
inequalities given in Eq.~(\ref{BE-inequality}), it has to
be entangled. It was shown numerically
in~\cite{akbari-ijqi-17} that the inequalities defined in
Equation~(\ref{BE-inequality}) can be used to detect the
entanglement present in the states $\sigma_b$ defined in
Eq.~(\ref{BE-mtrx}) for  $ 0<b<\frac{1}{\sqrt{17}}$.  Hence,
the protocol briefed above is able to detect the
entanglement of this family of  PPT entangled states in $2
\otimes 4$ dimensions.
\begin{figure}[t]
\begin{center}
\includegraphics[angle=0,scale=1]{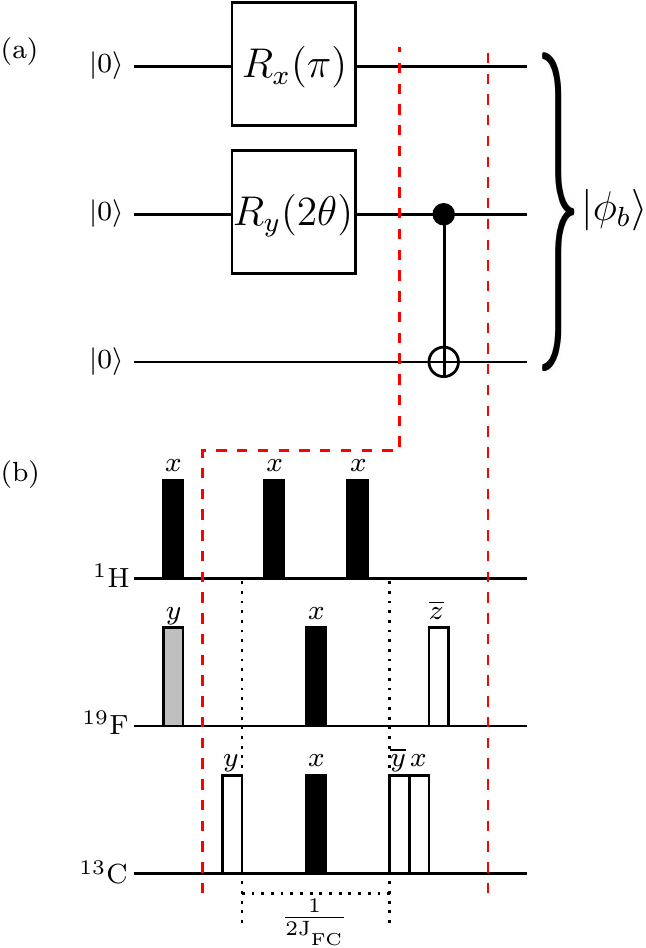}
\end{center}
\caption{(a) Quantum circuit to prepare  $ \vert
\phi_b\rangle $ from $ \vert 000 \rangle$ pseudopure state.
(b) NMR pulse sequence for quantum circuit given in (a).
Blank rectangles represent $ \frac{\pi}{2} $ RF pulses,
while black rectangles represent $ \pi $ spin-selective
rotations. The gray rectangle represents a rotation through
$ \theta=\;$Cos$^{-1}\sqrt{1+b}/\sqrt{2} $.  The phase of
each RF pulse is written above the respective pulse. A bar
over a phase implies negative phase, while the free
evolution time interval is given by $ (\rm
2J^{}_{FC})^{-1}$.} \label{ckt_ch5} 
\end{figure}

\section{Experimental Detection of 2$ \otimes $4 Bound Entanglement} 
\label{NMR-Implementation} Next is to proceed toward
procedure followed to experimentally prepare several
different states from the family of states given by
Eq.~(\ref{BE-mtrx}), detect them by measuring the
observables defined in Eq.~(\ref{BE-obs}) and check if we
observe a violation of the inequalities defined in
Eq.~(\ref{BE-inequality}).  i
\begin{figure}[t]
\begin{center}
\includegraphics[angle=0,scale=1]{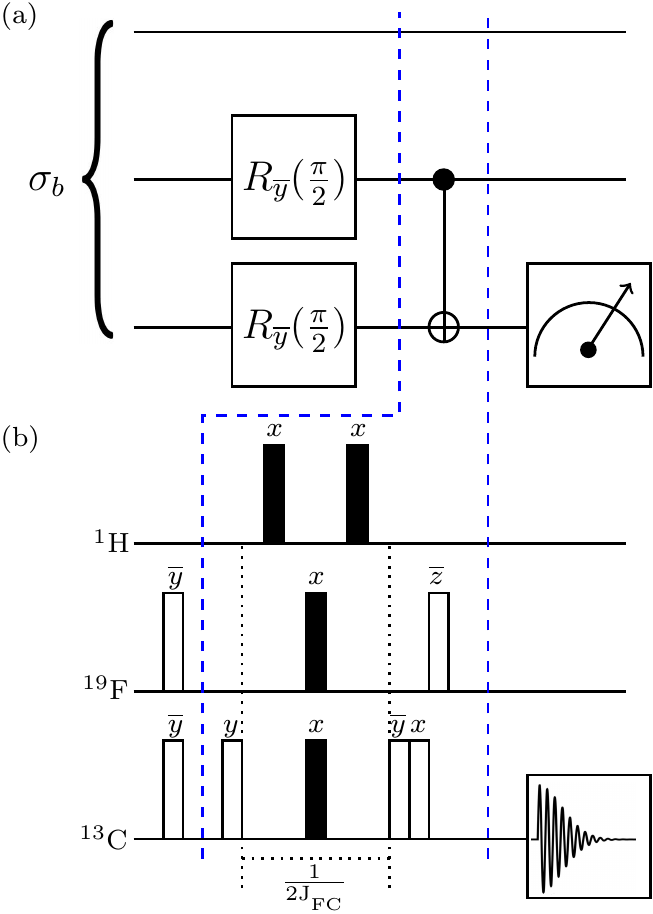}
\end{center}
\caption{(a) Quantum circuit to map $ \sigma_b $ to the state $ \sigma^{\prime}_b $ such that $ \langle B_1 \rangle_{\sigma_b}= \langle I_{3z} \rangle_{\sigma_b^{\prime}} $. (b) NMR pulse sequence to achieve the quantum circuit in (a). The unfilled rectangles denote $ \frac{\pi}{2} $ RF pulses, while the filled rectangles represent $\pi$ spin-selective RF pulses. The phase of each RF pulse is written above the respective pulse. A bar over a phase implies negative phase and the free evolution time interval is given by $ \rm \frac{1}{2J^{}_{FC}} $.} 
\label{mapping_ckt_ch5}
\end{figure}
In order to prepare the PPT entangled family of states in $2
\otimes 4$ dimensions using three qubits,  three spin-1/2
nuclei ($^1$H, $^{19}$F and $^{13}$C) were chosen to encode
the three qubits in a $^{13}$C-labeled sample of
diethylfluoromalonate dissolved in acetone-D6. See
Sec-\ref{Mapping} for NMR parameters, PPS preparation and
state mapping details.  Three dedicated channels for $ ^1
$H, $ ^{19} $F and $ ^{13} $C nuclei were employed having $
\frac{\pi}{2} $ RF pulse durations of 9.33 $\mu $s, 22.55
$\mu $s and 15.90 $\mu$s at the power levels of 18.14 W,
42.27 W and 179.47 W respectively.

The experimentally prepared bound entangled states in the current study were directly detected using the protocol discussed in Sec.~\ref{theory} and full QST \citep{leskowitz-pra-04} was also performed in each case to verify the results.\\ 

The next step was to experimentally prepare the PPT entangled family of states given in Eq.~(\ref{BE-mtrx}) (each with a fixed value of the parameter $b$) and to achieve this we utilized the method of temporal averaging~\citep{cory-physD-98}. The family of states $\sigma_b$ is an incoherent mixture of several pure states as given in Eq.(\ref{BE-1}), and the quantum circuit to prepare one such nontrivial state ($ \vert \phi_b \rangle $) is given in Fig.\ref{ckt_ch5}(a), where $ R_x(\pi) $ represents a local unitary rotation through an angle $ \pi $ with a phase $x$.  After experimentally preparing the state, one can measure the desired observable in Eq.~(\ref{BE-obs}), by mapping the state onto the Pauli basis operators.\\

The quantum circuit to achieve this is shown in Fig.\ref{mapping_ckt_ch5}(a), and this circuit maps the state $ \sigma_b \rightarrow \sigma_b^{\prime}$  such that $ \langle B_1 \rangle_{\sigma_b} = \langle I_{3z} \rangle_{\sigma_b^{\prime}} $.  The motivation for such a mapping~\citep{singh-pra-16,singh-pra-18,singh-qip-18} relies on the fact that in an NMR scenario, the expectation value  $ \langle I_{z} \rangle $, can be readily measured \citep{ernst-book-90}. The crux of the temporal averaging technique relies on the fact that the five states composing the PPT entangled state are generated via five different experiments. The states of these experiments are then added with appropriate probabilities to achieve the desired PPT entangled state. All the five states, appearing in Eq.(\ref{BE-1}), \ie\; $ \vert \phi_b \rangle $, $ \vert \psi_1 \rangle $, $ \vert \psi_2 \rangle $, $ \vert \psi_3 \rangle $ and the separable PPS state $ \vert 011 \rangle $ were experimentally prepared with state fidelities $ \geq $ 0.96. It is worthwhile to note here that $ \vert \phi_b \rangle  $ is a generalized biseparable state while $ \vert \psi_1 \rangle $ and $ \vert \psi_3 \rangle $ are LOCC equivalent biseparable states with maximal entanglement between the first and third qubits and $ \vert \psi_2 \rangle $ is a state belonging to the GHZ class. For the experimental demonstration of the 
\begin{figure}[H]
\begin{center}
\includegraphics[angle=0,scale=1]{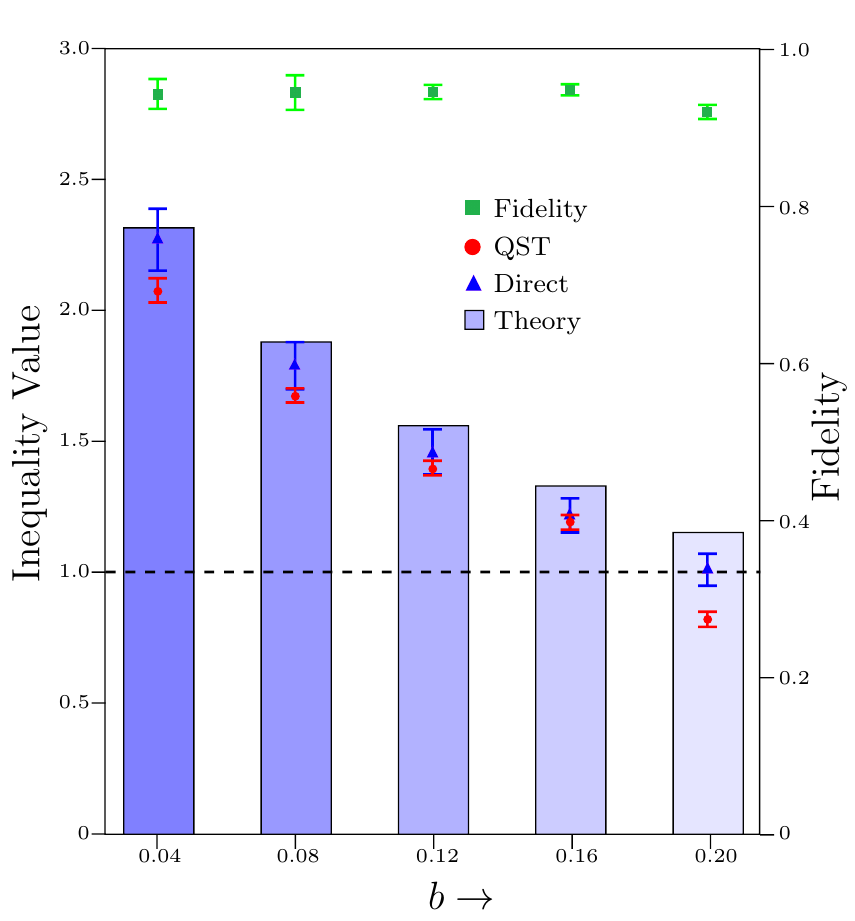}
\end{center}
\caption{Bars represent theoretically expected values, red circles are the values obtained via QST and blue triangles are the direct experimental values for the inequality appearing in Eq.(\ref{BE-obs}). Green squares are the mean experimental fidelities. Horizontal black dashed line is the reference line for states in Eq.(\ref{BE}) violating inequality of Eq.(\ref{BE-obs}).} 
\label{resultplot_ch5}
\end{figure}
\noindent detection protocol discussed in Sec.\ref{theory} values of $ b=0.04,\; 0.08,\; 0.12,\; 0.16 \; \rm and \; 0.20 $ were chosen and thereby  prepared five different PPT entangled states.  The quantum circuit as well as the NMR pulse sequence to prepare $ \vert \phi_b \rangle  $ is shown in Fig.(\ref{ckt_ch5}). Other states in Eq.(\ref{BE-1}) have similar circuits as well as pulse sequences and are not shown here.  The tomograph for one such experimentally prepared PPT entangled state, with $ b=0.04 $ and fidelity $ \rm F=0.968 $, is shown in Fig.(\ref{tomo_ch5}).\\
 
In order to measure the expectation values of the observables appearing in Eq.(\ref{BE-obs}) our earlier work~\citep{singh-pra-16,singh-pra-18} was utilized.  The idea is to unitarily map the state $ \sigma_b $ to a state say $ \sigma_b^{\prime} $, such that $ \langle  \mathbb{O} \rangle_{\sigma_b}= \langle  I_{iz} \rangle _{\sigma_b^{\prime}} $ where $ \mathbb{O} $ is one of the observables to be measured in the state $ \sigma_b $.  This is achieved by measuring $ I_{iz} $ on $ \sigma_b^{\prime} $.
\begin{table} [h]
\caption{Experimentally measured values of 
the inequality in Eq.~(\ref{BE-inequality}) 
showing maximum violation for five different PPT entangled states.}
\label{table_ch5}
\vspace*{12pt}
\centering
\renewcommand{\arraystretch}{1.5}
\begin{tabular}{c | c | c | c | c }
\hline
Obs. $\rightarrow$ &  &\multicolumn{3}{c}{Inequality value from: } \\

State(F) $\downarrow$ & $b$ & Theory & QST & Experiment\\
\hline
$\sigma_{b_1}$(0.946$\pm$0.019) & 0.04 & 2.311 &
2.061$\pm$0.046 & 2.269$\pm$0.118 \\
$\sigma_{b_2}$(0.947$\pm$0.022) & 0.08 & 1.876 &
1.660$\pm$0.027  & 1.784$\pm$0.090 \\
$\sigma_{b_3}$(0.949$\pm$0.009) & 0.12 & 1.557 &
1.382$\pm$0.028  & 1.451$\pm$0.086 \\
$\sigma_{b_4}$(0.953$\pm$0.007) & 0.16 & 1.327 &
1.179$\pm$0.028  & 1.213$\pm$0.065 \\
$\sigma_{b_5}$(0.925$\pm$0.009) & 0.20 & 1.150 &
0.807$\pm$0.029 & 1.007$\pm$0.061 \\
\hline
\end{tabular}
\end{table}


\begin{figure}[H]
\begin{center}
\includegraphics[angle=0,scale=1]{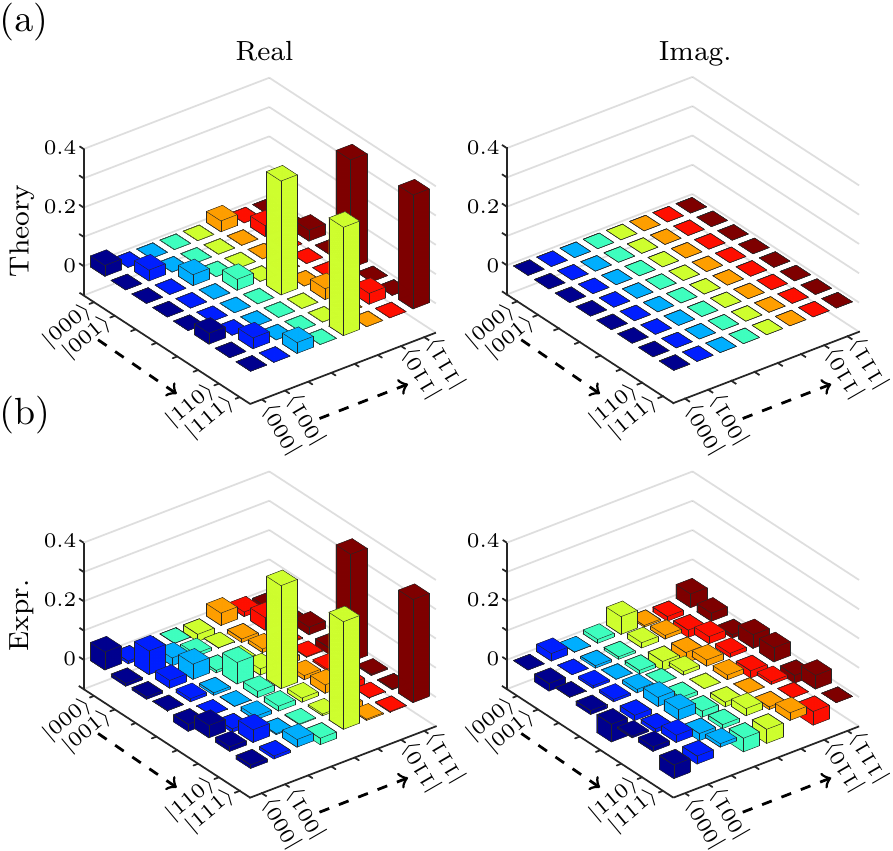}
\end{center}
\caption{Real and imaginary parts of the tomograph of the (a) theoretically expected and (b) experimentally reconstructed density operator for PPT entangled state with $b=0.04$ and state fidelity F=0.968.}
\label{tomo_ch5} 
\end{figure}

As an example, one can find the expectation value $ \langle B_1 \rangle_{\sigma_b} $ using the quantum circuit given in Fig.\ref{mapping_ckt_ch5}(a)  and the NMR pulse sequence given in Fig.\ref{mapping_ckt_ch5}(b) is implemented, followed by  a measurement of the spin magnetization of the third qubit.  Such a normalized magnetization of a qubit in the mapped state is indeed proportional to the expectation value of the $z$-spin angular momentum of the qubit~\citep{ernst-book-90}.\\

Experimentally measured values of the inequality given in
Eq.~(\ref{BE-inequality}) with maximum violation are reported
in Table-\ref{table_ch5}. For all five states with
different $b$ values, full QST was also performed and the
observables \ref{BE-obs} were analytically computed from the reconstructed density operators.  All the experimental results, tabulated in Table-\ref{table_ch5}, are plotted in Fig.~(\ref{resultplot_ch5}). All the experiments were performed several times to ensure the reproducibility of the experimental results as well as to estimate the errors reported in Table~\ref{table_ch5}. It was observed that the experimental values, within experimental error limits, agree well with theoretically expected values and validate the success of the detection protocol in identifying the PPT entangled family of states. The direct QST based measurements of the state also validate our experimental results.
\section{Conclusions}\label{concludingRemarks}
The characterization of bound entangled states is useful
since it sheds light on the relation between intrinsically
quantum phenomena such as entanglement and nonlocality. The
detection of bound entangled states is theoretically a hard
task and there are as yet no simple methods to characterize
all such states for arbitrary composite quantum systems. The
structure of PPT entangled states is rather complicated and
does not easily lead to a simple parametrization in terms of
a noise parameter. Work described in this chapter reports
the experimental creation of a family of PPT entangled
states of a qubit-ququart system  and the implementation of
a detection protocol involving local measurements to detect
their bound entanglement. Five different states which were
parametrized by a real parameter `$ b $', were
experimentally prepared (with state fidelities $ \geq $
0.95) to represent the PPT entangled family of states. All
the experiments were repeated several times to ensure the
reproducibility of the experimental results and error
estimation. In each case it was observed that the detection
protocol successfully detected the PPT entanglement of the
state in question within experimental error limits.  The
results were further substantiated via full QST for each
prepared state.  It would be interesting to create the PPT
entangled family of states using different pseudopure
creation techniques and in higher dimensions. The results of
this chapter are contained in \href{https://doi.org/10.1016/j.physleta.2019.02.027}{\rm Phys. Lett. A}, (2019), doi:10.1016/j.physleta.2019.02.027.

\chapter{Experimental Implementation of
Navascu\'es-Pironio-Ac\'{\i}n Hierarchy to Detect Quantum
Non-Locality}\label{NPA}

\section{Introduction}
This chapter reports the experimental investigations of the
non-local character of quantum correlations in a
hierarchy-based protocol.  It is well established that
quantum computation has a computational advantage over its
classical counterpart and the main resources utilized for
quantum computation are superposition, entanglement and
other quantum correlations \cite{nielsen-book-02}.
Entanglement plays a key role in several quantum
computational tasks \textit{e.g.} quantum cryptography
protocols \citep{ekert-prl-91}, quantum teleportation
\citep{bennett-prl-93}, quantum super dense coding
\citep{bennett-prl-92}, measurement-based quantum
computation \citep{briege-nature-09} and quantum key
distribution schemes \citep{barrett-prl-05, acin-njp-06}.
Creating entangled states in an experiment and certifying
the presence of entanglement in such states is of utmost
interest \citep{guhne-pr-09, horodecki-rmp-09} and
importance from the practical as well as the foundational
aspects of quantum physics. Most of the known entanglement
detection schemes rely on experimental quantum state
reconstruction \citep{horodecki-rmp-09}. It has been shown
that quantum state reconstruction is not cost-effective with
respect to experimental and computational resources
\citep{haffner-nature-05} and further, the detection of
entanglement of a known state is computationally a hard
problem \citep{brandao-prl-12} and scales exponentially with
the number of qubits. Methods to detect entanglement have
used violation of Bell-type inequalities
\citep{huber-pra-11, jungnitsch-prl-11}, entanglement
witnesses \citep{lewenstein-pra-00,guhne-jmo-03},
expectation values of the Pauli operators
\citep{zhao-pra-13, miranowicz-pra-13} as well as dynamical
learning techniques \citep{behrman-qic-08}. Although a
number of schemes exist for entanglement detection, but most
of them lack generality.\\

The motivation of the investigation is to experimentally
implement a non-local correlation and thereby devise an
entanglement detection  protocol which can readily be
generalized to higher numbers of qubits as well as to
multi-dimensional quantum systems. A promising direction is
to experimentally observe the violation of Bell-type
inequalities \citep{bell-book, bell-ppf-64, chsh-prl-69}.
This is particularly suitable as such inequalities have
recently been proposed as a general method for certifying
the non-local nature of the experimentally observed
correlations in a device-independent manner
\citep{npa-prl-07, pironio-siam-10,flavio-prx-17}. Work
described in this chapter details demonstration of the
experimental implementation of the
Navascu\'es-Pironio-Ac\'{\i}n (NPA) hierarchy to certify the
non-local correlations arising from local measurements
\citep{npa-prl-07, pironio-siam-10,flavio-prx-17} on a
three-spin system.\\

Consider a joint probability distribution $ P_{\alpha\beta} $. The question addressed in Ref.\citep{npa-prl-07} is that can there be a quantum description of $ P_{\alpha\beta} $ \ie can one have a quantum state $ \rho $, acting on the joint Hilbert space $ \mathcal{H}^{}_A \otimes \mathcal{H}^{}_B $, and the local measurement operators $ E_{\alpha}=\tilde{E}_{\alpha}\otimes I $ and $ E_{\beta}=I\otimes\tilde{E}_{\beta} $ such that
\begin{equation}\label{probabilitydistr}
P_{\alpha\beta}=\mathrm{Tr}(E_{\alpha}^{}E_{\beta}^{}.\rho)
\end{equation}
Here $ \tilde{E}_{\alpha} $ and $ \tilde{E}_{\beta} $ are local projection operators. This question can be used to design a test for the detection of non-locality from the actual probability distribution $ P_{\alpha\beta} $. In order to answer the above question, in general, one may need to search over all physical $ \rho $ and projection operators $ E_{\mu} $ which makes the problem computationally hard. A few attempts have been made to solve this problem and the first one to find the maximum violation of Clauser-Horne-Shimony-Holt (CHSH) inequality \citep{chsh-prl-69} by a quantum description was Tsirelson \citep{tsirelson-lpm-80}. Such attempts were limited to the simplest scenarios and were nowhere near generalization. Notable work has been done by Landau \citep{landau-fp-88} and Wehner \citep{wehner-pra-06}, where they have shown that the test of whether the experimental correlations arise from quantum mechanical description of nature or not, can be transformed to a semi-definite program (SDP). Solving such an SDP can reveal the local or non-local nature of the observed correlations.
\section{Brief Review of NPA Hierarchy}
\label{Theory_ch6}
In order to define SDP, consider projectors corresponding to
outcomes belonging to same measurement $M$ as E$ _{\nu} $
and E$_{\mu} $. The projectors : 
\begin{enumerate}[(i)]
\item are orthogonal \ie \; $ E_{\nu}E_{\mu}=0$ for $ \nu $,
$ \mu \in$ M, $ \mu\neq\nu $
\item sum to identity \ie \; $ \sum_{\mu\in M} E_{\mu}=I $
\item obey $ E_{\mu}^2=E_{\mu}^{\dagger}=E_{\mu} $
\item obey the commutation rule (for projectors on
subsystems A and B) as:
\begin{equation}
\label{projector_rules}
 [ E_{\alpha},E_{\beta}]=0 
\end{equation}
\end{enumerate}
It was assumed in Ref. \citep{npa-prl-07} that such a $ \rho
$ exists that satisfies Eq.(\ref{probabilitydistr}) and
(\ref{projector_rules}), and then they looked for the
implications as follows: It was observed that by taking
products of projection operators $E_{\mu} $ and linear
superposition of such products, one may define new operators
which may neither be projectors anymore nor Hermitian. Let $
S=\lbrace S_1,S_2,....,S_n \rbrace$ be a set of $n$ such
operators. There exists an $ n \times n $ matrix associated
with every such set $S$ and defined as 

\begin{equation}\label{moment-matrix}
\Gamma_{ij}=Tr(S_i^{\dagger}S_j \rho)
\end{equation}
$\Gamma $ is Hermitian and satisfies
\begin{equation}\label{sdpCons1}
\sum_{i,j} c_{ij}\Gamma_{ij}=0 \;\;\;  if \;\;\;\sum_{i,j} c_{ij}S_i^{\dagger}S_j=0
\end{equation}
\begin{equation}\label{sdpCons2}
\sum_{i,j}
c_{ij}\Gamma_{ij}=\sum_{\alpha,\beta}d_{\alpha\beta}P_{\alpha\beta}
\;\;\;  if \;\;\; \sum_{i,j}
c_{ij}S_i^{\dagger}S_j=\sum_{\alpha,\beta}d_{\alpha\beta}E_{\alpha}E_{\beta}
\end{equation}
Further, it can easily be proved that $ \Gamma $ is positive
semi-definite \ie $ \Gamma \geq 0 $ \citep{npa-prl-07}. So,
if a joint probability distribution P$ _{\alpha\beta} $ has
a quantum description \ie there exists a state $ \rho $ and
local measurement operators satisfying
Eq.(\ref{probabilitydistr}) and (\ref{projector_rules})
respectively, then finding such a state is equivalent to
finding the matrix $ \Gamma \geq 0 $ satisfying linear
constraints similar to Eq.~(\ref{sdpCons1}) and
Eq.~(\ref{sdpCons2}) and this amounts to solving an SDP
problem.\\

Having introduced the NPA hierarchy, one can move on to the protocol outlined in Ref. \citep{flavio-prx-17} that has been exploited in the current experimental study. The joint probability distribution considered is assumed to arise from local measurements on a separate state $ \rho_N $. The state $ \rho_N $ is shared among $N$ parties, each of them can
perform `$m$' measurements and each such measurement can
have `$d$' outcomes. Measurement by i$ ^{th}$ party is
represented by $ M_{x_i}^{a_i} $ with $ x_i\in \lbrace
0,...,m-1 \rbrace $ being the measurement choice and $
a_i\in \lbrace 0,...,d-1 \rbrace $ being the corresponding
outcome. By observing the statistics generated by measuring
all possible $ M_{x_i}^{a_i} $, one may write the empirical
values for the joint probability distributions
\begin{equation}\label{cond-Prob} p(a_1,...,a_N\vert
x_1,...,x_N)=Tr(M_{x_1}^{a_1}\otimes ...\otimes
M_{x_N}^{a_N}\rho_N)
\end{equation}
The correlations observed by measuring $ M_{x_i}^{a_i} $
locally, get encoded in the conditional probability
distributions having the form (\ref{cond-Prob}). Similar
expressions can be written for the reduced state probability
distribution which may arise from local measurements on a
reduced system.
\begin{equation}\label{cond-Red-Prob}
p(a_{i_1},...,a_{i_k}\vert
x_{i_1},...,x_{i_k})=Tr(M_{x_{i_1}}^{a_{i_1}}\otimes...\otimes
M_{x_{i_k}}^{a_{i_k}}\rho_{i_1.....i_k}) \nonumber
\end{equation}
with $ 0\leq i_1 <...<i_k<N $, \;$  1\leq k <N $ and $
\rho_{i_1.....i_k} $ is the reduced density operator
obtained from $\rho_N$ by tracing out an appropriate subsystem. Since we are dealing with dichotomic measurements
on qubits, it will be useful to introduce the concept of
correlators and their expectation values as follows
\begin{equation}
\langle M_{x_{i_1}}^{a_{i_1}}\otimes...\otimes
M_{x_{i_k}}^{a_{i_k}} \rangle = \sum(-1)^{\sum_{l=1}^k
a_{i_l}}p(a_{i_1},...,a_{i_k}\vert x_{i_1},...,x_{i_k})
\end{equation}
The index $k$ here dictates the order of the correlator
while $0 \leq i_1 < ... < i_k < N$ with $x_{i_j}\in \{
0,m-1\}$ and $1\leq k \leq N$. For $ k=2 $ the correlator
will be of second order of form $\langle
M_{x_{i_1}}^{(i_1)} M_{x_{i_2}}^{(i_2)}\rangle$ while for $
k=N $ one can have the full body correlator. It will be seen
later that these correlators in the simplest case turn out
to be multi-qubit Pauli operators entering the moment matrix
(Eq.(\ref{2QGamma})).
\subsection{Modified NPA Hierarchy}\label{m-NPA}
Having discussed the main features of NPA hierarchy
\citep{npa-prl-07}, the method for the detection of
non-local correlation is described as follows. Consider a
set $ O=\lbrace O_i \rbrace $ with $ 1 \leq i \leq k $ and $
O_i $ are some product of the measurement operators $
\lbrace \rm M_{x_i}^{a_i} \rbrace $ or their linear
combinations. One can associate a $k \times k$ matrix with
$O$ defined by Eq.(\ref{moment-matrix}) as $
\Gamma_{ij}=Tr(O_i^{\dagger}O_j.\rho_N ) $. For a given
choice of measurements on a separable state (a) $ \Gamma $
will be a positive semi-definite matrix, (b) Matrix elements
of $ \Gamma $ satisfy the linear constraints similar to
Eq.(\ref{sdpCons1})-(\ref{sdpCons2}), (c) Some of the matrix
element of $ \Gamma $ can be obtained by experimentally
measuring the probability distribution and (d) Some of the $
\Gamma $ matrix entries corresponds to unobservables.
Keeping these facts in mind, one can design a hierarchy
based test to see if a given set of correlations can arise
from an actual quantum realization by performing local
measurements on a separable state. One can define a set $
O_{\nu} $ consisting of products of `$ \nu $' local
measurement operators or linear superpositions of such
products. Once $ O_{\nu} $ is defined, one can look for
associated $ \Gamma \geq 0 $ satisfying constraints similar
to Eq.(\ref{sdpCons1})-(\ref{sdpCons2}) to see if a given
set of correlations can arise from actual local measurements
on a separable state. If no solution is obtained to such an
SDP then this would imply that the given set of correlations
cannot arise by local measurements on a separable quantum
state and hence the correlations are non-local. One can
always find a stricter set of constraints by increasing the
value of $ \nu $ \;\ie\; testing the nature of correlations
at the next level of the hierarchy.\\

In the experimental demonstration, as suggested in
Ref.\citep{flavio-prx-17}, the set of commutating
measurements have been used to design the SDP \ie an
additional constraint is introduced on the entries of $
\Gamma $ such that local measurements also commutate. This
additional constraint considerably reduces the original
computationally-hard problem \citep{flavio-prx-17}. All the
ideas developed till now can be understood with an example.
Consider $N=2$, two dichotomic measurements per party at the
hierarchy level $\nu =2 $. Let the measurement be labeled as
$ A_x $ and $ B_y$ with $ x,y=0,1 $. Set of operators is $
O_2=\lbrace I, A_0, A_1, B_0, B_1, A_0A_1, A_0B_0, A_0B_1,
A_1B_0, A_1B_1, B_0B_1 \rbrace $. One can see the
corresponding moment matrix $\Gamma $ can be written as 
\begin{footnotesize}
\begin{equation}
\label{2QGamma}
\hspace{-2cm}
 \Gamma=\left(
\begin{array}{lllllllllll}
1 & \textcolor{red}{\langle A_0 \rangle} &
\textcolor{black}{\langle A_1 \rangle} &
\textcolor{blue}{\langle B_0 \rangle} &
\textcolor{black}{\langle B_1 \rangle} &
\textcolor{yellow}{v_1} & \color[rgb]{0,0.58,0}{\langle A_0
B_0 \rangle} & \textcolor{purple}{\langle A_0 B_1 \rangle} &
\textcolor{orange}{\langle A_1 B_0 \rangle} &
\textcolor{black}{\langle A_1 B_1 \rangle} &
\textcolor[rgb]{0.49,0.62,0.75}{v_2} \\

 \textcolor{red}{\langle A_0 \rangle} & 1 &
\textcolor{yellow}{v_1} & \color[rgb]{0,0.58,0}{\langle A_0
B_0 \rangle} & \textcolor{purple}{\langle A_0 B_1 \rangle} &
\textcolor{black}{\langle A_1 \rangle} &
\textcolor{blue}{\langle B_0 \rangle} &
\textcolor{black}{\langle B_1 \rangle} &
\color[rgb]{0.55,0,0}{v_3} & \textcolor{green}{v_4} &
\textcolor{magenta}{v_5}\\

 \textcolor{black}{\langle A_1 \rangle} &
\textcolor{yellow}{v_1^{\ast}} & 1 &
\textcolor{orange}{\langle A_1 B_0 \rangle} &
\textcolor{black}{\langle A_1 B_1 \rangle} &
\textcolor{red}{v_6} & \color[rgb]{0.55,0,0}{v_3^{\ast}} &
\textcolor{green}{v_4^{\ast}} & \textcolor{blue}{\langle B_0
\rangle} & \textcolor{black}{\langle B_1 \rangle} &
\color[rgb]{0.55,0.55,0}{v_7}\\

\textcolor{blue}{\langle B_0 \rangle} &
\color[rgb]{0,0.58,0}{\langle A_0 B_0 \rangle} &
\textcolor{orange}{\langle A_1 B_0 \rangle} & 1 &
\textcolor[rgb]{0.49,0.62,0.75}{v_2} &
\color[rgb]{0.55,0,0}{v_3} & \textcolor{red}{\langle A_0
\rangle} & \textcolor{magenta}{v_5} &
\textcolor{black}{\langle A_1 \rangle} &
\color[rgb]{0.55,0.55,0}{v_7} & \textcolor{black}{\langle
B_1 \rangle}\\

\textcolor{black}{\langle B_1 \rangle} &
\textcolor{purple}{\langle A_0 B_1 \rangle} &
\textcolor{black}{\langle A_1 B_1 \rangle} &
\textcolor[rgb]{0.49,0.62,0.75}{v_2^{\ast}} & 1 &
\textcolor{green}{v_4} & \textcolor{magenta}{v_5^{\ast}} &
\textcolor{red}{\langle A_0 \rangle} &
\color[rgb]{0.55,0.55,0}{v_7^{\ast}} &
\textcolor{black}{\langle A_1 \rangle} &
\textcolor{blue}{v_8}\\

\textcolor{yellow}{v_1^{\ast}} & \textcolor{black}{\langle
A_1 \rangle} & \textcolor{red}{v_6^{\ast}} &
\color[rgb]{0.55,0,0}{v_3^{\ast}} &
\textcolor{green}{v_4^{\ast}} & 1 &
\textcolor{orange}{\langle A_1 B_0 \rangle} &
\textcolor{black}{\langle A_1 B_1 \rangle} &
\textcolor[rgb]{0.24,0.7,0.44}{v_9} &
\textcolor{purple}{v_{10}} & \color[rgb]{0,1,1}{v_{11}} \\

\color[rgb]{0,0.58,0}{\langle A_0 B_0 \rangle} &
\textcolor{blue}{\langle B_0 \rangle} &
\color[rgb]{0.55,0,0}{v_3} & \textcolor{red}{\langle A_0
\rangle} & \textcolor{magenta}{v_5} &
\textcolor{orange}{\langle A_1 B_0 \rangle} & 1 &
\textcolor[rgb]{0.49,0.62,0.75}{v_2} &
\textcolor{yellow}{v_1} & \color[rgb]{0,1,1}{v_{12}} &
\textcolor{purple}{\langle A_0 B_1 \rangle} \\

\textcolor{purple}{\langle A_0 B_1 \rangle} &
\textcolor{black}{\langle B_1 \rangle} &
\textcolor{green}{v_4} & \textcolor{magenta}{v_5^{\ast}} &
\textcolor{red}{\langle A_0 \rangle} &
\textcolor{black}{\langle A_1 B_1 \rangle} &
\textcolor[rgb]{0.49,0.62,0.75}{v_2^{}\ast} & 1 &
\color[rgb]{0,1,1}{v_{13}} & \textcolor{yellow}{v_1} &
\color[rgb]{0,0.58,0}{v_{14}} \\

\textcolor{orange}{\langle A_1 B_0 \rangle} &
\color[rgb]{0.55,0,0}{v_3^{\ast}} & \textcolor{blue}{\langle
B_0 \rangle} & \textcolor{black}{\langle A_1 \rangle} &
\color[rgb]{0.55,0.55,0}{v_7} &
\textcolor[rgb]{0.24,0.7,0.44}{v_9^{\ast}}  &
\textcolor{yellow}{v_1^{\ast}} &
\color[rgb]{0,1,1}{v_{13}^{\ast}} & 1 &
\textcolor[rgb]{0.49,0.62,0.75}{v_2} &
\textcolor{black}{\langle A_1 B_1 \rangle} \\

\textcolor{black}{\langle A_1 B_1 \rangle} &
\textcolor{green}{v_4^{\ast}} & \textcolor{black}{\langle
B_1 \rangle} & \color[rgb]{0.55,0.55,0}{v_7^{\ast}} &
\textcolor{black}{\langle A_1 \rangle} &
\textcolor{purple}{v_{10}^{\ast}} &
\color[rgb]{0,1,1}{v_{12}^{\ast}}
&\textcolor{yellow}{v_1^{\ast}} &
\textcolor[rgb]{0.49,0.62,0.75}{v_2^{}\ast} & 1 &
\textcolor{orange}{v_{15}}\\

\textcolor[rgb]{0.49,0.62,0.75}{v_2^{\ast}}&
\textcolor{magenta}{v_5^{}\ast} &
\color[rgb]{0.55,0.55,0}{v_7^{\ast}} &
\textcolor{black}{\langle B_1 \rangle} &
\textcolor{blue}{v_8^{\ast}} &
\color[rgb]{0,1,1}{v_{11}^{\ast}} &
\textcolor{purple}{\langle A_0 B_1 \rangle} &
\color[rgb]{0,0.58,0}{v_{14}^{\ast}} &
\textcolor{black}{\langle A_1 B_1 \rangle} &
\textcolor{orange}{v_{15}^{\ast}} & 1
\end{array}
\right)
\end{equation}
\end{footnotesize}
\noindent while following are the unassigned variables\\
\linebreak
\begin{footnotesize}
$
\begin{array} {llll}
v_1= \langle A_0A_1 \rangle, & v_2=\langle B_0B_1 \rangle,&
v_3=\langle A_0A_1B_0 \rangle,& v_4=\langle A_0A_1B_1
\rangle,\\
v_5= \langle A_0B_0B_1 \rangle, & v_6=\langle A_1A_0A_1
\rangle,& v_7=\langle A_1B_0B_1 \rangle,& v_8=\langle
B_1B_0B_1 \rangle,\\
v_9= \langle A_1A_0A_1B_0 \rangle, & v_{10}=\langle
A_1A_0A_1B_1 \rangle,& v_{11}=\langle A_1A_0B_0B_1 \rangle,&
v_{12}=\langle A_0A_1B_0B_1 \rangle, \\
v_{13}= \langle A_0A_1B_1B_0 \rangle, & v_{14}=\langle
A_0B_1B_0B_1 \rangle, & v_{15}=\langle A_1B_1B_0B_1 \rangle
&
\end{array}
$
\end{footnotesize}
\\

One may note that by introducing local measurements
commutativity \ie $ [A_0,A_1]=[B_0,B_1]=0  $ the matrix
elements, of the $ \Gamma $ matrix given by
Eq.(\ref{2QGamma}) were simplified. Particularly, the
following reduction in the number of variables can be
noticed : $v_i=v_i^{\ast}$ for $i\in [1,15]$ and
$v_6=\langle A_0 \rangle$, $v_8=\langle B_0 \rangle$,
$v_9=v_14=\langle A_0B_0 \rangle$, $v_{10}=\langle A_0B_1
\rangle$,  $v_{15}=\langle A_1B_0 \rangle$ and also
$v_{11}=v_{12}=v_{13}$. For a visual representation, the
variables that become identical because of the commutativity
constraints are represented by the same color in
Eq.(\ref{2QGamma}). The hence generated SDP will check if
the set of observed correlations $ \rm \lbrace \langle A_x
\rangle, \langle B_y \rangle, \langle A_xB_y \rangle \rbrace
$ are local. This can be achieved by substituting the
experimental values of the correlators in $ \Gamma $ matrix
and leaving the unobservables as variables. SDP will
optimize over such variables to see if a given set of
correlations are local or non-local. It has been shown
\citep{navascues-njp-08, pironio-siam-10} that this method
converges \ie\; if a given set of correlations are non-local
then the SDP will fail at a finite number of steps $\nu $.\\

\section{Tripartite Non-Local Correlation Detection}\label{tripartite-sdp} 
To experimentally demonstrate the detection of correlations which can not arise from local measurements on a separable state, a
three-qubit system was used. It has been shown
\cite{dur-pra-00} that a genuine three-qubit system can be
entangled in two inequivalent ways. CHSH scenario
\citep{chsh-prl-69} deals with (2, 2, 2) case \ie $N=2$,
$m=2$ and $d=2$. Any correlation violating CHSH inequality
exhibits non-local nature in a sense that in principle one
cannot write a local hidden variable theory which can
reproduce the observed statistics. In the current
experimental study, the scenario is (3,2,2) \ie\; three
parties with two dichotomic observables per party. The
measurements of three parties are labeled as $ A_x $ , $ B_y
$  and $ C_z $ with $ x,\;y,\;z \in [0,1] $. One can
construct set $ O_2 $ for three parties the way it was done
in the previous section for $ N=2 $. As detailed in Ref.
\citep{flavio-prx-17} to detect non-local correlations
arising from W state one needs to perform local measurements
$ M_0^{(i)}=\sigma_x $ and $ M_1^{(i)}=\sigma_z $ for all
three parties for the observables entering the moment matrix
associated with $ O_2 $ defined above. Here $ \sigma_{x/y/z}
$ are the spin-half Pauli operators. Also for GHZ type state
the measurements to generate the statistics were chosen as $
M_0^{(i)}=\sigma_x $ and $ M_1^{(i)}=\frac{\sigma_z +
\sigma_x}{\sqrt{2}} $ . A full body correlator is also
introduced while detecting non-local correlations generated
by GHZ state as such states are not suitable for detection
of non-local correlation using fewer body correlators
\citep{flavio-prx-17}.
\subsection{NMR Implementation of Non-Local Correlations Detection Scheme}\label{NMRImplementation-Ch6}
In order to experimentally demonstrate the detection of
non-local correlations, NMR hardware was used. Further, a
three nuclear spin-$\frac{1}{2}$ ensemble was utilized to
initialize the quantum system in prerequisite state on which
local measurements were performed. As already stated there
are only two inequivalent classes \citep{dur-pra-00}, under
local operations and classical communications (LOCC), of
genuine tripartite entanglement viz W-class and GHZ-class.
So the system was initialized in the representative states
of both these classes, to be tested experimentally. In order
to test the non-locality present in experimental expectation
values of the correlators following steps were followed for
a given state:

\begin{itemize}
\item Quantum system was initialized in one of the genuine
tripartite pure states.
\item It was assumed that the correlations observed from
local measurements on such states will fail SDP formulated
in Sec.-\ref{Theory_ch6} at the second level of the modified
NPA hierarchy.
\item At the second level ($ \nu=2 $) of the hierarchy, the
expectation values of all the correlators were measured
experimentally in the state under investigation.
\item Once all the observables of the moment matrix $ \Gamma
$ Eq. (\ref{moment-matrix}) were measured experimentally,
they were fed in the matrix $ \Gamma $, then the rest of the
unobservable entries were left as variables to be optimized
via SDP to achieve $ \Gamma \geq 0 $ under linear
constraints similar to Eq.(\ref{sdpCons1})-(\ref{sdpCons2})
as well as commutativity relaxation constraints $
[A_0,A_1]=[B_0,B_1]=[C_0,C_1]=0 $ to NPA hierarchy.  
\item Above formulated SDP was solved using codes available
at \cite{NPACode-github} by modifying them for (3,2,2)
scenario.
\end{itemize}


\subsection{NMR Experimental Set-up and System Initialization}
For the experimental realization $ ^{13}C $ labeled
diethylflouromalonate sample dissolved in acetone-D6 in
liquid state NMR is used. Three spin-$\frac{1}{2}$ nuclei
\ie\; $ ^1H $, $ ^{19}F $ and $ ^{13}C $ encode the qubit 1, qubit 2 and qubit 3 respectively. The free Hamiltonian of three qubit system in the rotating frame is given by
\citep{ernst-book-90}
\begin{equation}\label{NMR-Hamiltonian}
H=-\sum_{i=1}^3 \omega_i I_{iz}+2\pi\sum_{i,j=1}^3 J_{ij}I_{iz}I_{jz}
\end{equation}
with indices $i,j $=1, 2 or 3 represent the qubit
number, $ \omega_i $ is the respective chemical shift, $
I_{iz} $ being the $z$-component of spin angular momentum
and $ J_{ij} $ is the scalar coupling constant. System was initialized in the pseudopure state (PPS) $ \vert 000
\rangle $ using spatial averaging technique
\citep{cory-physD-98, mitra-jmr-07}
\begin{equation}
\rho_{_{PPS}}^{}=\frac{1-\epsilon}{2^3}\mathbb{I}_8+\epsilon\vert 000 \rangle\langle 000 \vert \nonumber
\end{equation} 
where $ \epsilon\sim 10^{-5} $ is the room temperature
thermal magnetization and $ \mathbb{I}_8 $ is 8$ \times $8
identity operator. Details of the experimental parameters,
state preparation and state mapping can be found in
Sec.-\ref{Mapping}. State mapping is used to measure the
desired correlators which happens to be Pauli operators in
the current demonstration. Quantum circuits and NMR pulse
sequences to prepare the W and GHZ states are given in
Ref.\citep{dogra-pra-15}.


\subsection{Non-Locality Detection by Experimentally Measuring the Moments/Correlators}
At the second level of the modified NPA hierarchy in (3, 2, 2) scenario the set $ O_2=\lbrace \mathbb{I}_8, A_0, A_1, B_0, B_1, C_0, C_1, A_0A_1, A_0B_0, A_0B_1, A_0C_0, A_0C_1, A_1B_0, A_1B_1, A_1C_0, A_1C_1,$
\linebreak
$ B_0B_1, B_0C_0, B_0C_1, B_1C_0, B_1C_1, C_0C_1 \rbrace $. 
\begin{figure}[H]
\begin{center}
\includegraphics[scale=1]{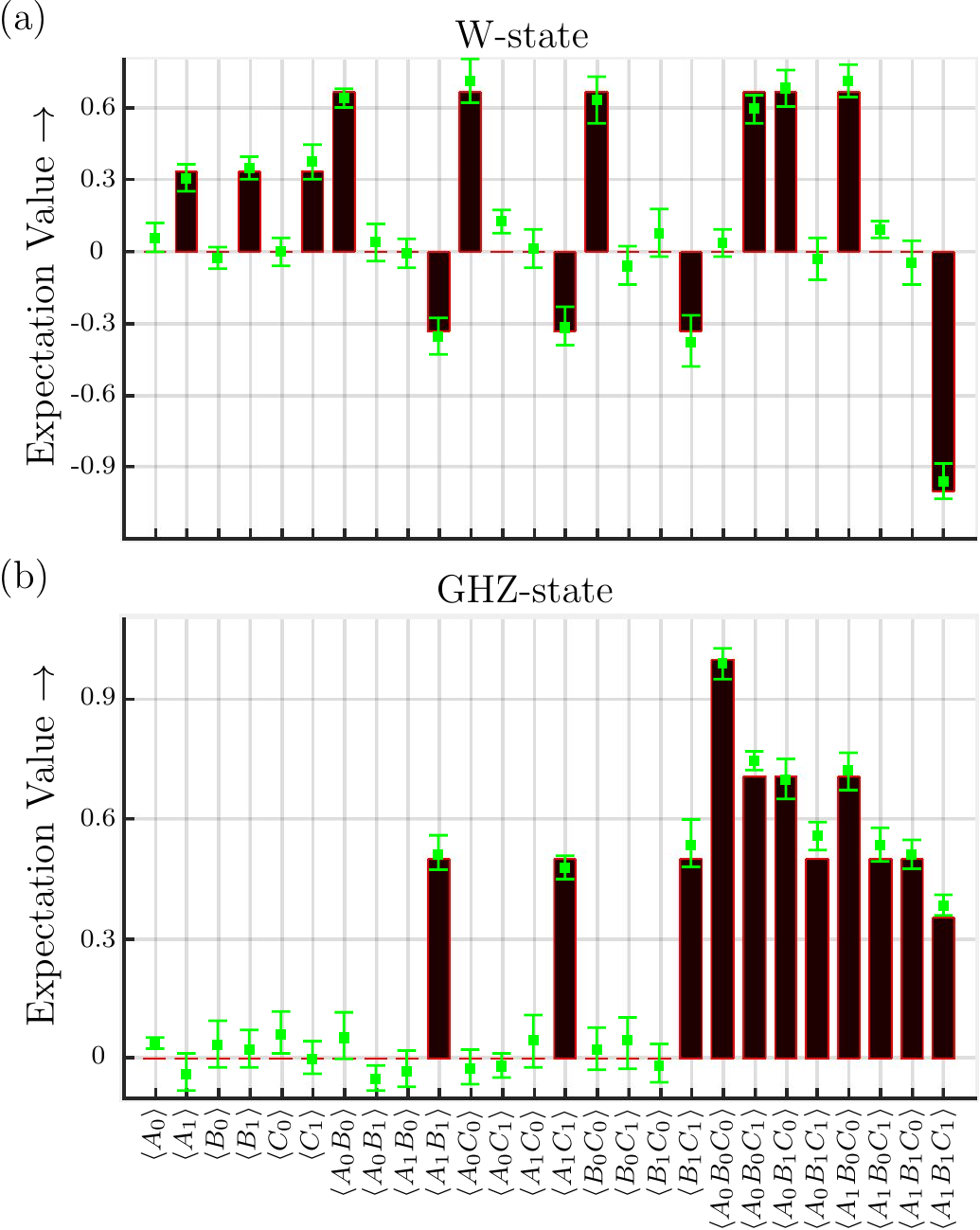}
\end{center}
\caption{Bar plots for the observable moments of the moment
matrix $\Gamma$ for (a) W-state and (b) GHZ-states. Bars
represent theoretically expected values while green squares
are the experimentally observed values.}
\label{gammaplots}
\end{figure}
\noindent The moment matrix in this case is a 22 $ \times $ 22 matrix
with all diagonal entries as 1. Further, the matrix has 26
observable moments while rest of the moments enter the
moment matrix as unobservables and were left as variables to
be optimized in SDP as detailed in Sec.-\ref{NMRImplementation-Ch6}. As an example, the moment/correlator $ \Gamma_{4,12} $, in the case of W state, is an observable $ \sigma_{1x} \sigma_{2x} \sigma_{3z} $ while the moment/correlator $ \Gamma_{1,18} $ is $ -i\sigma_{2y} $ which is not an observable and hence entered the moment matrix as a SDP variable. The next task was to find the expectation values of the correlators in the state under investigation. In NMR experiments the observed signal is proportional to the $z$-magnetization of the ensemble which indeed is proportional to the expectation value of the Pauli $z$-spin angular momentum operator in the given state. Hence the direct observable in typical NMR experiments is the Pauli $z$-operator expectation values of the nuclear spins. In recent works \citep{singh-pra-16,singh-pra-18}, schemes were developed to find the expectation values of any desired Pauli operators in the given state. This was achieved by mapping the state $ \rho\rightarrow \rho_l=U_l.\rho.U_{l}^{\dagger} $ followed by $z$-magnetization measurement. It has been shown in \citep{singh-pra-16,singh-pra-18} that the expectation value of $ \sigma_{iz} $ in state $ \rho_l $ is indeed the expectation value of the desired Pauli operator in the state $ \rho $. The explicit forms of the unitary operators $ U_l $, as well as quantum circuits and NMR pulse sequences, for two and three qubit Pauli spin operators are given in Refs. \citep{singh-pra-16} \& \citep{singh-pra-18} respectively.\\

As stated earlier the information regarding local/non-local nature of the observed correlations gets encoded in the measured correlators $ \lbrace \langle A_x \rangle,\;\langle B_y \rangle, \;\langle C_z \rangle,\; \langle A_x B_y \rangle,
\linebreak
\langle A_x C_z \rangle,\; \langle B_y C_z \rangle,\;
\langle A_x B_y C_z \rangle \rbrace $. The hence formulated SDP in both the cases, \ie W as well GHZ state, failed to find $ \Gamma \geq 0 $ at the second level of the modified NPA hierarchy. This confirmed that the observed correlations can not arise from the local measurements on a separable state and hence the states are genuinely entangled. A bar plot for the observable moments of the moment matrix $\Gamma$ for W-state and GHZ-states is depicted in Fig(\ref{gammaplots}). In both the cases the SDP was also formulated directly from experimentally reconstructed density matrices using full QST. This further verified and supported the results of modified NPA protocol obtained by the direct measurements of the correlators. It is interesting to note here the experimental protocol demonstrated here was on pure states but the scheme is also capable of detecting non-locality of states which are convex sum of white noise and pure states up to a certain degree of mixedness \citep{flavio-prx-17}.

\section{Conclusions} \label{remarks} 
Modified NPA hierarchy was used to detect the non-local
nature of the correlations performing local measurements by
means of a semi-definite program. A set comprising of
products and/or linear superpositions of such products was
defined and an associated positive semi-definite moment
matrix was also defined. Non-local correlation detection
protocols require measuring some correlator experimentally,
to generate the statistics to be tested. Once the moment
matrix embedded with the empirical data is obtained, the
semi-definite program optimizes, under some linear
constraints on the entries of the moment matrix, to see if
the observed correlations can arise from local measurements
on a separable state. The protocol has been tested
experimentally on three-qubit W and GHZ states utilizing NMR
hardware. In both the cases, the SDP successfully detected
the non-local nature of the observed correlation, as the
resulting SDP was unfeasible at the second level of the
modified hierarchy. These results were also verified by
direct full quantum state tomography. It would be
interesting to see the performance of the protocol in higher
dimension as well as more number of parties on an actual
physical system, as the structure of the entanglement
classes is much more complex in such cases.\\

The subsystems involved in our experiments reside on the same molecule and therefore, strictly speaking it is not possible to achieve a space-like separation between the events occurring in the different subsystem spaces. Therefore, the term ``local'' here pertains to subsystems and non-local implies something that goes across subsystems \ie involves operators that are go beyond subsystems and refer to joint measurements. This word of caution is important and therefore we explicitly mention it here.

\chapter{Summary and Future Outlook}\label{Summary}
\vspace{-0.8cm}
This thesis is a step further in the direction of experimental detection of quantum correlations including entanglement and discord. Most of the existing quantum correlation detection protocols require the state information beforehand to yield the detection results. Also, most such protocols are cost-intensive on experimental as well as computational resources. The thesis begins with the experimental investigation of the quantum entanglement in arbitrary bipartite states. In order to witness the entanglement the concept of entanglement witness (EW) was utilized. The key feature of the detection protocol is that it does not require any prior state information. A set of local measurements was chosen in such a way that, after measurement, they helped in the construction of EW and thereby entanglement detection utilizing semi-definite programming (SDP). It was demonstrated that only a three measurement setting sufficed to detect the entanglement in the case of maximally entangled Bell states. The protocol was also tested on a two-parameter class of qubit-qutrit entangled states and simulations suggest that only four measurement settings can successfully detect the entanglement. This work is an extension of the simulations on qutrit-qutrit entanglement detection \cite{szangolies-njp-15} and is a promising candidate for higher-dimensional bipartite entanglement detection utilizing EW through SDP. Its worth mentioning that the a critical step is to strategically choose the set of
measurement settings and this in-turn ensure the optimal
entanglement detection.

The next experimental investigation was focused on the detection of quantum correlations possessed by separable mixed states \eg non-classical correlations (NCC). A non-linear positive map was successfully implemented to detect the NCC present in a two-qubit state. The key feature of the experimental implementation was that the detection was achieved in a \textit{single-shot} NMR experiment due to the non-destructive nature of NMR measurements. The detection capabilities of the positive-map were also explored by letting the state evolve. It was observed that at the transverse relaxation time, \ie $T_2$ scale, the map was unable to detect NCC which otherwise was detected by direct quantum discord (QD) calculations. Nevertheless, this appeared to be attributed to the fact that the positive map utilized very low state information to yield `\textit{yes}/\textit{no}' answer on the status of NCC while QD requires full quantum state tomography (QST). Scheme looks promising for the experimental exploration of mixed states quantum correlations in higher-dimensional as well as multipartite quantum systems.

Entanglement detection as well as characterization in random three-qubit states was also investigated experimentally. The detection protocol was tested on seven representative states of six SLOCC-inequivalent classes and twenty randomly generated states. The entanglement measure 3-tangle was utilized to differentiate between genuine three-qubit entangled states, \ie GHZ and W class of states, in a single experiment. Only four experimental settings proved to be sufficient for the successful classification of the entanglement in the three-qubit pure generic states. Further, concurrence based three-qubit entanglement classification protocol for the most general three-qubit states was also implemented successfully. It would be interesting to explore the entanglement classification in more than three-qubit states, utilizing Pauli witness operators, as the entanglement characterization is a challenging and non-trivial task.

Mixed states possess a more subtle type of entanglement, called the bound entanglement(BE), which is undistillable  into EPR pairs via LOCC operations. Quantum states possessing BE doesn't violate PPT (positive under partial transposition) criterion. The thesis also explored a single-parameter class of qubit-ququart BE states with the aim of their detection using minimum experimental settings. The qubit-ququart states were mapped on to three-qubit states as in both the cases the underlying Hilbert space dimension is eight. BE detection method used the measurement of only three Pauli observables to see the violation of standard quantum limit (SQL). BE was successfully detected in several representative states from the single-parameter BE class of states and QST was used to establish the PPT nature of the experimentally created states. Key feature was that the BE was detected in the states lying near the entangled/separable boundary by directly measuring the Pauli observables while QST failed to detect BE.

There are entangled states whose measurement statistics can be simulated by a local hidden variable model. Bell-type inequalities \eg  Clauser-Horne-Shimony-Holt (CHSH) inequality in particular is suitable for experimental entanglement detection by observing the violation of SQL. CHSH inequality is limited to $(N=2,m=2,d=2)$ scenario \ie two-parties, two-measurement settings per party with two outcomes for each measurement setting. Experimental investigation for a general case $(N,m,d)$ was considered utilizing Navascu\'es-Pironio-Ac\'{\i}n (NPA) hierarchy. Protocol was tested on genuine three-qubit entangled states \ie GHZ and W states. It was demonstrated that the non-local nature of the quantum correlations was detected at the second level of NPA hierarchy. This method is particularly useful as it can easily be implemented in multipartite as well as higher dimensional cases.
\appendix
\chapter{Semi-Definite Program to Detect Entanglement in Random Two-Qubit States}\label{Append-A}
Following is the MATLAB script, used to detect entanglement using pre-chosen set of measurements, utilizing SDP defined in Eq.(\ref{SDP_Def}) and data given in Sec-\ref{SDP_example}. YALMIP \citep{Lofberg2004} and SeDuMi \citep{strum-oms-99} packages are required to be installed before running the following code. Following code as well as the \textbf{``mkstate''} function code are available at\\
\href{https://sites.google.com/site/amandeepsidhuiiserm/codes?authuser=0}{https://sites.google.com/site/amandeepsidhuiiserm/codes?authuser=0}.
\begin{lstlisting}[style=Matlab-editor, basicstyle=\mlttfamily\scriptsize]
% C,P and Q are initialize as SDP variables in YALMIP
C=sdpvar(16,1);
P=sdpvar(4,4);
Q=sdpvar(4,4);
% Below is the Entanglement Witness defined using only three observables 
% "mkstate" is a function to genrate Pauli operators
W=C(1,1)*mkstate('+1II',0)+(1/2)*(C(2,1)*mkstate('+1XX',0)+C(3,1)*mkstate('+1YY',0)+C(4,1)*mkstate('+1ZZ',0));
M=[1;-0.490;0.487;0.479;0;0;0;0;0;0;0;0;0;0;0;0]; % M vector built from experimental data 
cf=C'*M; % This is the cost function to be minimize
prcn=1e-5; % This will decide how negative "cf" is sufficient to establish entanglement presence
F=[W==P+pt_nonorm(Q,1,2), P>=0, Q>=0, trace(W)==1]; % "F" is the set of constraints
optimize(F,cf); % YALMIP SDP optimization of "cf" under constraints "F" using SeDuMi as solver
if value(cf)+prcn<0
    disp('-----------------------------------------------');
    disp('The State is Entangled!');
    disp('Number of Randome Local Measurements used=3');
    disp('-----------------------------------------------');
else
    C=sdpvar(16,1);
    P=sdpvar(4,4);
    Q=sdpvar(4,4);
    W=C(1,1)*mkstate('+1II',0)+(1/2)*(C(2,1)*mkstate('+1XX',0)+C(3,1)*mkstate('+1YY',0)+C(4,1)*mkstate('+1ZZ',0)+C(5,1)*mkstate('+1XY',0));
    %the M vector built from experimental data
    cf=C'*M; %this is cost function to be minimize
    F=[W==P+pt_nonorm(Q,1,2), P>=0, Q>=0, trace(W)==1]; % set of constraints
    optimize(F,cf); % YALMIP SDP usind SeDuMi as solver
if value(cf)+prcn<0
    disp('-----------------------------------------------');
    disp('The State is Entangled!');
    disp('Number of Randome Local Measurements used=4');
    disp('-----------------------------------------------');
else
    C=sdpvar(16,1);
    P=sdpvar(4,4);
    Q=sdpvar(4,4);
    W=C(1,1)*mkstate('+1II',0)+(1/2)*(C(2,1)*mkstate('+1XX',0)+C(3,1)*mkstate('+1YY',0)+C(4,1)*mkstate('+1ZZ',0)+C(5,1)*mkstate('+1XY',0)+C(6,1)*mkstate('+1XZ',0));
    %the M vector built from experimental data
    cf=C'*M; %this is cost function to be minimize
    F=[W==P+pt_nonorm(Q,1,2), P>=0, Q>=0, trace(W)==1]; % set of constraints
    optimize(F,cf); % YALMIP SDP usind SeDuMi as solver
if value(cf)+prcn<0
    disp('-----------------------------------------------');
    disp('The State is Entangled!');
    disp('Number of Randome Local Measurements used=5');
    disp('-----------------------------------------------');
else
    C=sdpvar(16,1);
    P=sdpvar(4,4);
    Q=sdpvar(4,4);
    W=C(1,1)*mkstate('+1II',0)+(1/2)*(C(2,1)*mkstate('+1XX',0)+C(3,1)*mkstate('+1YY',0)+C(4,1)*mkstate('+1ZZ',0)+C(5,1)*mkstate('+1XY',0)+C(6,1)*mkstate('+1XZ',0)+C(7,1)*mkstate('+1YX',0));
    %the M vector built from experimental data
    cf=C'*M; %this is cost function to be minimize
    F=[W==P+pt_nonorm(Q,1,2), P>=0, Q>=0, trace(W)==1]; % set of constraints
    optimize(F,cf); % YALMIP SDP usind SeDuMi as solver
if value(cf)+prcn<0
    disp('-----------------------------------------------');
    disp('The State is Entangled!');
    disp('Number of Randome Local Measurements used=6');
    disp('-----------------------------------------------');
else
    C=sdpvar(16,1);
    P=sdpvar(4,4);
    Q=sdpvar(4,4);
    W=C(1,1)*mkstate('+1II',0)+(1/2)*(C(2,1)*mkstate('+1XX',0)+C(3,1)*mkstate('+1YY',0)+C(4,1)*mkstate('+1ZZ',0)+C(5,1)*mkstate('+1XY',0)+C(6,1)*mkstate('+1XZ',0)+C(7,1)*mkstate('+1YX',0)+C(8,1)*mkstate('+1YZ',0));
    %the M vector built from experimental data
    cf=C'*M; %this is cost function to be minimize
    F=[W==P+pt_nonorm(Q,1,2), P>=0, Q>=0, trace(W)==1]; % set of constraints
    optimize(F,cf); % YALMIP SDP usind SeDuMi as solver
if value(cf)+prcn<0
    disp('-----------------------------------------------');
    disp('The State is Entangled!');
    disp('Number of Randome Local Measurements used=7');
    disp('-----------------------------------------------');
else
    C=sdpvar(16,1);
    P=sdpvar(4,4);
    Q=sdpvar(4,4);
    W=C(1,1)*mkstate('+1II',0)+(1/2)*(C(2,1)*mkstate('+1XX',0)+C(3,1)*mkstate('+1YY',0)+C(4,1)*mkstate('+1ZZ',0)+C(5,1)*mkstate('+1XY',0)+C(6,1)*mkstate('+1XZ',0)+C(7,1)*mkstate('+1YX',0)+C(8,1)*mkstate('+1YZ',0)+C(9,1)*mkstate('+1ZX',0));
    %the M vector built from experimental data
    cf=C'*M; %this is cost function to be minimize
    F=[W==P+pt_nonorm(Q,1,2), P>=0, Q>=0, trace(W)==1]; % set of constraints
    optimize(F,cf); % YALMIP SDP usind SeDuMi as solver
if value(cf)+prcn<0
    disp('-----------------------------------------------');
    disp('The State is Entangled!');
    disp('Number of Randome Local Measurements used=8');
    disp('-----------------------------------------------');
else
    C=sdpvar(16,1);
    P=sdpvar(4,4);
    Q=sdpvar(4,4);
    W=C(1,1)*mkstate('+1II',0)+(1/2)*(C(2,1)*mkstate('+1XX',0)+C(3,1)*mkstate('+1YY',0)+C(4,1)*mkstate('+1ZZ',0)+C(5,1)*mkstate('+1XY',0)+C(6,1)*mkstate('+1XZ',0)+C(7,1)*mkstate('+1YX',0)+C(8,1)*mkstate('+1YZ',0)+C(9,1)*mkstate('+1ZX',0)+C(10,1)*mkstate('+1ZY',0));
    %the M vector built from experimental data
    cf=C'*M; %this is cost function to be minimize
    F=[W==P+pt_nonorm(Q,1,2), P>=0, Q>=0, trace(W)==1]; % set of constraints
    optimize(F,cf); % YALMIP SDP usind SeDuMi as solver
if value(cf)+prcn<0
    disp('-----------------------------------------------');
    disp('The State is Entangled!');
    disp('Number of Randome Local Measurements used=9');
    disp('-----------------------------------------------');
else
    C=sdpvar(16,1);
    P=sdpvar(4,4);
    Q=sdpvar(4,4);
    W=C(1,1)*mkstate('+1II',0)+(1/2)*(C(2,1)*mkstate('+1XX',0)+C(3,1)*mkstate('+1YY',0)+C(4,1)*mkstate('+1ZZ',0)+C(5,1)*mkstate('+1XY',0)+C(6,1)*mkstate('+1XZ',0)+C(7,1)*mkstate('+1YX',0)+C(8,1)*mkstate('+1YZ',0)+C(9,1)*mkstate('+1ZX',0)+C(10,1)*mkstate('+1ZY',0)+C(11,1)*mkstate('+1XI',0));
    %the M vector built from experimental data
    cf=C'*M; %this is cost function to be minimize
    F=[W==P+pt_nonorm(Q,1,2), P>=0, Q>=0, trace(W)==1]; % set of constraints
    optimize(F,cf); % YALMIP SDP usind SeDuMi as solver
if value(cf)+prcn<0
    disp('-----------------------------------------------');
    disp('The State is Entangled!');
    disp('Number of Randome Local Measurements used=10');
    disp('-----------------------------------------------');
else
    C=sdpvar(16,1);
    P=sdpvar(4,4);
    Q=sdpvar(4,4);
    W=C(1,1)*mkstate('+1II',0)+(1/2)*(C(2,1)*mkstate('+1XX',0)+C(3,1)*mkstate('+1YY',0)+C(4,1)*mkstate('+1ZZ',0)+C(5,1)*mkstate('+1XY',0)+C(6,1)*mkstate('+1XZ',0)+C(7,1)*mkstate('+1YX',0)+C(8,1)*mkstate('+1YZ',0)+C(9,1)*mkstate('+1ZX',0)+C(10,1)*mkstate('+1ZY',0)+C(11,1)*mkstate('+1XI',0)+C(12,1)*mkstate('+1YI',0));
    %the M vector built from experimental data
    cf=C'*M; %this is cost function to be minimize
    F=[W==P+pt_nonorm(Q,1,2), P>=0, Q>=0, trace(W)==1]; % set of constraints
    optimize(F,cf); % YALMIP SDP usind SeDuMi as solver
if value(cf)+prcn<0
    disp('-----------------------------------------------');
    disp('The State is Entangled!');
    disp('Number of Randome Local Measurements used=11');
    disp('-----------------------------------------------');
else
    C=sdpvar(16,1);
    P=sdpvar(4,4);
    Q=sdpvar(4,4);
    W=C(1,1)*mkstate('+1II',0)+(1/2)*(C(2,1)*mkstate('+1XX',0)+C(3,1)*mkstate('+1YY',0)+C(4,1)*mkstate('+1ZZ',0)+C(5,1)*mkstate('+1XY',0)+C(6,1)*mkstate('+1XZ',0)+C(7,1)*mkstate('+1YX',0)+C(8,1)*mkstate('+1YZ',0)+C(9,1)*mkstate('+1ZX',0)+C(10,1)*mkstate('+1ZY',0)+C(10,1)*mkstate('+1XI',0)+C(11,1)*mkstate('+1YI',0)+C(12,1)*mkstate('+1YI',0)+C(13,1)*mkstate('+1ZI',0));
    %the M vector built from experimental data
    cf=C'*M; %this is cost function to be minimize
    F=[W==P+pt_nonorm(Q,1,2), P>=0, Q>=0, trace(W)==1]; % set of constraints
    optimize(F,cf); % YALMIP SDP usind SeDuMi as solver
if value(cf)+prcn<0
    disp('-----------------------------------------------');
    disp('The State is Entangled!');
    disp('Number of Randome Local Measurements used=12');
    disp('-----------------------------------------------');
else
    C=sdpvar(16,1);
    P=sdpvar(4,4);
    Q=sdpvar(4,4);
    W=C(1,1)*mkstate('+1II',0)+(1/2)*(C(2,1)*mkstate('+1XX',0)+C(3,1)*mkstate('+1YY',0)+C(4,1)*mkstate('+1ZZ',0)+C(5,1)*mkstate('+1XY',0)+C(6,1)*mkstate('+1XZ',0)+C(7,1)*mkstate('+1YX',0)+C(8,1)*mkstate('+1YZ',0)+C(9,1)*mkstate('+1ZX',0)+C(10,1)*mkstate('+1ZY',0)+C(10,1)*mkstate('+1XI',0)+C(11,1)*mkstate('+1YI',0)+C(12,1)*mkstate('+1YI',0)+C(13,1)*mkstate('+1ZI',0)+C(14,1)*mkstate('+1IX',0));
    %the M vector built from experimental data
    cf=C'*M; %this is cost function to be minimize
    F=[W==P+pt_nonorm(Q,1,2), P>=0, Q>=0, trace(W)==1]; % set of constraints
    optimize(F,cf); % YALMIP SDP usind SeDuMi as solver
if value(cf)+prcn<0
    disp('-----------------------------------------------');
    disp('The State is Entangled!');
    disp('Number of Randome Local Measurements used=13');
    disp('-----------------------------------------------');
else
    C=sdpvar(16,1);
    P=sdpvar(4,4);
    Q=sdpvar(4,4);
    W=C(1,1)*mkstate('+1II',0)+(1/2)*(C(2,1)*mkstate('+1XX',0)+C(3,1)*mkstate('+1YY',0)+C(4,1)*mkstate('+1ZZ',0)+C(5,1)*mkstate('+1XY',0)+C(6,1)*mkstate('+1XZ',0)+C(7,1)*mkstate('+1YX',0)+C(8,1)*mkstate('+1YZ',0)+C(9,1)*mkstate('+1ZX',0)+C(10,1)*mkstate('+1ZY',0)+C(10,1)*mkstate('+1XI',0)+C(11,1)*mkstate('+1YI',0)+C(12,1)*mkstate('+1YI',0)+C(13,1)*mkstate('+1ZI',0)+C(14,1)*mkstate('+1IX',0)+C(15,1)*mkstate('+1IY',0));
    %the M vector built from experimental data
    cf=C'*M; %this is cost function to be minimize
    F=[W==P+pt_nonorm(Q,1,2), P>=0, Q>=0, trace(W)==1]; % set of constraints
    optimize(F,cf); % YALMIP SDP usind SeDuMi as solver
if value(cf)+prcn<0
    disp('-----------------------------------------------');
    disp('The State is Entangled!');
    disp('Number of Randome Local Measurements used=14');
    disp('-----------------------------------------------');
else
    C=sdpvar(16,1);
    P=sdpvar(4,4);
    Q=sdpvar(4,4);
    W=C(1,1)*mkstate('+1II',0)+(1/2)*(C(2,1)*mkstate('+1XX',0)+C(3,1)*mkstate('+1YY',0)+C(4,1)*mkstate('+1ZZ',0)+C(5,1)*mkstate('+1XY',0)+C(6,1)*mkstate('+1XZ',0)+C(7,1)*mkstate('+1YX',0)+C(8,1)*mkstate('+1YZ',0)+C(9,1)*mkstate('+1ZX',0)+C(10,1)*mkstate('+1ZY',0)+C(10,1)*mkstate('+1XI',0)+C(11,1)*mkstate('+1YI',0)+C(12,1)*mkstate('+1YI',0)+C(13,1)*mkstate('+1ZI',0)+C(14,1)*mkstate('+1IX',0)+C(15,1)*mkstate('+1IY',0)+C(16,1)*mkstate('+1IZ',0));
    %the M vector built from experimental data
    cf=C'*M; %this is cost function to be minimize
    F=[W==P+pt_nonorm(Q,1,2), P>=0, Q>=0, trace(W)==1]; % set of constraints
    optimize(F,cf); % YALMIP SDP usind SeDuMi as solver
if value(cf)+prcn<0
    disp('-----------------------------------------------');
    disp('The State is Entangled!');
    disp('Number of Randome Local Measurements used=15');
    disp('-----------------------------------------------');
else
    disp('-----------------------------------------------');
    disp('The State is Separable!');
    disp('-----------------------------------------------');
end
end
end
end
end
end
end
end
end
end    
end    
end    
end
\end{lstlisting}
Following is the output and SDP runtime parameters yielded by above written MATLAB code. 
\begin{lstlisting}[style=Matlab-editor,  basicstyle=\mlttfamily\scriptsize]
SeDuMi 1.32 by AdvOL, 2005-2008 and Jos F. Sturm, 1998-2003.
Alg = 2: xz-corrector, theta = 0.250, beta = 0.500
Put 17 free variables in a quadratic cone
eqs m = 24, order n = 11, dim = 51, blocks = 4
nnz(A) = 69 + 0, nnz(ADA) = 576, nnz(L) = 300
 it :     b*y       gap    delta  rate   t/tP*  t/tD*   feas cg cg  prec
  0 :            1.77E+00 0.000
  1 :   7.32E-01 3.86E-01 0.000 0.2178 0.9000 0.9000   0.77  1  1  1.3E+00
  2 :   3.81E-01 1.03E-01 0.000 0.2661 0.9000 0.9000   2.07  1  1  2.6E-01
  3 :   4.80E-01 4.09E-03 0.000 0.0398 0.9900 0.9900   1.07  1  1  7.8E-03
  4 :   4.78E-01 8.64E-05 0.000 0.0211 0.9900 0.9900   1.01  1  1  1.6E-04
  5 :   4.78E-01 1.86E-06 0.000 0.0215 0.9900 0.9900   1.00  1  1  3.5E-06
  6 :   4.78E-01 3.99E-08 0.000 0.0215 0.9900 0.9900   1.00  1  1  7.6E-08
  7 :   4.78E-01 8.58E-10 0.000 0.0215 0.9900 0.9900   1.00  1  1  1.6E-09
  8 :   4.78E-01 1.84E-11 0.000 0.0215 0.9900 0.9900   1.00  1  1  3.5E-11

iter seconds digits       c*x               b*y
  8      0.5  11.1  4.7800000001e-01  4.7800000001e-01
|Ax-b| =   3.6e-11, [Ay-c]_+ =   6.7E-12, |x|=  4.2e+00, |y|=  1.3e+00

Detailed timing (sec)
   Pre          IPM          Post
3.090E-01    4.680E-01    7.200E-02    
Max-norms: ||b||=1, ||c|| = 1,
Cholesky |add|=0, |skip| = 0, ||L.L|| = 1.03389.
-----------------------------------------------
The State is Entangled!
Number of Randome Local Measurements used=3
-----------------------------------------------
\end{lstlisting}

\chapter{NMR Pulse Program for Sequential Measurements}\label{SeqFID}
Below is the 2D NMR pulse program used in sequential measurements, in order to detect NCC using Eqn.(\ref{SeqFID_eq}), introduced in Sec-\ref{chapter_ncc} and Sec-\ref{mapvalue}.
\begin{lstlisting}[style=Matlab-editor, basicstyle=\mlttfamily\scriptsize]
;zg
;avance-version (25/04/17)
;1D sequence
; Modified by: Amandeep Singh 25/04/2017
; For Sequential FID 
;$CLASS=HighRes
;$DIM=2D
;$TYPE=
;$SUBTYPE=
;$COMMENT= For Sequential FID

;#define AMAN
#include <Avance.incl>
#include <Grad.incl>
#include <De.incl>

"d13=0.000581395" ;=1/8J
"d14=d13*8"       ;=2/J
"acqt0=-p1*2/3.1416"
"d2=10u"


1 ze
2 30m 
d1


;----------------NCC State Prep. + CH BLOCK ----------------
;-----------------------------------------------------------
; Sequence to Prepare Sigma State from Thermal state from 1 to 6
; STEP 1 of sequence
    d2 pl2:f2
11  p2:f2 ph2 ; ph2=0(+x)

; STEP 2 of sequence
    d2 UNBLKGRAD
12  p16:gp1
    d16
;   d2 BLKGRAD

; STEP 3 of sequence
    d2 pl1:f1
13  p1:f1 ph3 ; ph3=1
;   d2*0.5

; STEP 4 of sequence
14  d13
; Refocusing during evolution
    p1*2:f1 ph1
    d2 pl2:f2
    p2*2:f2 ph1
    d13
; Correcting phase introduced due to refocusing;
    p2*2:f2 ph1
    d2 pl1:f1  
    p1*2:f1 ph1
    d2*0.5

; STEP 5 of sequence
15  p1:f1 ph2 ; ph2=0 (+x)
    d2*0.5

; STEP 6 of sequence
16  p1*0.5:f1 ph5 ; ph5=3 (-y)
;-----------------------------------------------------------

;Delay to see capability of CH to detect sigma state
    d10*3

;-----------------------------------------------------------
; Steps 7 to 11 are implementation of CH on Sigma state
; STEP 7 (1 of CH Sequence)
71  d2*0.5
    p1:f1 ph4 ;ph4=2
    d2*0.5
    p1:f1 ph5 ;ph5=3
    d2*0.5
    p1:f1 ph2 ;ph2=0
    d2*0.5
    p1:f1 ph3 ;ph3=1

; STEP 8  (2 of CH Sequence)
82  d13*2
; Refocusing during evolution
    p1*2:f1 ph1
    d2 pl2:f2
    p2*2:f2 ph1
    d13*2
; Phase correction introduced by refocusing
    p2*2:f2 ph1
    d2 pl1:f1
    p1*2:f1 ph1

; STEP 9  (3 of CH Sequence)
93  d2*0.5  
    p1*2:f1 ph5 ;ph5=3
    d2*0.5
    p1:f1 ph2 ;ph2=0
    d2*0.5
    p1*0.5:f1 ph5 ;ph5=3

; STEP 10  (4 of CH Sequence)
104  d13*2
; Refocusing during evolution
    p1*2:f1 ph1
    d2 pl2:f2
    p2*2:f2 ph1
    d13*2
; Phase correction introduced by refocusing
    p2*2:f2 ph1
    d2 pl1:f1
    p1*2:f1 ph1

; STEP 11  (5 of CH Sequence)
115 d2*0.5
    p1:f1 ph5 ;ph5=3
    d2*0.5
    p1:f1 ph2 ;ph2=0
    d2*0.5
    p1*1.5:f1 ph3 ;ph3=1
    d2*0.5
    p1:f1 ph4 ;ph4=2
    d2*0.5
    p1*2:f1 ph3 ;ph3=1
    d2*0.5
    p1:f1 ph2 ;ph2=0
    d2*0.5
;-------------- Acquisition Pulse---------------------------
    p1:f1 ph3 ;ph3=1
    go=5
;-----------------------------------------------------------

3  10u pl2:f2

;=======================================
; Refocusing during evolution
    p2*2:f2 ph1
    d2 pl1:f1
    p1*2:f1 ph1
    d14*104.8
; Phase correction introduced by refocusing
    p1*2:f1 ph1
    d2 pl2:f2
    p2*2:f2 ph1
;========================================

;----------Undo the pi/2 Acq Command id FID#1---------------

; Undoing was't required as undo pulse 
; cancels 1st pi/2 pulse of CNOT gate
   
;----------------------------CNOT---------------------------
;   p1:f1 ph3 ;ph3=1
    d13*2
; Refocusing during evolution
    p1*2:f1 ph1
    d2 pl2:f2
    p2*2:f2 ph1
    d13*2
; Phase correction introduced by refocusing
    p2*2:f2 ph1
    d2 pl1:f1
    p1*2:f1 ph1
  
    d2*0.5
    p1:f1 ph5 ;ph5=3
    d2*0.5
    p1:f1 ph2 ;ph2=0
    d2*0.5  
  
;-----------------------------------------------------------
  vd ; this is variable delay list with single entry of 1us 

;---------------------FID writing to ser file---------------
5 go=3 ph31
  30m wr #0 if #0 ivd
lo to 3 times td1 

exit

ph1=1
;ph1=0 2 2 0 1 3 3 1 
ph2=0
ph3=1
ph4=2
ph5=3
ph31=1
;ph31=0 2 2 0 1 3 3 1


;pl1 : f1 channel - power level for pulse (default)
;p1 : f1 channel -  high power pulse
;d1 : relaxation delay; 1-5 * T1
;ns: 1 * n, total number of scans: NS * TD0
;d13=1/8J
;d16=200u (Gradient delay)
;d2=10u
;d10= delay to relax sigma state (multiples of a=2/J)
;td1: number of experiments = number of delays in vd-list
;FnMODE: undefined
;define VDLIST
;this pulse program produces a ser-file (PARMOD = 2D)



;$Id: zg,v 1.11 2017/04/25 17:49:31 ber Exp $
\end{lstlisting}

\renewcommand{\bibname}{References} 
\bibliographystyle{phreport}
\bibliography{masterBIB}

\end{document}